\newif\ifprint

\printfalse
\ifprint
	\documentclass[a4paper,twoside,openright]{report}
\else
	\documentclass[a4paper]{report}
\fi

\newcommand{\thesistitle}{Modelling and Verifying an Object-Oriented
Concurrency Model in GROOVE}
\newcommand{\thesisauthor}{Claudio Corrodi}
\newcommand{\thesisemail}{clcorrod@ethz.ch}
\newcommand{\thesisstart}{October 10, 2014}
\newcommand{\thesisend}{April 10, 2015}
\newcommand{\thesissupervisor}{Christopher M. Poskitt \& Alexander Heußner}
\title{\thesistitle}
\author{\thesisauthor}

\usepackage{layout}
\usepackage[
	includehead,
	a4paper,
	textwidth=345pt,
	textheight=598pt,
]{geometry}

\ifprint
\usepackage{rotating}
\else
\usepackage{pdflscape}
\fi
\usepackage{adjustbox}

\usepackage{calc}
\makeatletter
\newlength{\landscapetextwidth}
\newlength{\landscapetextheight}
\newlength{\landscapefigurewidth}
\newlength{\landscapefigureheight}
\newlength{\figurewidth}
\newlength{\figureheight}

\makeatother

\makeatletter
\newenvironment{sidewaystab}%
{%
\newgeometry{%
	includehead,%
	includefoot,%
	textwidth=\textwidth,%
	textheight=\textheight%
}%
\resetHeadWidth%
\ifprint%
\begin{sidewaystable}%
\else%
\begin{landscape}%
\begin{table}%
\fi%
}%
{%
\ifprint
\end{sidewaystable}%
\else
\end{table}%
\end{landscape}%
\fi
\vfill
\clearpage
\aftergroup\restoregeometry
\resetHeadWidth
}%
\makeatother

\makeatletter
\newenvironment{sidewaysfig}%
{%
\newgeometry{%
	includehead,%
	includefoot,%
	textwidth=\landscapetextheight,%
	textheight=\landscapetextwidth%
}%
\resetHeadWidth%
\pagestyle{empty}
\ifprint%
\begin{sidewaysfigure}%
\else%
\begin{landscape}%
\begin{figure}%
\fi%
}%
{%
\ifprint
\end{sidewaysfigure}%
\else
\end{figure}%
\end{landscape}%
\fi
\vfill
\clearpage
\aftergroup\restoregeometry
\resetHeadWidth
}%
\makeatother

\usepackage[final]{microtype}
\usepackage{fancyhdr}

\makeatletter
\newcommand{\resetHeadWidth}{\fancy@setoffs}
\makeatother

\ifprint

\renewcommand{\sectionmark}[1]{\markright{\thesection\ \ #1}}
\else

\renewcommand{\sectionmark}[1]{\markright{}}
\fi

\fancypagestyle{titlepagestyle}
{%
	\fancyhfoffset{1cm}
	\lhead{\includegraphics[height = 1.2cm]{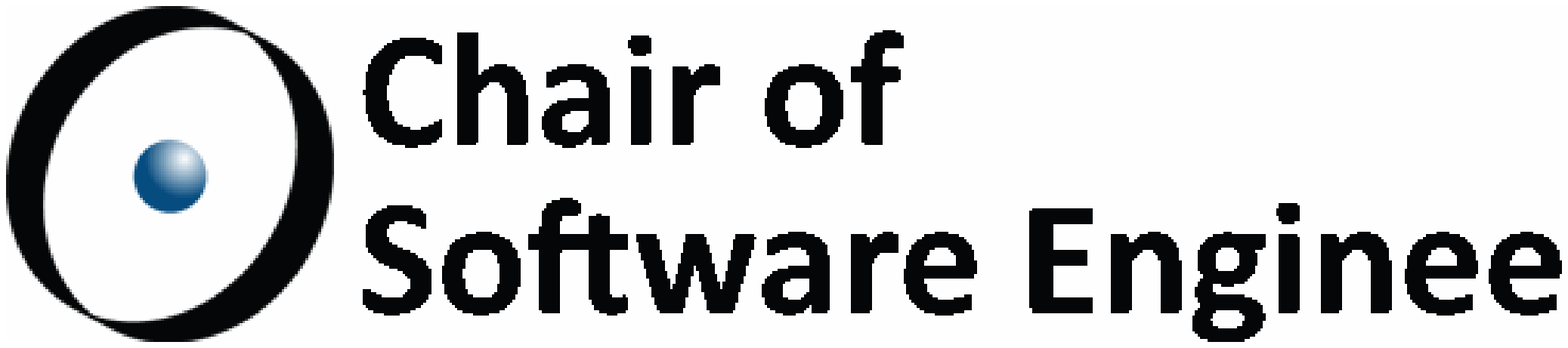}}
	\lfoot{\includegraphics[height = 1.2cm]{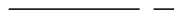}}
	\rhead{}
	\cfoot{}
	\rfoot{\includegraphics[height = 1.2cm]{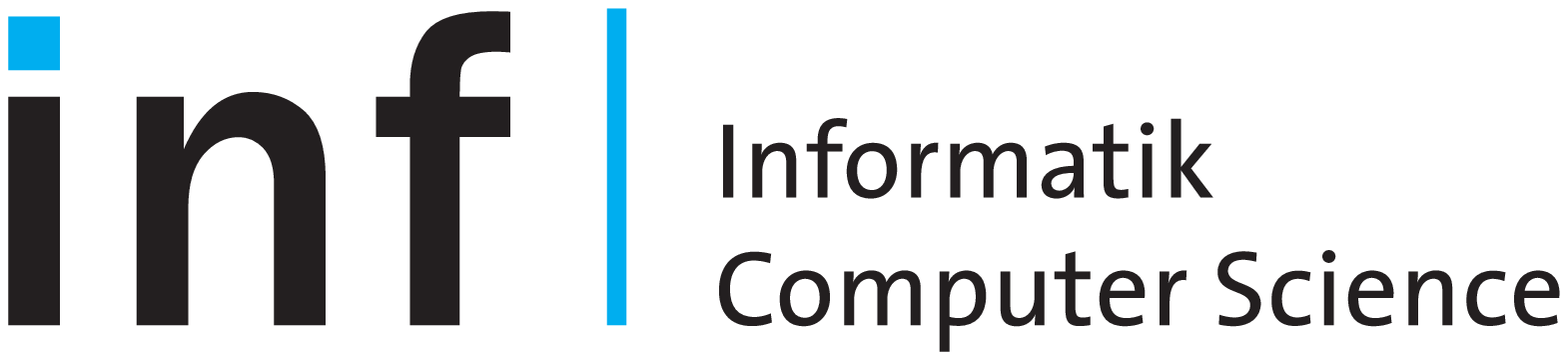}}
	
}

\usepackage{booktabs}

\usepackage{multicol}

\usepackage[inline]{enumitem}

\usepackage{subcaption}

\usepackage{todonotes}

\usepackage{xspace}

\usepackage{hyperref}
\hypersetup{%
}
\ifprint
\hypersetup{%
	pdfborderstyle={/S/U/W 1}
}
\else
\hypersetup{%
	pdfborderstyle={/S/U/W 1},
	pdftitle = \thesistitle,
	pdfauthor = \thesisauthor,
}
\fi

\usepackage[toc,nonumberlist]{glossaries}
\ifprint
\glsdisablehyper
\else
\glsdisablehyper
\fi

\renewcommand*{\CustomAcronymFields}{%
	name={\noexpand\textsc{\the\glsshorttok}},% name is abbreviated form
	description={\the\glslongtok},% description is long form
	first={\noexpand\emph{\the\glslongtok}\space(\noexpand\textsc{\the\glsshorttok})},%
	firstplural={\noexpand\emph{\the\glslongtok\noexpand\acrpluralsuffix}\space(\noexpand\textsc{\the\glsshorttok})},%
	text={\noexpand\textsc{\the\glsshorttok}},%
	plural={\noexpand\textsc{\the\glsshorttok\noexpand\acrpluralsuffix}}%
}
\SetCustomStyle
\newacronym{cpm}{cpm}{Concurrent Processor Model}
\newcommand{\cpm}{\gls{cpm}\xspace}

\newacronym{groove}{groove}{GRaphs for Object-Oriented VErification}
\newcommand{\groove}{\gls{groove}\xspace}

\newacronym{cpmo}{cpm+oo}{\textsc{cpm} with Object Orientation}
\newcommand{\cpmo}{\gls{cpmo}\xspace}

\newacronym{scoop}{scoop}{Simple Concurrent Object-Oriented Programming}
\newcommand{\scoop}{\gls{scoop}\xspace}

\newglossaryentry{coreScoop}{name={\textsc{c}ore\textsc{scoop}},description={A subset of \acrshort{scoop} focusing on concurrency features}}
\newcommand{\corescoop}{\gls{coreScoop}\xspace}

\newacronym{lts}{lts}{Labelled Transition System}
\newcommand{\lts}{\gls{lts}\xspace}

\newacronym{gps}{gps}{Graph Production System}
\newcommand{\gps}{\gls{gps}\xspace}

\newacronym{ltl}{ltl}{Linear Temporal Logic}
\newcommand{\ltl}{\gls{ltl}\xspace}

\newacronym{ctl}{ctl}{Computational Tree Logic}
\newcommand{\ctl}{\gls{ctl}\xspace}

\newacronym{dpo}{dpo}{Double-Pushout}
\newcommand{\dpo}{\gls{dpo}\xspace}

\newacronym{spo}{spo}{Single-Pushout}
\newcommand{\spo}{\gls{spo}\xspace}

\newacronym{gts}{gts}{Graph Transformation System}
\newcommand{\gts}{\gls{gts}\xspace}

\newacronym{antlr}{antlr}{ANother Tool for Language Recognition}
\newcommand{\antlr}{\gls{antlr}\xspace}

\newacronym{ebnf}{ebnf}{Extended Backus-Naur Form}

\newcommand{\statespace}{state-space\xspace}
\newcommand{\statespaces}{state-spaces\xspace}
\newcommand{\Statespace}{State-space\xspace}

\newacronym{gxl}{gxl}{Graph eXchange Language}
\newcommand{\gxl}{\gls{gxl}\xspace}

\newcommand{\nonseparate}{non-separate\xspace}

\newcommand{\xml}{\textsc{xml}\xspace}

\newacronym{gcd}{gcd}{Grand Central Dispatch}

\makeglossaries

\newcommand{\acr}[1]{\textsc{\glsentryshort{#1}}}

\usepackage[T1]{fontenc}
\usepackage[utf8]{inputenc}
\usepackage{url}
\usepackage{breakurl}
\usepackage[maxnames=4]{biblatex}
\bibliography{references.bib}
\DeclareBibliographyCategory{gam}
\addtocategory{gam}{Heussner-PCM15a}

\definecolor{darkred}{rgb}{0.55, 0.0, 0.0}

\usepackage{upquote}
\usepackage{listings}

\lstdefinestyle{basestyle}
{%
	frame=single,
	tabsize=2,
	numbers=left,
	framexleftmargin=1.5em,
	xleftmargin=2.0em,
	xrightmargin=0.5em,
	basicstyle=\footnotesize\ttfamily,
	upquote=true,
	columns=fixed,
	showstringspaces=false,
	numberstyle=\ttfamily\tiny,
	numbersep=5pt,
	extendedchars=true,
	breaklines=true,
	showtabs=false,
	showspaces=false,
	showstringspaces=false,
	identifierstyle=\ttfamily,
	keywordstyle=\ttfamily\bfseries\color[rgb]{0,0,0.6},
	keywordstyle=[2]\ttfamily\bfseries\color{darkred},
	commentstyle=\it\ttfamily\color[rgb]{0,0.4,0},
	stringstyle=\it\ttfamily\color[rgb]{0.4,0,0},
	captionpos=b,
	escapeinside={(*@}{@*)}
}

\lstdefinestyle{eiffellistingstyle}
{%
	style = basestyle,
	language = Eiffel,
	deletekeywords={separate}, % remove from Eiffel keywords
	morekeywords = {across, alias, all, and, as, check, class, interface, creation, create, debug, deferred, do, else, elseif, end, ensure, expanded, export, external, False, feature, from, frozen, if, implies, indexing, infix, inherit, inspect, invariant, is, like, local, loop, not, obsolete, old, once, or, prefix, redefine, rename, require, rescue, retry, select, some, strip, then, True, undefine, unique, until, variant, when, xor, Result, Current, Void, attached, detachable, agent},
	morekeywords=[2]{separate},
	morecomment=[l]--
}

\lstnewenvironment{eiffelcode}[1][]
{%
	\lstset{style=eiffellistingstyle,#1}
}{}

\lstnewenvironment{antlrcode}[1][]
{%
	\lstset{style=eiffellistingstyle,language={},#1}
}{}

\def\eiffelinline{\lstinline[style=eiffellistingstyle,basicstyle=\normalsize\ttfamily]}
\def\eif{\eiffelinline}

\newcommand{\eiffelfile}[2][]{\lstinputlisting[style=eiffellistingstyle,#1]{#2}}

\lstdefinestyle{latexlistingstyle}
{%
	style = basestyle,
	language=TeX,
	morekeywords={align,begin}
}

\lstnewenvironment{latexcode}[1][]
{%
	\lstset{style=latexlistingstyle,#1}
}{}

\definecolor{groove_blue}{RGB}{0,0,255}
\definecolor{groove_green}{RGB}{0,178,0}
\definecolor{groove_red}{RGB}{255,0,0}
\definecolor{groove_remark}{RGB}{255,140,0}
\newcommand{\gr}[1]{\textsf{#1}}
\newcommand{\grbf}[1]{\gr{\textbf{#1}}}
\newcommand{\grit}[1]{\gr{\textit{#1}}}

\newcommand{\grbl}[1]{\textcolor{groove_blue}{\gr{#1}}}
\newcommand{\grgr}[1]{\textcolor{groove_green}{\gr{#1}}}
\newcommand{\grre}[1]{\textcolor{groove_red}{\gr{#1}}}

\usepackage{groovetikz}
\usepackage{tikzscale}

\newcommand{\myclearpage}{\clearpage{\pagestyle{empty}\cleardoublepage}}
\makeatletter
\newenvironment{tikzbox}%
{%
\begin{adjustbox}{max width=\textwidth, max height=\totalheight}
}%
{%
\end{adjustbox}
}%

\newenvironment{sidewaystikzbox}%
{%
\begin{adjustbox}{max width=\landscapefigurewidth, max height=\landscapefigureheight}
}%
{%
\end{adjustbox}
}%
\makeatother

\usepackage[group-separator={,}]{siunitx}

\usepackage{pdfpages}

\usepackage{varioref}

\usepackage{csquotes}

\begin{document}

\pagenumbering{roman}

\begin{titlepage}
\thispagestyle{titlepagestyle}

\begin{center}
\vspace*{3.2cm}
\begin{huge}
	\thesistitle\\
\end{huge}
\vspace{0.5cm}
\begin{Large}
	Master's Thesis\\
\end{Large}
\vfill
\thesisauthor\\
ETH Zurich \\
\thesisemail \\
\vspace*{0.8cm}
\vspace{0.3cm}
\thesisstart\space--\space\thesisend
\vfill
\end{center}

\begin{flushleft}
\hspace*{5pt}Supervised by: \\
\hspace*{5pt}\thesissupervisor\\
\hspace*{5pt}Prof. Bertrand Meyer
\end{flushleft}
\vspace*{0.8cm}
\end{titlepage}
\setcounter{page}{2}

\pagestyle{empty}
\cleardoublepage

\newpage
\begin{center}
	\vspace*{2in}
	{\it In memory of Renato.}
\end{center}
\cleardoublepage

\pagestyle{empty}
\section*{Abstract}
\glsunset{scoop}
\glsunset{groove}
\scoop is a programming model and language that allows concurrent programming
at a high level of abstraction. Several approaches to verifying \scoop
programs have been proposed in the past, but none of them operate directly on
the source code without modifications or annotations.

We propose a fully automatic approach to verifying (a subset of) \scoop
programs by translation to graph-based models. First, we present a graph
transformation based semantics for \scoop.  We present an implementation of
the model in the state-of-the-art model checker \groove, which can be used to
simulate programs and verify concurrency and consistency properties, such as
the impossibility of deadlocks occurring or the absence of postcondition
violations. Second, we present a translation tool that operates on \scoop
program code and generates input for the model.  We evaluate our approach by
inspecting a number of programs in the form of case studies.

\glsreset{scoop}
\glsreset{groove}

\cleardoublepage

% Declaration of Originality
%\includepdf[pages=-,lastpage=1,pagecommand={\thispagestyle{empty}}]{misc/Declaration-Originality.eps}
\newgeometry{%
	includehead,%
	includefoot,%
	textwidth=\paperwidth,%
	textheight=\paperheight%
}%
\begin{center}
	\makebox[\textwidth]{
		\includegraphics[width=\paperwidth]{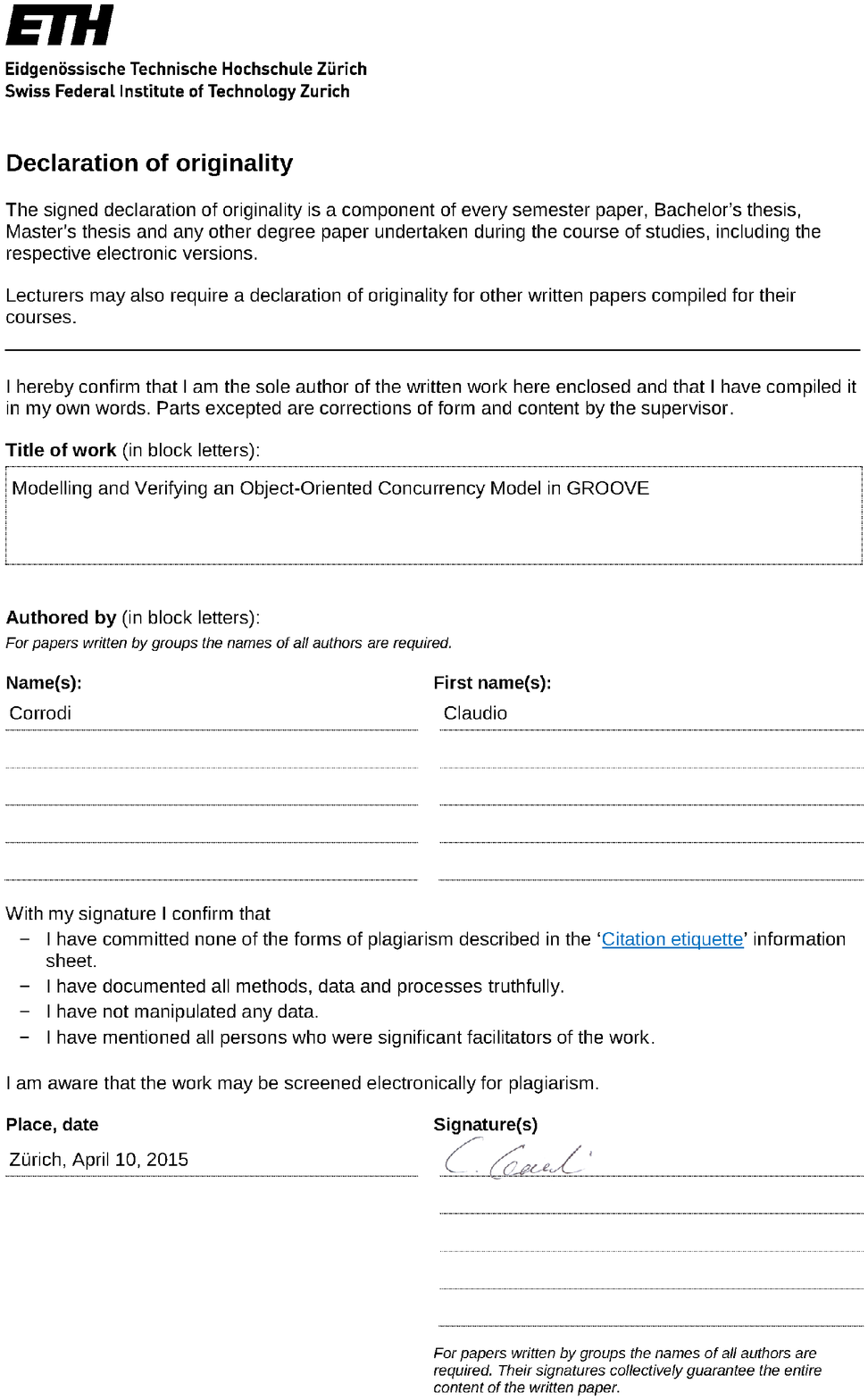}
	}
\end{center}
\restoregeometry

\pagestyle{plain}
\newgeometry{
	includehead,
	a4paper,
	textwidth=345pt,
	textheight=600pt,
}
\tableofcontents
\restoregeometry
\cleardoublepage

\section*{Acknowledgments}

I would like to thank Chris Poskitt, Alexander Heu{\ss}ner, and Bertrand Meyer
for supervising this thesis, giving me the opportunity to work on this
project, and providing helpful input in meetings and conference calls. In
particular, I would like to thank Chris Poskitt for providing advice and
support in countless discussions, meetings, and email conversations during the
past six months. I would also like to thank Benjamin Morandi for his input and
suggestions.
\\ \\
\textit{Claudio Corrodi, April 2015}

\cleardoublepage
\section*{Author's Declaration}
\nocite{Heussner-PCM15a}

I declare that the work and results presented in this thesis are my own,
except where otherwise stated. Parts of this thesis have been published in
\cite{Heussner-PCM15a}, where I am a contributing author. Details regarding
the use of the results of this publication can be found in
Section~\ref{section:introduction:related_work}.
\\ \\
\textit{Claudio Corrodi, April 2015}

\cleardoublepage

\pagestyle{fancy}
\pagenumbering{arabic}

\chapter{Introduction}
\label{chapter:intro}
\glsunset{groove}

In this chapter, we start with the motivation of this thesis and describe our
contributions, before we present the research hypothesis and list the goals
that we want to meet in order to consider this thesis a success. An overview
of the thesis structure follows, before we close this chapter with a
description of previously published work present in this thesis.

\section{Motivation}

With the shift to multiprocessor and multicore systems, concurrent and
parallel programming becomes an important part of object-oriented software
development. While object-oriented models and languages allow programming at a
high level of abstraction when writing sequential programs, they often rely on
low-level constructs for concurrency, such as locks, semaphores, and threads.
These constructs are very error prone and difficult to use correctly.
\glsreset{scoop}\scoop is a concurrency model and language that extends Eiffel
with concurrency mechanisms. The model hides low-level constructs in its
implementation, and instead provides the user with simple to use constructs
that allows concurrency to be expressed at a high level of abstraction. In
particular, lock management and thread creation is no longer expressed
explicitly.

It is still possible to introduce concurrency bugs with the high-level
constructs from \scoop, most prominently deadlock. Naturally, these bugs are
difficult to detect, as they may not occur in every program execution.  With
program verification approaches, it is possible to prove the correctness of
implementations and make sure that no concurrency related bugs exist for the
modelled semantics and input program.

Currently, there exist several formalisations of
\scoop~\cite{West:2015:ERO:2688500.2688545,MorandiSNM13,CSPSCOOP2007,Scoop:Beyond_Contracts}.
They do not focus on verification but rather resolving language ambiguities.
They can not be used for model checking, due to the \statespace explosion
problem inherent to concurrency models.  In addition, existing approaches for
the verification of \scoop
programs~\cite{DBLP:journals/corr/CaltaisM14,ModularSCOOP} focus on deadlock
prevention and work on either annotated source code or with manually
translated model input.

In this thesis, we propose an alternative approach to verify \scoop programs.
First, we present a graph-based model that focuses on the concurrency
mechanisms in \scoop, leaving out advanced object-oriented features. This
leads to a compact model with a strong formal foundation.  Second, we add
object-oriented features from \scoop to the model, obtaining an expressive
formalisation that allows representing \scoop programs more directly as model
input. The models are implemented using the state-of-the-art model checker
\groove. We then present a translation tool
that works directly on \scoop source code. With this tool, we are able to
translate a subset of \scoop programs and generate input for the model
checker. By putting these parts together, we provide a fully automatic tool
that allows verification by model checking. We focus on verification of
properties like deadlock or pre- and postcondition violations. By focusing on
the core of \scoop and abstracting away from internals of the formalisations,
we are able to reduce the \statespace sizes. We discuss why our abstractions
and optimisations do not change the expressiveness of the modelled \scoop
subset.

\section{Research Hypothesis and Contributions}
\label{section:introduction:research_hypothesis}

The research hypothesis is as follows.
\begin{displayquote}
	A subset of valid \scoop programs can be modelled using a graph
	transformation system. These programs can, without modification of the
	source code, be automatically translated to input graphs for the
	transformation system.  Using verification by model checking, it is possible
	to verify a number of properties such as absence of deadlock or absence of
	precondition violations for a given input program.
\end{displayquote}

To satisfy the hypothesis, we specify the following goals for this project:
\begin{itemize}
	\item Provide a formalisation of a subset of \scoop as a graph-based
		model using the \groove toolkit.
	\item Create a translation tool that operates on \scoop
		source code and generates input graphs for the model.
	\item Make informal soundness arguments for the translation and model.
	\item Provide a simple tool that allows verification of certain
		properties like deadlock freedom or absence of precondition failures with
		a single step by specifying \scoop source code and model parameters.
	\item Evaluate the created translation tool and graph model by inspecting
		a number of \scoop programs that use its concurrency features, as well as
		a thorough discussion of the characteristics and performance of the
		toolchain.
\end{itemize}

\section{Thesis Overview}

Chapter~\ref{chapter:scoop} gives an overview of the \scoop model and its
primary implementation.

Chapter~\ref{chapter:gts} briefly describes the theoretical background of this
thesis and gives a detailed description of
\groove, the main tool used to implement the
graph models.

Chapter~\ref{chapter:cpm} presents a formal model that focuses on the
fundamental concurrency mechanisms of \scoop and gives an in-depth description
of its implementation in \groove.

Chapter~\ref{chapter:cpmo} builds on the previous one by extending the model
with object-oriented features and discussing the extended model in detail.

Chapter~\ref{chapter:translation} discusses the design and implementation of a
translation tool that translates \scoop programs to input graphs for the model
described in Chapter~\ref{chapter:cpmo}. A discussion of future work with
respect to Chapters~\ref{chapter:cpmo} and~\ref{chapter:translation} closes
this chapter.

Chapter~\ref{chapter:case_studies} evaluates our implementations, with a focus
on the model described in Chapter~\ref{chapter:cpmo}. We discuss a number of
examples in depth, where we present source code, generated model input, and
obtained results. We look at the performance of our model in conjunction with
\groove from several angles.

Chapter~\ref{chapter:conclusion} concludes this thesis by summarising our
contributions and revisiting the research hypothesis.

The tools and programs that were written during the course of this project can
be found online at \cite{onlinerepo}.

\section{Published Work}
\label{section:introduction:related_work}

This thesis contains results published in the following paper, where the
author of this thesis is a contributing author.
\printbibliography[heading=none,category=gam]
\noindent This thesis contains results from the paper as follows.
\begin{itemize}
	\item Chapter~\ref{chapter:cpm}: The \glsreset{cpm}\cpm (which was developed
		with Poskitt and Heu{\ss}ner) is presented in the paper. The
		detailed description of its \groove
		implementation is my own work.
	\item Chapter~\ref{chapter:cpmo}: A brief overview of \cpmo has been given
		in the paper in Section 5, which I contributed to. In this thesis, we
		present the model in more detail. The augmented model in this chapter is
		my own work.
	\item Chapter~\ref{chapter:case_studies}: Programs presented in the paper
		are reused in this thesis. The \cpmo model has undergone several changes
		since writing the paper, and the results presented in this thesis are
		based upon more recent revisions of the \cpmo model.
\end{itemize}

\myclearpage

\chapter{SCOOP: An Object-Oriented Concurrency Model}
\label{chapter:scoop}

\glsreset{scoop}\scoop \cite{Nienaltowski07a} is a programming model that
provides concurrent, asynchronous, object-oriented programming. Its main
implementation is an extension to the Eiffel programming language and is
distributed with the
EiffelStudio\footnote{\url{https://www.eiffel.com/eiffelstudio/}, accessed
April 10, 2015} software. Several formalisation of the model exist, with the
most recent by Morandi \cite{Morandi2014}, which we consider in this work.

Basic knowledge of the Eiffel programming language, along with concepts like
Design By Contract, is assumed. A general introduction to the language and its
concepts can be found in \cite{Meyer:2009:TCL:1529941}.

In this chapter, we give an overview of the \scoop model and introduce a
running example for the remainder of this thesis.

\section{The SCOOP Model}

The goal of \scoop is to enable concurrent programming at a high level of
abstraction, without relying on low-level constructs like locks, semaphores,
and threads. To achieve this, \scoop adds a new keyword \eif{separate} to the
Eiffel language, which allows expressing concurrency relations between objects
as follows.

\scoop introduces the notion of a \emph{processor}, which is an abstract thread
of execution that is able to execute instructions sequentially. A processor is
the \emph{handler} of a number of objects, and object references can point
to objects that are handled by the same processor (\emph{\nonseparate}
references) or objects that are potentially handled by different processors
(\emph{separate} references). The set of objects handled by a given processor
is called a \emph{region}.

In the source code, one can annotate types (in particular in feature
declarations and formal arguments) as \eif{separate}, expressing that the
reference points to an object potentially handled by a different processor.

With the concept of processors, the semantics of feature calls are different.
If a client executes the call \eif{a.f(b1, b2, \ldots)}, with target \eif{a}
and arguments \eif{b1, b2, \ldots}, then the following cases can be
distinguished:
\begin{itemize}
	\item If the target \eif{a} is handled by the current processor, then the
		call is applied immediately.
	\item If the target \eif{a} is handled by a different processor, then the
		client logs the call with the supplier. The call is then enqueued in the
		\emph{request queue} of the handler of the supplier and processed at some
		point in the future.
\end{itemize}
In the second case, depending on whether the call is a \emph{command} (a call
that does not return a value) or a \emph{query}, the client waits for the
supplier to execute the request. In the first case, the client can continue
execution without waiting. In the second case, the client needs the result
(e.g.\ as a value in an assignment, or as a value for parameter passing), and
therefore waits until the supplier returns the value, making the call
sequential.

In order to avoid data races, \scoop only allows calls on separate targets
which are formal arguments of the enclosing routine. When executing a routine,
the \scoop runtime waits for exclusive access to the request queues of the
handlers of the separate arguments. Once the request queues are locked, the
routine starts executing and, since no other processor has access to the
locked request queues, the requests logged by the routine are guaranteed to be
executed in order and without interleaving requests from other processors.
Shared memory, another source of data races, does not exist in \scoop, since
object data can only be modified using procedures and not directly accessed
from outside (e.g.\ a statement like \eif{foo.id := 0} is forbidden if \eif{id}
is an attribute).

Contracts, i.e.\ class invariants and routine pre- and postconditions are an
integral part of Eiffel. Preconditions are Boolean assertions that must hold
before the body is executed. If a precondition does not hold, a runtime error
occurs. In \scoop, the semantics of preconditions change. While statements
involving \nonseparate objects behave like before, expressions involving
separate objects can become \emph{wait conditions}. If a wait condition does
not hold, the processor simply waits until it holds instead of generating a
runtime error. For example, a consumer might have a precondition in its
\eiffelinline{consume} routine that states that the inventory must have an
item ready, as seen in Listing~\ref{listing:groove:wait_condition}. Since the
inventory is separate and its state can be modified through requests from
other processors, the consumer simply waits until the inventory is not empty
anymore. Locks are acquired before the preconditions and wait conditions are
evaluated, but released if a wait condition does not hold yet, giving other
processors the possibility to enqueue request on the handlers of the targets.
Wait conditions are a powerful and expressive synchronization mechanism. The
lack of explicit locking makes this kind of synchronization particularly easy
to use.

\begin{eiffelcode}[
	label=listing:groove:wait_condition,
	caption={\eif{CONSUMER.consume} routine implementation.}
]
class CONSUMER
feature
	consume (a_inventory: separate INVENTORY)
			-- Consume the item held in `a_inventory'.
		require
			full_inventory: a_inventory.has_item
		do
			consumed_item := a_inventory.item
			a_inventory.remove
		end

	-- remaining class code omitted
end
\end{eiffelcode}

\section{A Running Example}
Throughout this report, we will use a running example to demonstrate the
contributions of this project, in particular translation to and verification
with our formal models in \groove.

The \emph{Dining Philosophers Problem} is a well known problem that involves
several entities interacting and is well suited for demonstrating concurrency
models. In this problem, a number of philosophers sit at a round table, with a
fork in between each pair of adjacent philosophers. The philosophers each
perform two activities in a loop: thinking and eating. In order to eat, a
philosopher needs to pick up both the left and the right fork before eating,
and put them down afterwards. The goal is to devise an algorithm that abides
these rules and does not get stuck in a deadlock.

Listing~\ref{listing:dp_intro_philosopher} shows the \eif{PHILOSOPHER} class
of a \scoop implementation (which we adapted from an implementation in the
EVE \cite{eve} source code repository) of the problem. During his time at the
table (feature \eif{live}), a philosopher eats \eif{times_to_eat} times.
Notice how there is no code handling picking up and putting down the forks.
Instead, this is done implicitly: the \eif{eat} routine takes two objects of
type \eif{separate FORK} as arguments. Once a philosopher is inside the
(empty) \eif{eat} body, he has exclusive access to the processors handling the
left and right forks, simulating picking up both forks and thus not allowing
other philosophers getting access to the forks. Since both forks are arguments
in the same routine, their respective processors are locked atomically, which
guarantees that no deadlock can occur.

While \scoop allows concurrent programming at a high level of abstraction, it
can still be difficult to spot problems related to concurrency. For example,
an unexperienced \scoop programmer may have implemented the \eif{eat} method
as shown in Listing~\ref{listing:dp_intro_bad_eat}. In this implementation, a
philosopher first picks up the left fork, and then the right one. An execution
may take place where each philosopher picks up its left fork and waits for
the right one to become available, which never happens; the program is stuck
and a deadlock has occurred.

\eiffelfile[
		label={listing:dp_intro_philosopher},
		caption={Implementation of a philosopher in \scoop.},
		escapechar=\#
	]{listings/dining_philosophers/intro_philosopher.e}

\eiffelfile[
		label={listing:dp_intro_bad_eat},
		caption={Implementation of the \eif{eat} feature that can result in a
		deadlock.}
	]{listings/dining_philosophers/intro_bad_eat.e}

\section{Related Work}

A first description of \scoop appeared in 1993
\cite{Meyer:1993:SCO:162685.162705} and an updated description was published in
1997 \cite{Meyer:1997:OSC:261119}. A prototype of \scoop has been implemented
between 2005 and 2008 at ETH Zürich, and an implementation maintained by
Eiffel Software\footnote{\url{https://www.eiffel.com/}, accessed April 10, 2015}
is currently distributed with the EiffelStudio IDE.

Since its introduction, several formalisations of \scoop have been proposed~\cite{Brooke:Scoop:CSP:Model,MorandiSNM13,Scoop:Beyond_Contracts,West:2015:ERO:2688500.2688545,Nienaltowski07a}.
We consider the work done by Morandi et al.~\cite{MorandiSNM13,Morandi2014} in
this thesis.

\myclearpage

\chapter{Graph Transformation Systems \& GROOVE}
\label{chapter:gts_groove}\label{chapter:gts}\label{chapter:groove}
\glsreset{groove}
In the course of this project, we have been working with the \groove
\cite{Rensink2003d} toolkit, which is a set of tools based on a strong formal
foundation in \glspl{gts} that can be used for modelling, simulation, and
verification. In this section, we give a short informal introduction to the
\gts theory \groove is based on and then discuss the \groove toolkit in
detail. We showcase its features by providing a \gts for our running example,
the dining philosophers problem.

A graph transformation is, informally speaking, the process of altering an
input graph to get an output graph by using rules that describe the
manipulation.  There are a number of different approaches to graph
transformation, which provide a wide range of semantics of rule applications.
From an operational standpoint, the approaches differ in how rules are defined
and in the situations in which they can be matched and applied. One such
approach is the algebraic approach, discussed in
\cite{Ehrig:2006:FAG:1121741}, which is used by the \groove toolkit.

\section{The Algebraic Approach}

In the algebraic approach to graph transformation systems, pushout
constructions (from category theory) are at the core and are used to allow
gluing graphs together. The two main approaches, \dpo and \spo, allow for a
compact and abstract representation of graph transformations. What follows is
an informal overview of these approaches.

\begin{figure}
	\centering
	\begin{tikzbox}
		\includegraphics{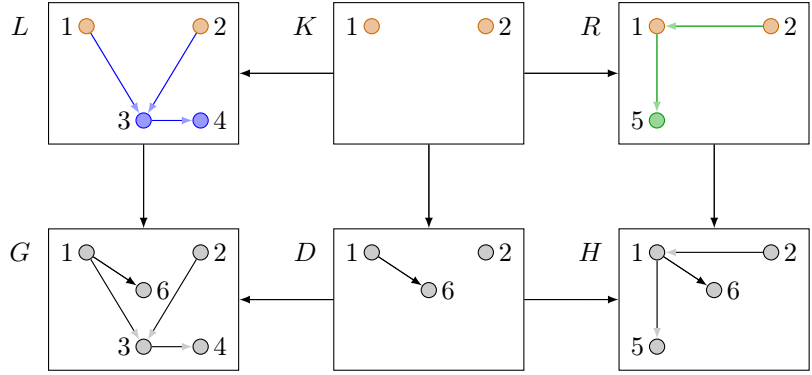}
	\end{tikzbox}
	\caption[\dpo Rule Application]{Rule application in the \dpo approach.}
	\label{fig:dpo}
\end{figure}

\paragraph{The \dpo Approach}
In the \dpo approach, a graph transformation is described as a \emph{rule}
consisting of three graphs, $L$, $K$, and $R$. The graph $K$ describes the
interface of the rule, i.e.\ the parts of the graph to be matched and
preserved. The left-hand side $L \backslash K$ of the rule describes the part
that is to be deleted, and the right-hand side $R \backslash K$ the part that
is to be created. An application of the rule described by $L$, $K$, and $R$ on
the graph $G$, shown by example in Figure~\ref{fig:dpo}, is performed by
applying the following steps.
\begin{enumerate}
	\item Find a \emph{morphism} from $L$ to $G$, that is, find a
		structure-preserving mapping from nodes and edges in $L$ to nodes and
		edges in $G$. In Figure~\ref{fig:dpo}, this mapping is expressed by node
		identifiers, where nodes in $L$ are mapped to nodes in $G$ with the same
		identifier.
	\item Construct a graph $D$ from $G$ by removing the matched edges and nodes
		in $L \backslash K$ from $G$. The combination of $L$ and $D$ at the
		interface nodes and edges from $K$ (in our example nodes $1$ and $2$) is
		called \emph{glueing} and results in $G$.
\item With a similar combination of $D$ and $R$ using the interface $K$, the
	output graph $H$ is obtained.
\end{enumerate}

\paragraph{The \spo Approach}

\begin{figure}
	\centering
	\begin{tikzbox}
		\includegraphics{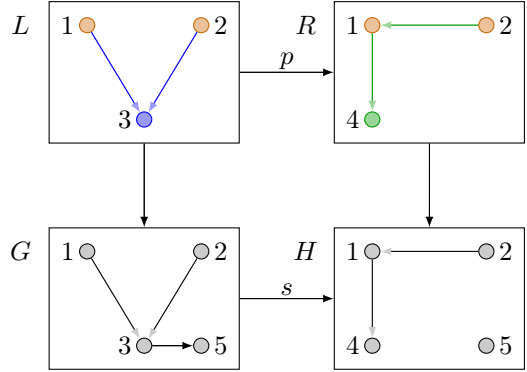}
	\end{tikzbox}
	\caption[\spo Rule Application]{Rule application in the \spo approach.
	Orange nodes denote the common interface $K$, blue edges and nodes the parts
	of $L$ that are to be deleted, and green edges and nodes of $R$ the ones that
	are to be added. Note that the edge between nodes 3 and 5 in $G$ is a
	dangling edge after the deletion of $L\backslash K$ in $G$, and is therefore
	deleted as well.}
	\label{fig:spo}
\end{figure}

In the \spo approach, specifying $K$ is omitted. Instead, a rule consists
only of the left-hand side $L$ and the right-hand side $R$. An example
application in the \spo approach is shown in Figure~\ref{fig:spo}. The
following steps are necessary to apply a rule in the \spo approach.
\begin{enumerate}
	\item Obtain the common interface $K = L \cap R$.
	\item Find a morphism from $L$ to $G$, as in the \dpo approach.
	\item Delete $L\backslash K$ from $G$, and join $R\backslash K$, using the
		common interface as glueing nodes and edges. If there are \emph{dangling
		edges} (i.e.\ edges that have a source or a target, but not both)
		remaining, delete them as well.
\end{enumerate}

They key difference between \spo and \dpo is that in the \spo approach,
dangling edges are allowed in the final graph, which is not possible in the
\dpo approach. Figure~\ref{fig:spo} shows a situation where a dangling edge
(the edge between nodes $3$ and $4$ in $G$) remains after deleting $L
\backslash K$, which is then deleted as well.

\groove allows configuring whether applications which delete dangling edges
are allowed. If so, dangling edges simply get deleted after the application to
make sure the resulting construct is a valid graph. Otherwise, when only cases
without dangling edges are allowed, \spo and \dpo are equivalent form an
operational point of view.  In our models, we never allow applications with
dangling edges, which requires us to specify all edges incident to a deleted
node on the left-hand side of a rule, ensuring that no edges get deleted ``by
accident''.

\section{GROOVE}

The \groove toolkit is written in Java and consists of a number of components.

The \textbf{Simulator} is a GUI tool that provides features to create and edit
\glspl{gps}. It is particularly useful for designing a system as it provides
immediate feedback on how the system behaves. One can apply rules to start
graphs and explore a \lts either by manually choosing rules to apply one after
another, or by automatically exploring the \statespace for a certain amount of
applications.

With a finished \gps, using the Simulator to model check various start graphs
can become cumbersome and automating the task becomes difficult. For this
scenario, the \textbf{Generator} was created, which is a command-line tool
that explores the \statespace of a given \gps. Like the Simulator, the
Generator can use different strategies for exploration, such as
breadth-first-search or depth-first-search. The Generator also allows
specifying \ltl and \ctl formulae
(a thorough
discussion of \ltl and \ctl can be found in \cite{Huth:2004:LCS:975331})
and searching for counterexamples.  The
Generator provides various metrics such as the size of the \lts, feedback about
\ltl and \ctl formulae, and profiling information.

Other components that can be used as standalone applications are included in
the above two tools. The \textbf{Model Checker} can be used to verify \ltl and
\ctl properties for labelled transition systems created by the Generator, but
is included in the Generator as well. The \textbf{Viewer} is a simple GUI tool
that can render graphs from a \gps and is used as part of the Simulator.

\subsection{Graph Production Systems}

\groove stores its \acrlong{gps} in \verb|.gps| folders. Such a
folder consists of the following components (stored as individual files).
\begin{itemize}
	\item Production rules are stored as \verb|.gpr| files and encode graph
		transformations in the sense of the \spo approach. They are
		rendered as a single graph using colour codes to distinguish left-hand
		side, interface, and right-hand side, as well as other properties of the
		rule.
	\item Type graphs are stored as \verb|.gpy| files. If active, \groove only
		allows using rules and start graphs that conform to them.  Multiple type
		graphs can be active at a time.
	\item Start graphs are stored as \verb|.gst| files and represent stating
		points for the exploration.
	\item The \verb|system.properties| file contains a number of configuration
		entries, most notably whether dangling edges should be allowed, the name
		of the active start graph, the active type graphs, and the exploration
		strategy to be used.
		
		An important system property is whether rules can be matched injectively
		or not. If so, distinct nodes that are matched from a source must have a
		distinct node in the target graph. Otherwise, multiple nodes in the rule
		can be mapped to the same node in the target graph. The configuration of
		this property can be overridden for individual rules.
\end{itemize}

The individual files conform to the \gxl file format, which is an \xml format
that specifies graph information. It is used in \groove to store
individual graphs and associated properties in the files mentioned above.
Using an \xml representation of \glspl{gps} makes pre- and postprocessing of
\groove input and output respectively very accessible and easy to handle.

To illustrate how the various components of a \gps work together, we model the
dining philosophers problem as a simple \gps in \groove. Note that the
representation in this section is unrelated to \scoop or the formal models we
introduce in Chapters~\ref{chapter:cpm} and~\ref{chapter:cpmo}, and
instead is a standalone model of the problem.

\subsection{Type Graphs}

Type graphs determine the form of other graphs in the system, in particular
the form of rules and start graphs. While the feature is optional, it is
rather useful when working on a system, as graphs that do not conform to the
specified type graph are highlighted in the Simulator, which helps to detect
typos and other errors.

\begin{figure}
	\centering
	\begin{tikzbox}
		\includegraphics{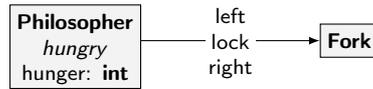}
	\end{tikzbox}
	\caption[Dining philosophers type graph]{Type graph of the dining
	philosophers \gts.}
	\label{fig:gts:type_graph_dp}
\end{figure}

Figure~\ref{fig:gts:type_graph_dp} shows the type graph for a \groove
solution to the dining philosophers problem.  It specifies that a philosopher
can be \grit{hungry} (using an optional node flag) and has a \gr{hunger}
integer value attached.  The only edges in this system are edges from
philosophers to forks. Not only has a philosopher edges to its \gr{left} and
\gr{right} forks, but it can also have a lock on them, expressed by the
\gr{lock} edge, indicating that a philosopher has picked up the forks.

\subsection{Graph Representation}

The \groove Simulator augments graph representations from a simple directed
graph with edge labels to a more compact, readable format. As mentioned
earlier, rules are represented as one single graph with nodes and edges of
different kinds of nodes and edges (in particular, readers, erasers, creators,
embargoes, and conditional creators).

Figure~\ref{fig:groove:dp_start} shows the start graph for a configuration of
the dining philosophers problem with four philosophers. On the left-hand side,
the start graph is shown as a directed graph with labelled edges. In the
middle, the condensed form that \groove uses is shown, where self-edges are
collapsed into the nodes, and on the right-hand side the graph is rendered in
\groove with internal node identifiers (note that they are not a part of the
model). We use the \groove representations of graphs throughout this report,
as the resulting graphs are intuitive and contain the same information as
before.  We occasionally enable node identifiers in order to be able to refer
to individual nodes easily.

In this model of the dining philosophers problem, each node that has
a self-edge labelled \verb|type:Philosopher|---we say ``a node of type
\verb|Philosopher|'' in this case---has two outgoing edges to nodes of type
\verb|Fork|, one of them labelled \verb|left| and the other one labelled
\verb|right|. In addition, philosophers contain an integer value denoting the
amount of times a philosopher wants to eat, encoded by the self-edges labelled
\verb|let:hunger=2|. The goal is to find rules that model the behaviour of the
philosophers, namely grabbing the forks, eating, and putting them back down.

\begin{sidewaysfig}
	\centering
	\begin{subfigure}[t]{0.33\textwidth}
		\centering
		\begin{tikzbox}
			\includegraphics{figures/groove_start_dp_1.tikz}
		\end{tikzbox}
		\caption{Dining Philosophers start graph as a directed graph with edge
		labels.}
	\end{subfigure}
	\begin{subfigure}[t]{0.33\textwidth}
		\centering
		\begin{tikzbox}
			\includegraphics{figures/groove_start_dp_2.tikz}
		\end{tikzbox}
		\caption{The same start graph, as rendered by \groove.}
	\end{subfigure}
	\begin{subfigure}[t]{0.33\textwidth}
		\centering
		\begin{tikzbox}
			\includegraphics{figures/groove_start_dp_3.tikz}
		\end{tikzbox}
		\caption{Start graph rendered in \groove with node identifiers.}
	\end{subfigure}
	\caption[Start Graph Comparison]{Comparison of a start graph
		as a directed graph with edge labels (left) and as rendered in \groove
		(middle and right). In \groove, self-edges are collapsed and displayed
		inside the node. In addition, certain values are rendered differently,
		e.g.\ \gr{type:} prefixes are omitted but the following value is printed
		in bold.}
	\label{fig:groove:dp_start}
\end{sidewaysfig}

\begin{figure}
	\centering
	\begin{tikzbox}
		\includegraphics{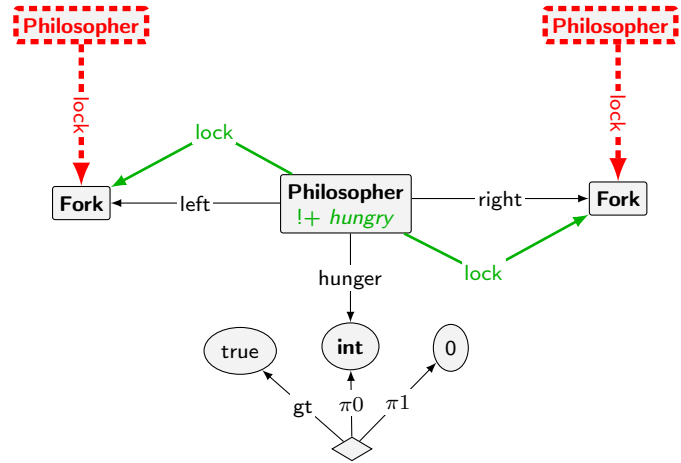}
	\end{tikzbox}
	\caption[Rule \texttt{pick\_up}]{Picking up forks. A philosopher with
		positive hunger \grgr{locks} both forks if none of them is locked by
		another philosopher. The \textit{\grgr{hungry}} flag is created to
		indicate that the philosopher has not eaten yet during this round.}
	\label{fig:groove:dp_rule_pick_up}
\end{figure}

\begin{figure}
	\centering
	\begin{tikzbox}
		\includegraphics{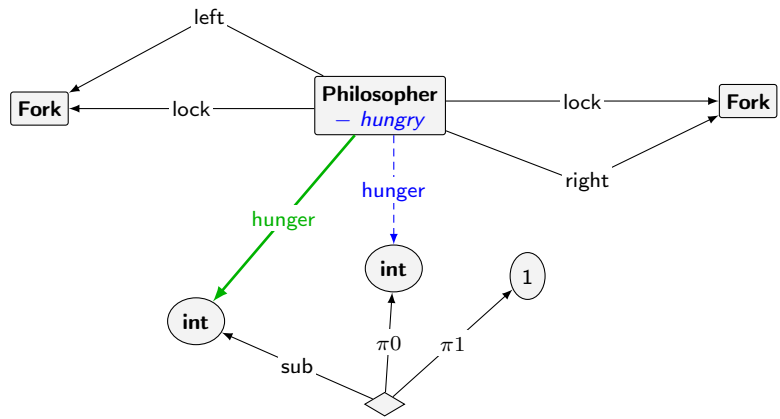}
	\end{tikzbox}
	\caption[Rule \texttt{eat}]{A philosopher eats if he is hungry. In the
		process, the \textit{\grbl{hungry}} flag (self-edge with label
		\gr{flag:hungry}) is removed, and a product node is used to decrease
		the hunger value by one.}
	\label{fig:groove:dp_rule_eat}
\end{figure}

\begin{figure}
	\centering
	\begin{tikzbox}
		\includegraphics{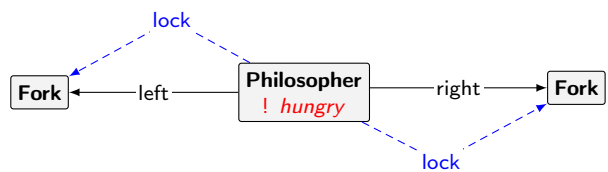}
	\end{tikzbox}
	\caption[Rule \texttt{put\_down}]{A philosopher puts down his forks if he
		has them in his hands (\grbl{lock} edges) and is not hungry (embargo on
		the \textit{\grre{hungry}} flag).}
	\label{fig:groove:dp_rule_put_down}
\end{figure}

Figures~\ref{fig:groove:dp_rule_pick_up}, \ref{fig:groove:dp_rule_put_down},
and~\ref{fig:groove:dp_rule_eat} show the rules for picking up the forks,
eating, and putting down the forks. These rules showcase various kinds of node
and edge types, in particular the following.
\begin{description}
	\item[Reader] Edges and nodes that are displayed in black are the ones that
		are present in both sides of the rule. These are matched and preserved
		when applying the rule.
	\item[Creator] Creator edges and nodes are only present in the right-hand
		side of the rule, which means that they are not required for the rule to
		match but will be created upon application.
	\item[Embargo] These edges and nodes express a negative application
		condition. The rule only matches, if there is no match for these edges and
		nodes in the source graph. For example, in the pick up rule
		(Figure~\ref{fig:groove:dp_rule_pick_up}) , we express with embargo nodes
		and edges that a philosopher should only lock both forks if they are not
		already locked.
	\item[Eraser] Finally, eraser edges and nodes (dashed blue) are elements
		that are only present on the left-hand side of the rule, which means that
		they are required for matching but will be deleted when the rule is
		applied.
	\item[Operations] Arithmetic operations can be performed using product nodes
		(rhombus shaped nodes), which take a number of arguments (via $\pi$ edges)
		and point to a result node (operation edge, such as \verb|gt| for greater
		than). Figure~\ref{fig:groove:dp_rule_pick_up} shows how the greater than
		operation can be used to enforce that the rule only matches if the hunger
		value is greater than zero.
\end{description}

\begin{figure}
	\centering
	\begin{tikzbox}
		\includegraphics{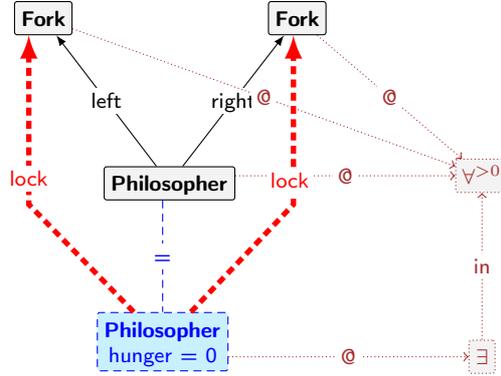}
	\end{tikzbox}
	\caption[Rule \texttt{leave}]{The rule for leaving the table matches, if
		all philosophers (expressed by the forks and the philosopher reader
		node) are in a state where the hunger is zero and they do not have locks
		on forks, in which case the \textbf{\grbl{Philosopher}} gets deleted.}
	\label{fig:groove:dp_rule_leave}
\end{figure}

More complex statements can be made with the help of \emph{quantifiers}.
Suppose the philosophers want to leave the table after having eaten the number
of times that they wanted to. Since philosophers are polite, they do not leave
until everyone at the table has finished.
Figure~\ref{fig:groove:dp_rule_leave} shows a rule that achieves this. We
first match all \grbf{Philosopher} nodes that have a left and a right fork
attached with the $\forall^{>0}$ quantifier (which denotes that in order for the
rule to apply, the subgraph ``at'' this quantifier (attached via \gr{@}
edges) has to match at least once, as opposed to the $\forall$ quantifier where the
rule can match zero occurrences of the subgraph). Then we require that there
must exists a \grbf{Philosopher} node (dashed blue, attached to the $\exists$
node via \gr{@} edge which in turn is nested inside the $\forall^{>0}$ quantifier),
which is in fact the same as the reader \grbf{Philosopher} node (expressed
with the \gr{=} edge) and where its hunger value is equal to zero.

Since one of the \grbf{Philosopher} nodes is an eraser node, the matching
philosophers get deleted upon rule application, modelling the collective
leaving of the table.

\subsection{Rule Priorities}

In a graph state where more than one rule is applicable, it may be desirable
to force a certain order when exploring the \statespace. \groove allows
controlling the order of rule applications. Using simple rule
priorities---integer values associated with a rule---a system applies rules
with higher priorities before rules with lower priorities.  For example, if
our philosophers do not like to eat alone, we could assign the rule
\gr{pick\_up} a higher priority than the rule \gr{eat}, which means that as
long as there are philosophers which are able to pick up their forks, they do
so, and once no additional philosopher can pick up its forks anymore are the
ones currently having the forks allowed to eat.

\groove also provides more advanced mechanisms for controlling rule
applications. In particular, control programs allow specifying complex
expressions with conditionals, loops, choices, and other control flow
mechanisms. Since we do not use control programs in this work, we do not
discuss them here. Instead, we refer the interested reader to
\cite{groove_user_manual}.

\subsection{Verification by Model Checking}
\begin{figure}
	\centering
	\begin{tikzbox}
		\includegraphics{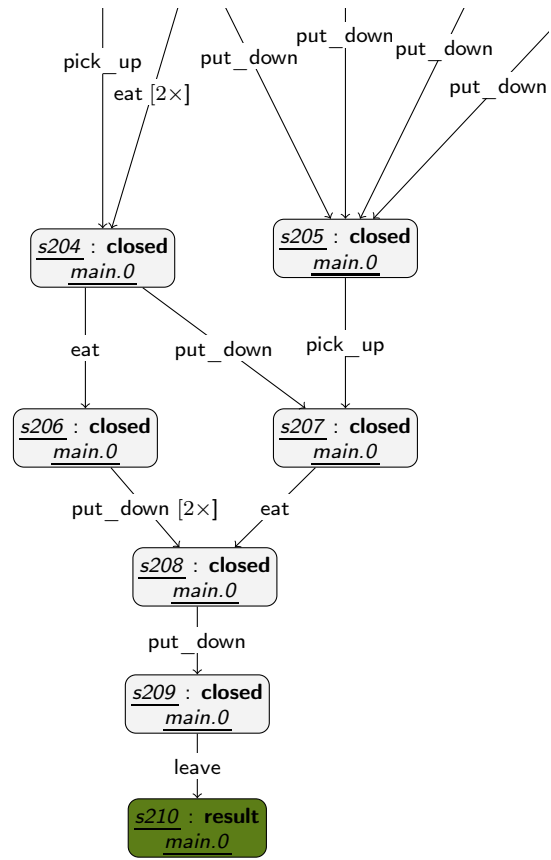}
	\end{tikzbox}
	\caption[\groove \lts excerpt]{Excerpt from \lts of the dining philosophers
		model. After the last philosopher has put down his forks (transition from
		$s208$ to $s209$), all philosophers leave and the simulation has reached a
		final state ($s210$) where no more rule applications are possible.}
	\label{fig:groove:dp_lts_excerpt}
\end{figure}

Now that we have modelled the dining philosophers problem, we can generate the
\statespace and inspect it. \groove has many options for \statespace
exploration, for example it can do breadth-first search, depth-first search,
random linear exploration, and other exploration types. A state is called
\emph{final}, if no modifying rule (a rule with erasers or creators) is
applicable anymore. If we want to show that the dining philosophers example
always results in the philosophers leaving the table (i.e.\ applying the
\texttt{leave} rule before being in a final state), we could do this by
generating the full \statespace and inspecting paths and rule applications in
the \lts, which is a graph where nodes represent states and edge labels denote
which rule leads to the outgoing state. An excerpt of an \lts generated with
our example can be seen in Figure~\ref{fig:groove:dp_lts_excerpt}.

Obviously, inspecting an \lts by hand is not feasible for larger \statespaces.
Fortunately, we can specify \ltl and \ctl formulae in \groove.  In our
example, we could verify that all executions end up with the \texttt{leave}
rule being evaluated with the \ltl formula \texttt{F leave}, expressing that,
starting from the initial configuration, eventually the rule \texttt{leave} is
applied. Since one can use arbitrary rules in \ltl and \ctl formulae, we can
create rules capturing certain properties, such as a generic deadlock, of a
system state and include them in the formulae.

\section{Related Work}

Our focus on \gts is limited to the theory relevant to \groove. An
introduction to the algebraic approach, in particular to the \dpo approach,
can be found in \cite{Poskitt13a}, and a thorough discussion of the algebraic
approach in \cite{Ehrig:2006:FAG:1121741}.

While we focus on the \groove features relevant to this thesis here, the
\groove User Manual \cite{groove_user_manual} provides a more detailed
description of \groove features. A set of best practices when working with
\groove is presented in \cite{DBLP:journals/eceasst/ZambonR14}.

Several papers \cite{Rensink2003d,Rensink2006} discussing \groove have been
published, and \groove is compared to other simulation and model checking
tools in \cite{eemcs17830}.

\myclearpage

\chapter{Towards a Concurrency Model for SCOOP}
\label{chapter:cpm}

As we have seen in Chapter~\ref{chapter:scoop}, \scoop is a rich programming
model that provides a framework for concurrent programming and is equipped
with advanced object-oriented features.  While this is great from a user
perspective, it also makes modelling of the complete language a difficult
task. To conquer this difficulty, we first isolate concurrency related
features from \scoop to obtain a subset of the model called \corescoop. This
subset of \scoop is formalised by the
\glsreset{cpm}\cpm~\cite{Heussner-PCM15a}, a \gts based formal model.  Thanks
to the modular and extensible nature of the model, more features from \scoop
are added and eventually become \cpmo as presented in
Chapter~\ref{chapter:cpmo}. In this chapter, an overview of \corescoop and a
detailed description of \cpm and its primary implementation in \groove is
given. In the next Chapter, we then present \cpmo, which adds object-oriented
features from \scoop to \cpm.

\section{CoreSCOOP}

We define a small subset of \scoop called \corescoop. In this subset, only
basic object-oriented features exist.  There are only three kinds of data:
integers, Booleans, and references to processors. A processor can execute a
simple method with statements that modify local data---such as assigning a sum
of two local integers to another local variable---as well as asynchronous
commands and synchronous queries, where the target must be a different
processor. A method can not call other methods on the same processor as there
are no local calls. Simple method calls can be simulated by inlining the
called method, but this does obviously not work for recursive calls.

The main part of \corescoop---handling queries and commands---remains as in
\scoop. To enqueue a feature request in some processors request queue, one has
to first obtain a lock to the queue of the target processor. While \scoop
handles locking implicitly by requiring separate targets to be controlled,
\corescoop handles locking explicitly, and locking can occur at any place in a
method.

\cpm is a formal model for \corescoop and follows the specification in
\cite{Morandi2014}. In the next section, we present this formalisation and its
primary implementation as a \gts in \groove.

\section{CPM}

\cpm is a \gts modelling the behaviour of \corescoop. It allows simulating
configurations with a number of processors, each one performing computations
on integer, Boolean, and reference values. We discuss the system, in
particular the production rules involved, in detail in the following sections,
by separating concerns in the following groups: control flow, system state,
and queries and operations. We then discuss how the rules are prioritized to
achieve the desired behaviour.

\subsection{Control Flow} 
\begin{figure}
	\centering
	\begin{tikzbox}
		\includegraphics{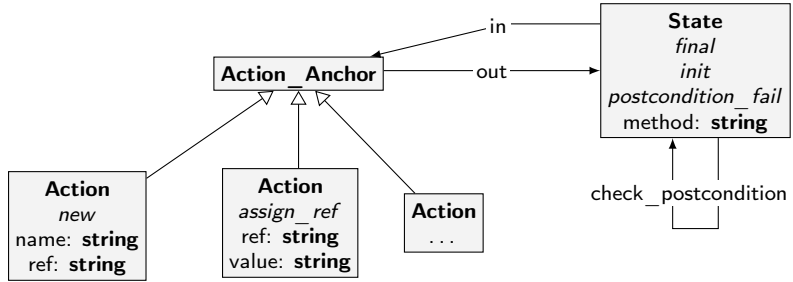}
	\end{tikzbox}
	\caption[Control flow type graph]{Control flow type graph.}
	\label{fig:cpm:type_graph_control_flow}
\end{figure}

In a \cpm start graph, methods are stored as
control flow subgraphs.
Figure~\ref{fig:cpm:type_graph_control_flow} shows the relevant subset of the
type graph of \cpm.
Methods start with an initial state (nodes of type
\textbf{State} with the \textit{init} flag), which is labelled with the method
name. From state nodes, outgoing edges (labelled \gr{in}) lead to action nodes,
which in turn have an edge (labelled \gr{out}) leading to state nodes. A final
state node does not have an outgoing edge and denotes the end of the method.
Action nodes contain information about the type of action (e.g.\ assignment,
processor creation, locking) and additional data relevant to the action (e.g.\
a query target or command parameters).

\begin{figure}
	\centering
	\begin{tikzbox}
		\includegraphics{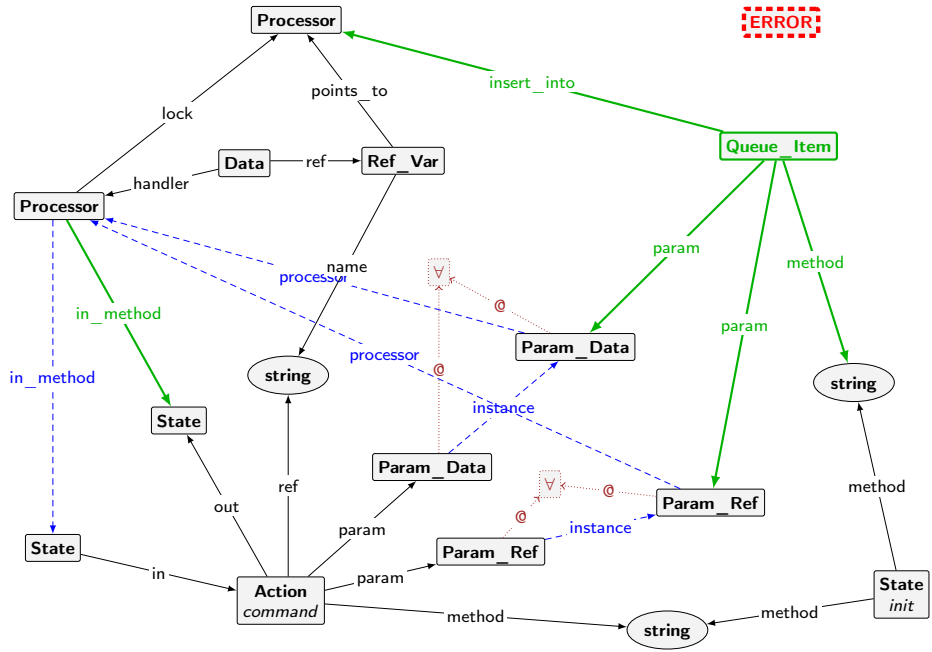}
	\end{tikzbox}
	\caption[Command action rule]{\gr{action\_Command} rule.}
	\label{fig:cpm:rule_action_command}
\end{figure}

\begin{figure}
	\centering
	\begin{tikzbox}
		\includegraphics{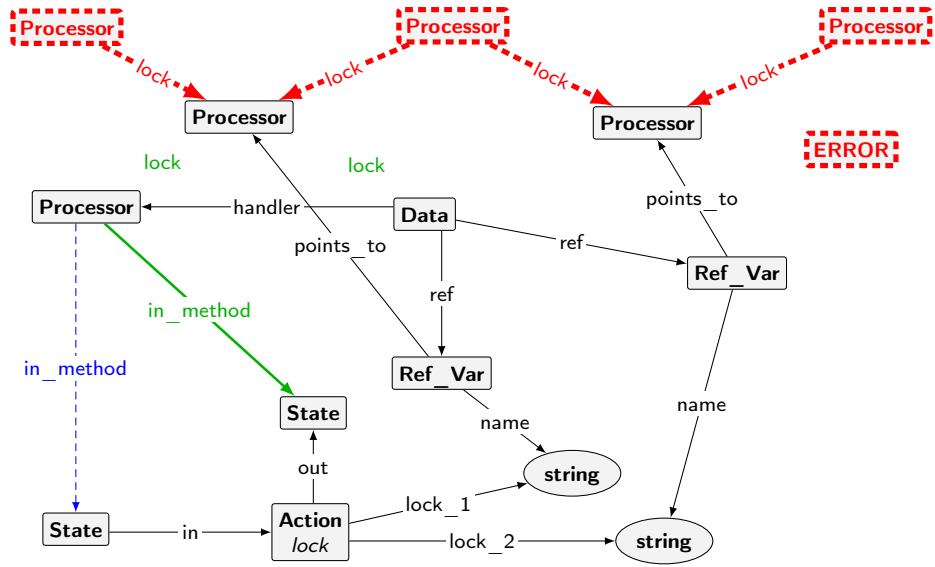}
	\end{tikzbox}
	\caption[Lock\_2 action rule]{\gr{action\_Lock\_2} rule.}
	\label{fig:cpm:rule_action_lock_2}
\end{figure}

\begin{figure}
	\centering
	\begin{tikzbox}
		\includegraphics{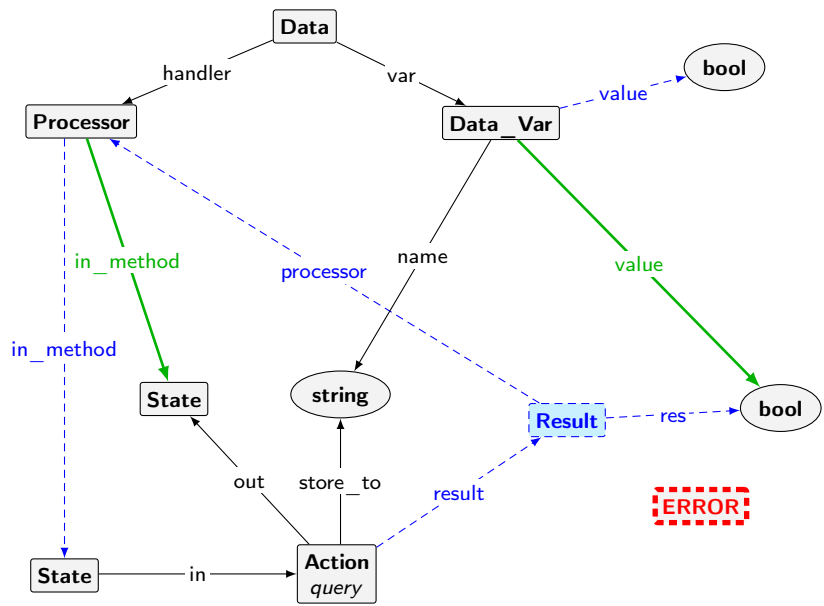}
	\end{tikzbox}
	\caption[Query action rule]{\gr{action\_Query} rule.}
	\label{fig:cpm:rule_action_Query}
\end{figure}

There are a number of relevant rules, namely the following.
\begin{description}
	\item[\gr{action\_\dots}] The \cpm \gts contains a number of action nodes.
		These nodes represent atomic units of work such as assignments, locking,
		commands, and so on, and can be compared to statements in a \scoop program
		(although there are explicit locking actions that do not have a
		counterpart in \scoop, where locking is done implicitly). Actions have the
		lowest priority, thus are applied when other rules, in particular
		scheduling rules for queues, can not be applied
		anymore. The following actions exists in \cpm:
		\begin{itemize}
			\item \gr{action\_Assign\_\ldots} group: These rules perform an
				assignment operation. The assignment operation has been split in a
				number of sub-rules in order to keep individual rules simple, as there
				are a number of different scenarios for assignments: assignments of
				references and primitive data, void assignments, assignments to fresh
				variables and used variables, assignments to the special \eif{Result}
				(return) value.
			\item \gr{action\_Command}: This action performs a command, and is shown
				in Figure~\ref{fig:cpm:rule_action_command}. Since commands are
				always asynchronous (and therefore executed on a different processor),
				a \grbf{Queue\_Item} is created and put on the processor that handles
				the target node. This enables queue management rules to be applied,
				which then eventually result in the target processor executing this
				particular request. Since the \gr{action\_Command} rule advances the
				\gr{in\_method} edge, the calling processor can continue execution
				(once action rules are enabled again).
			\item \gr{action\_Lock\_1} and \gr{action\_Lock\_2}: These actions
				acquire locks for one or two processors respectively, cf.\
				Figure~\ref{fig:cpm:rule_action_lock_2} for an illustration of the
				latter. Embargo nodes prevent the rules from being applied if a
				processor is already locked by another one.
			\item \gr{action\_Unlock}: The counterpart to the lock actions consists
				of a single rule, as unlocking multiple locks does not have to happen
				atomically.
			\item \gr{action\_Unlock\_Creator}:
				When creating new processors, the created processor is locked by the
				creating processor. By convention, the next action of the creator is a
				lock action of the created processor. As a result, the creating
				processor has to wait until the creation procedure, which contains a
				\gr{Unlock\_Creator} action at the end, removes this lock. This
				mechanism simulates the behaviour of \scoop, where creation
				procedures---even for separate objects---are executed sequentially.
			\item \gr{action\_New\_Attached} and \gr{action\_New\_Void}: These rules
				create a new processor and point the designated reference variable to
				the newly created processor. Again, this task has been split into two
				separate rules for readability reasons and to avoid excessive usage of
				quantifier nodes.
			\item \gr{action\_Query}: As opposed to the command action, the query
				rule only binds the result of the executed query to the target (by
				assigning the result to the \grbf{Data\_Var} matching the
				\gr{store\_to} edge, see Figure~\ref{fig:cpm:rule_action_Query}). The
				\gr{Queue\_Item}s are instead created by other rules (e.g.
				\gr{bexp\_Query} (Figure~\ref{fig:cpm:rule_bexp_query}) for Boolean
				queries, which are discussed in the system state section).
			\item \gr{action\_Test}: This rule performs a Boolean test by advancing
				only if the evaluated expression is true. The preceding state node has
				in certain situations two \gr{in} edges, each pointing to a test
				action node, where one action node points to a Boolean expression, and
				the other one to its negation, as illustrated in
				Figure~\ref{fig:cpm:start_graph_action_Test}. This implements an
				if-else branching mechanism, and guarantees that the processor can
				make progress.
			\item \gr{action\_TestPostcondition}: In case there is a configuration
				node that denotes that we want to check postconditions, this rule is
				applied when a processor is in a state preceding an action node with
				the \gr{test\_postcondition} flag. The rule matches if the test result
				evaluates to true, and puts the processor in a final state.
		\end{itemize}
	\item[\gr{config\_CheckPostcondition}] In case there is a
		\grbf{Configuration} node with the \grit{check\_postconditions} flag,
		postconditions will be checked. To do so, this rule follows a final state
		along the \grit{check\_postcondition} edge to another state node by
		redirecting the \gr{in\_method} edge of the relevant processor.
		Postconditions are the only situation where a state is followed directly
		by another state. The rule does not match if the configuration node is not
		present, providing an intuitive way to enable or disable postcondition
		checking.
\end{description}

\begin{figure}
	\centering
	\begin{tikzbox}
		\includegraphics{figures/cpm_start_graph_action_Test.tikz}
	\end{tikzbox}
	\caption[Test action start configuration]{Excerpt from a start graph
	illustrating usage of two action nodes to provide an if-else construct.}
	\label{fig:cpm:start_graph_action_Test}
\end{figure}

\begin{figure}
	\centering
	\begin{tikzbox}
		\includegraphics{figures/cpm_rule_bexp_query.tikz}
	\end{tikzbox}
	\caption[Boolean query expression rule]{\gr{bexp\_Query} rule.}
	\label{fig:cpm:rule_bexp_query}
\end{figure}

\subsection{System State}

\begin{figure}
	\centering
	\begin{tikzbox}
		\includegraphics{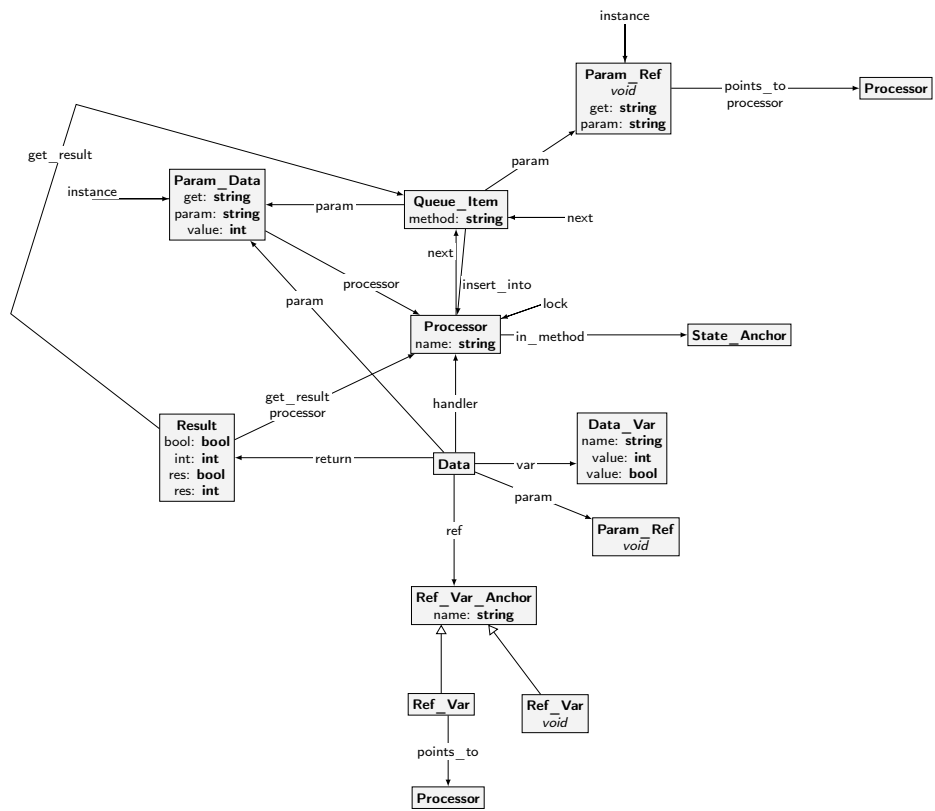}
	\end{tikzbox}
	\caption[System state type graph]{Type graph of the system state. Note that
	self edges are rendered by an arrow that leads from a label to a node (an
	example is the \gr{instance} edge of the \grbf{Param\_Ref} node). We render
	self-edges in this manner throughout this thesis.}
	\label{fig:cpm:type_graph_system_state}
\end{figure}

The system state is concerned with processors, queue management, and handling
of data. The relevant type graph is shown in
Figure~\ref{fig:cpm:type_graph_system_state}.

\begin{figure}
	\centering
	\begin{tikzbox}
		\includegraphics{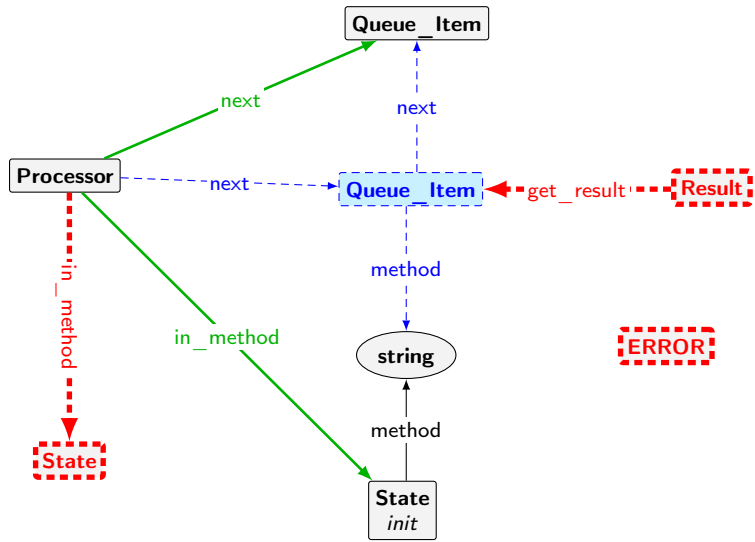}
	\end{tikzbox}
	\caption[Remove queue item rule]{\gr{queue\_Remove\_Command\_MultipleQueued}
	rule.}
	\label{fig:cpm:queue_remove_command_multiplequeued}
\end{figure}

Processors are at the core of \cpm states. During their lifetime, they
are either handling requests or they are idle. In the first case, they are
executing a method at a certain position, denoted by the \gr{in\_method} edge.
When requests are made by other processors, a \grbf{Queue\_Item} is created
which has a \gr{insert\_into} edge to the target processor, as can be seen in
the rules \gr{action\_Command} (Figure~\ref{fig:cpm:rule_action_command}) and
\gr{bexp\_Query} (Figure~\ref{fig:cpm:rule_bexp_query}). Once a queue item is
created, a number of rules come into action that are responsible for queue
management, namely the following:
\begin{description}
	\item[\gr{queue\_Insert\_EmptyBusy} and \gr{queue\_Insert\_NotEmpty}] These
		rules can be applied when a queue request has been made (with an
		\gr{insert\_into} edge), and their effect is to simply put the item at the
		end of the queue.
	\item[\gr{queue\_Remove\_ParamRef} and \gr{queue\_Remove\_ParamData}]
		Nodes representing parameters are attached to the queue item upon creating
		it. These two rules prepare the call by removing the connection between
		queue item and parameter node, and attaching the parameter to the
		processor's data node.  The next rules will handle the remaining part of
		the queue item.
	\item[\gr{queue\_Remove\_\ldots}] The remaining four rules in the
		\gr{queue\_Remove} group remove a query or a command request from the top
		of the queue and activate the processor to start execution at the given
		method. There are two rules for the case with one item on the queue and
		two rules for the case with more items on the queue.

		Figure~\vref{fig:cpm:queue_remove_command_multiplequeued} shows the rule
		\gr{queue\_Remove\_Command\_MultipleQueued}, which handles the case of 
		multiple queue items and the top item being a command request.
\end{description}

\subsection{Queries and Other Operations}

\begin{figure}
	\centering
	\begin{tikzbox}
		\includegraphics{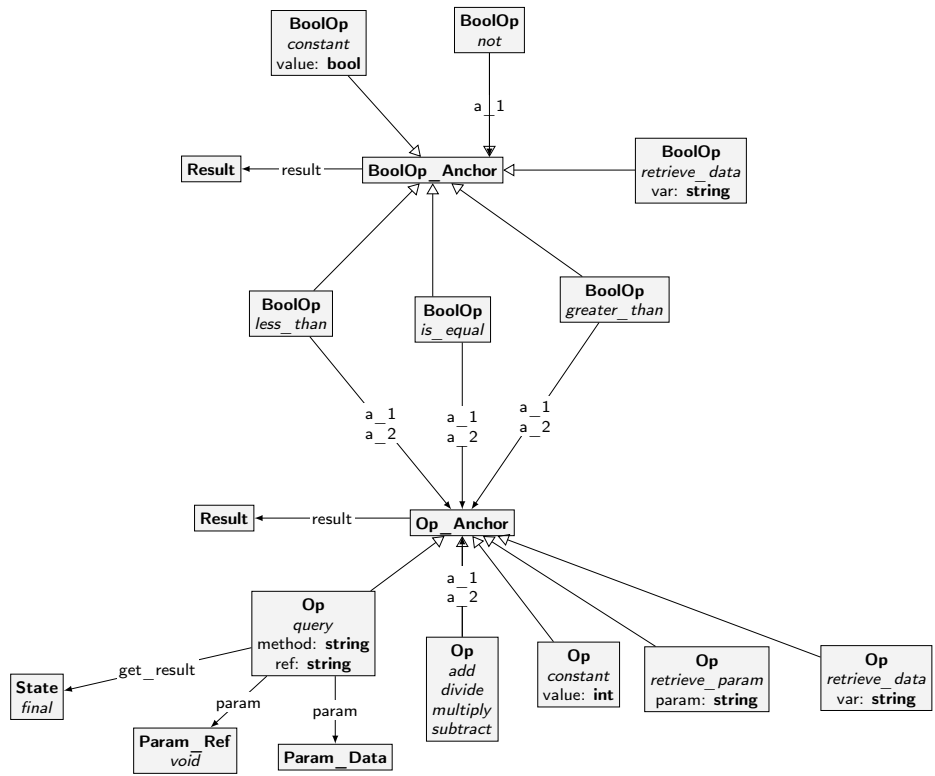}
	\end{tikzbox}
	\caption[Operations and queries type graph]{Type graph of operations and
		queries.}
	\label{fig:cpm:type_graph_ops}
\end{figure}

While there exist a query flag for action nodes, the rule that advances over
an action node only does so after the query has been evaluated and is
essentially an assignment operation where the right-hand side happens to be a
query. Similarly, other assignment operations also contain right-hand sides
that need to be evaluated before the assignment can be performed. For example,
in the assignment \gr{r\_1 := r\_2}, the reference on the right-hand side must
first be fetched. The group of rules in this subsection handle this, and they
have higher priorities than the action rules in order to make sure that
whenever an action requires arguments, they are fetched first. The type graph
for operations, shown in Figure~\ref{fig:cpm:type_graph_ops}, contains the
operation types.  Since the operation types are encoded using flags, it would
be possible to have, for example, an \grbf{Op} node with both the
\grit{constant} and the \grit{add} flag, as it is not possible to force having
exactly one flag. By convention, we do not support multiple flags for such a
node. We set up the type graph to reflect how the various node types are
intended to be used.

\begin{figure}
	\centering
	\begin{tikzbox}
		\includegraphics{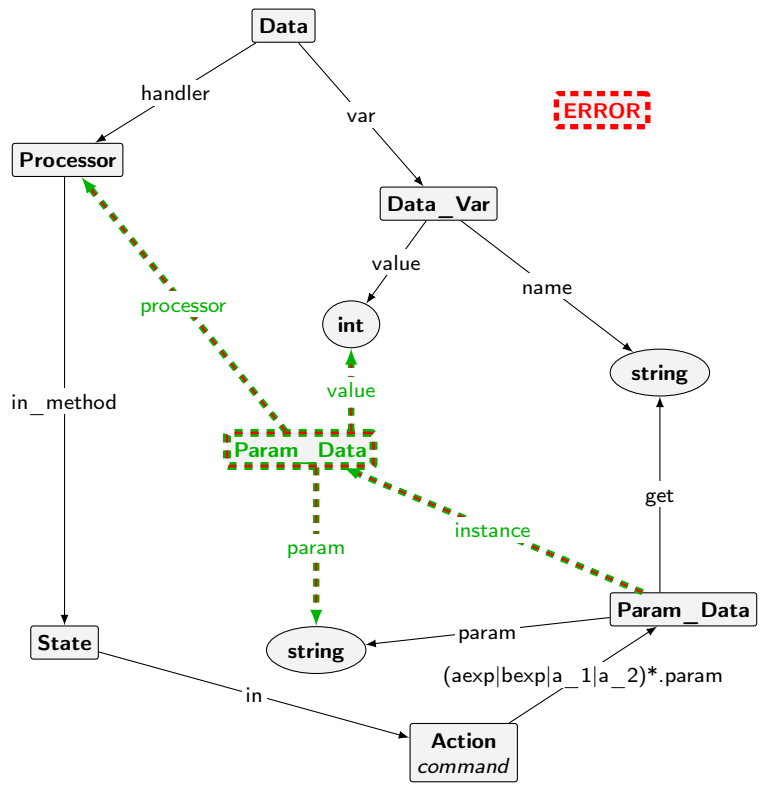}
	\end{tikzbox}
	\caption[Integer parameter fetching rule]{\gr{getparam\_Data} rule.}
	\label{fig:cpm:rule_getparam_data}
\end{figure}

Queries do not appear as action nodes themselves. Instead, they are attached
to an assignment action. Similarly, integer and Boolean operations are not
targets either, as they appear in either complex expressions or on the
right-hand side of an assignment. The relevant rules are the following:
\begin{description}
	\item[\gr{aexp\_\dots}] Arithmetic expression rules evaluate integer
		expressions, by creating a \grbf{Result} nodes and attaching them to
		\grbf{Op} nodes. The following rules exists for arithmetic expressions:
		\begin{itemize}
			\item \gr{aexp\_constant}: Creates a result with the value specified by
				the operation itself.
			\item \gr{aexp\_RetrieveParam}: Fetches a parameter from the data
				handled by the current processor.
			\item \gr{aexp\_RetrieveData}: Retrieves an integer data value from the
				current processor.
			\item \gr{aexp\_Multiply, aexp\_Divide, aexp\_Add, aexp\_Subtract}:
				Evaluates the binary operation and creates a result node.
		\end{itemize}
	\item[\gr{bexp\_\dots}] Analogously to the arithmetic expression rules,
		Boolean expression rules evaluate Boolean expressions. The following rules
		exist in \cpm:
		\begin{itemize}
			\item \gr{bexp\_constant, bexp\_RetrieveData}: Analogous to the
				arithmetic expression variants.
			\item \gr{bexp\_GreaterThan, bexp\_LessThan, bexp\_IsEqual, bexp\_not}:
				Evaluates the binary operation and creates a result node with a
				Boolean result.
		\end{itemize}
	\item[\gr{bexp\_Query}] The rule for Boolean queries creates a
		\grbf{Queue\_Item} which will be inserted into the queue of the target
		processor, as illustrated in Figure~\ref{fig:cpm:rule_bexp_query}. It is
		similar to the \gr{action\_Command} rule. The target processor will
		execute the request and once the result is available, it can be matched by
		the \gr{action\_Query} rule.
	\item[\gr{getparam\_Ref\_\ldots}] This group consists of rules for fetching
		method parameters for command actions. They perform the step of looking up
		the value of a reference or data variable and create a \grbf{Param\_Data}
		instance, as illustrated in Figure~\ref{fig:cpm:rule_getparam_data} for
		the integer case.
\end{description}

Queries and operations often require intermediate nodes. For example, we
attach a \grbf{Result} nodes to an \grbf{Op} after evaluating it. Once the
processor has used this result and moved past the state where it was required,
we can safely remove it in order to keep the graph clutter-free. The following
rules clean the graph in various situations.
\begin{description}
	\item[\gr{cleanup\_RemoveParamRef\_Attached}] 
	\item[\gr{cleanup\_RemoveParamRef\_Void}] 
	\item [\gr{cleanup\_DiscardParamData}]
		Once the processor is in a final state of a method, parameters are not
		required any more and are removed by these rules. In fact, we must remove
		them as otherwise the system may misbehave in subsequent method calls
		(e.g.\ if the next call has the same parameter names, two nodes of the same
		parameter exist and rules that match it have two possible applications).
	\item[\gr{cleanup\_exp\_DiscardResults\_Op}] 
	\item[\gr{cleanup\_exp\_DiscardResults\_BoolOp}] Once a processor moves past
		an action that has an \grbf{Op} or a \grbf{BoolOp} node attached, the
		corresponding result nodes are not required any more and are removed by
		these rules.
	\item[\gr{cleanup\_FinalState\_BoolQuery} and \gr{cleanup\_FinalState}]
		These rules are applied when a processor reaches the final state of a
		query or command respectively. The \gr{in\_method} edge is deleted, as
		well as associated edges and nodes related to the result value.
\end{description}

\subsection{Rule Priorities}

As mentioned earlier, the rules in \cpm are not applied with the same
priorities. This has various reasons. First of all, it enables control on what
rules are applied in which situations. For example, cleanup rules have
priorities such that they are performed before new actions are performed,
ensuring that no ``leftover'' nodes stay in the graph (e.g.\ parameter
instances from earlier commands and queries). If the cleanup rules do not
have higher priorities, the graph may end up in a state where an action rule
has multiple matches, in particular matches with old and invalid instance
nodes.

Another advantage of rule priorities is that they can be used to attack the
state-space explosion problem. By assigning fine grained priorities to the
rules that do not influence the modelled behaviour, we reduce the possible
interleavings in an execution. For example, it does not matter in which order
cleanup rules are applied, as once all matching cleanup rules are applied, the
system always returns to the same state. By assigning each cleanup rule a
unique priority value and thus forcing a fixed order, these local interleaving
scenarios are eliminated.

Of course one has to pay attention to which rules can have different
priorities and which ones need to have the same priorities. Action rules
generally should have the same priorities, since these nodes can create queue
items and we are interested in the interleavings with unique queue item
sequences.

Our approach is in line with Zambons and Rensinks paper
\cite{DBLP:journals/eceasst/ZambonR14} on best practices in \groove, where
they suggest to use some form of rule scheduling whenever possible. While they
mention that ``the use of control programs is usually preferred over
priorities'', we think that priorities are sufficient and easier to maintain
in our case.

Table~\ref{table:cpm:priorities} shows a list of all rules in \cpm and their
priorities.

\begin{table}
	\centering
	{\footnotesize
	\begin{tabular}{lr}\toprule
		\textbf{Rule} & \textbf{Priority}\\
		\toprule
		\gr{error\_Deadlock\_2Proc\_DiffState} & 100\\
		\gr{error\_Deadlock\_2Proc\_SameState} & \\
		\gr{error\_DivideByZero} & \\
		\gr{error\_NoSelfRef} & \\
		\gr{error\_PostconditionFail} \\
		\gr{error\_VoidCall} \\
		\midrule
		\gr{config\_CheckPostcondition} & 60\\
		\midrule
		\gr{cleanup\_RemoveParamRef\_Attached} & 50\\
		\gr{cleanup\_RemoveParamRef\_Void} & 49\\
		\gr{cleanup\_DiscardParamData} & 48\\
		\gr{cleanup\_exp\_DiscardResults\_Op} & 47\\
		\gr{cleanup\_exp\_DiscardResults\_BoolOp} & 46\\
		\midrule
		\gr{getparam\_Ref\_Attached} & 44\\
		\gr{getparam\_Ref\_Void} & 43\\
		\gr{aexp\_constant} & 41\\
		\gr{bexp\_constant} & 40\\
		\gr{aexp\_RetrieveParam} & 39\\
		\gr{aexp\_RetrieveData} & 38\\
		\gr{bexp\_RetrieveData} & 37\\
		\gr{bexp\_Query} & 36\\
		\gr{aexp\_Multiply} & 35\\
		\gr{aexp\_Divide} & 34\\
		\gr{aexp\_Add} & 33\\
		\gr{aexp\_Subtract} & 32\\
		\gr{bexp\_GreaterThan} & 31\\
		\gr{bexp\_LessThan} & 30\\
		\gr{bexp\_IsEqual} & 29\\
		\gr{bexp\_not} & 28\\
		\midrule
		\gr{queue\_Insert\_EmptyBusy} & 20\\
		\gr{queue\_Insert\_NotEmpty} & 19\\
		\gr{queue\_Remove\_ParamRef} & 18\\
		\gr{queue\_Remove\_ParamData} & 17\\
		\gr{queue\_Remove\_BoolQuery\_MultipleQueued} & 16\\
		\gr{queue\_Remove\_BoolQuery\_SingleQueued} & 15\\
		\gr{queue\_Remove\_Command\_MultipleQueued} & 14\\
		\gr{queue\_Remove\_Command\_SingleQueued} & 13\\
		\midrule
		\gr{cleanup\_FinalState\_BoolQuery} & 5\\
		\gr{cleanup\_FinalState} & 4\\
		\midrule
		\gr{action\_TestPostcondition} & 2\\
		\midrule
		\gr{action\_Assign} & 0\\
		\gr{action\_AssignRef\_Param\_Attached} & \\
		\gr{action\_AssignRef\_Param\_SameTarget} & \\
		\gr{action\_AssignRef\_Param\_Void} & \\
		\gr{action\_AssignRef\_Ref\_Attached} & \\
		\gr{action\_AssignRef\_Ref\_SameTarget} & \\
		\gr{action\_AssignRef\_Ref\_Void} & \\
		\gr{action\_Command} & \\
		\gr{action\_Lock\_1} & \\
		\gr{action\_Lock\_2} & \\
		\gr{action\_New\_Attached} & \\
		\gr{action\_New\_Void} & \\
		\gr{action\_Query} & \\
		\gr{action\_Test} & \\
		\gr{action\_Unlock} & \\
		\gr{action\_Unlock\_Creator} & \\
		\bottomrule
	\end{tabular}
	}
	\caption[Rule priorities]{Rule priorities. Note that an empty priority
	means that the rule has the same priority as the one above it, e.g.\ all
	error rules have priority 100.}
	\label{table:cpm:priorities}
\end{table}

\section{Dining Philosophers}
\subsection{Start Graph}

This section revisits our running example of the dining philosophers by
showing \cpm in action. A dining philosophers start configuration for \cpm is
shown in Figures~\vref{fig:cpm:start_graph_dphil_nodeadlock_application_make}
and~\vref{fig:cpm:start_graph_dphil_nodeadlock_other} (the graph has been split
up for readability, but both make up a single graph and are represented in
\groove as such).

Most of the nodes in
Figure~\ref{fig:cpm:start_graph_dphil_nodeadlock_application_make} belong to
the control flow graph of \gr{APPLICATION.make}, which is the root procedure
in this example. It roughly translates to the code in
Listing~\ref{listing:cpm:dphil}, with a simple loop that instantiates forks
and philosophers and connects them accordingly. In addition to the method
graph, there is also a \grbf{Processor} node. It is the handler of its data,
which consists of a number of reference and integer variables. Note that the
variables have generic names (such as \eif{v_1} for Boolean and integer data
and \eif{r_1} for references), and that there is a mapping from \cpm variable
names to the variable names of the code in the listing.

\begin{eiffelcode}[
		label=listing:cpm:dphil,
		caption={\eif{APPLICATION} class for the dining philosophers.}
	]
class APPLICATION
	feature
		make
			do
				-- variable mappings:
				--   v_1: i
				--   v_2: philosopher_count
				--   v_3: round_count
				--   r_1: first_fork
				--   r_2: left_fork
				--   r_3: right_fork
				--   r_4: a_philosopher
				from
					i := 1
					create first_fork.make
				until
					i > philosopher_count
				loop
					if i < philosopher_count then
						create right_fork.make
					else
						right_fork := first_fork
					end
					create a_philosopher.make (i, left_fork, right_fork, round_count)
					lock (a_philosopher)
					a_philosopher.live
					unlock (a_philosopher)
					left_fork := right_fork
					i := i + 1
			end
	
	i, round_count: INTEGER
	first_fork, left_fork, right_fork: separate FORK
	a_philosopher: separate_PHILOSOPHER
end

class FORK
	feature
		make
			do
			end
end

class PHILOSOPHER
	feature
		make (philosopher: INTEGER;
		      left, right: separate FORK;
					round_count: INTEGER)
			do
				-- variable mappings:
				-- p_1: philosopher
				-- p_2: left
				-- p_3: right
				-- p_4: round_count
				-- v_1: id
				-- v_2: times_to_eat
				-- r_1: left_fork
				-- r_2: right_fork
				id := philosopher
				left_fork := left
				right_fork := right
				times_to_eat := round_count
			end

		live
			do
				lock (left_fork, right_fork)
				-- eat
				unlock (left_fork)
				unlock (right_fork)
				times_to_eat := times_to_eat - 1
			end

		left_fork, right_fork: separate FORK
		id, times_to_eat: INTEGER
end
\end{eiffelcode}

Figure~\ref{fig:cpm:start_graph_dphil_nodeadlock_other} shows the control flow
graphs for forks and philosophers. Processors representing forks do not
execute any code after their creation procedure. They are created, the
\grit{unlock\_creator} action is performed, and afterwards they just exist in
the system, but do not execute more requests (as no other processor ever
performs a command or query on a fork). In the \gr{make} method of the
philosopher, the object is initialized by assigning the parameters to the
processor's reference variables and data variables.  Finally, the subgraph
representing the \gr{live} method is traversed to perform the main loop of the
philosophers. Most notably, this subgraph contains actions to lock
(representing atomically acquiring the forks and eating) and unlock the fork
processors.

\subsection{Rule Applications}

With the start graph presented in the previous section, we can now inspect the
behaviour of \cpm.
With the \groove Simulator, it is possible to follow the
state-space exploration visually. Applicable rules are pointed out to the user
and the part of the graph that matches is highlighted.
Figure~\vref{fig:cpm:exploration_dphil_nodeadlock_before_command} shows the
program right before the first creation procedure (a command) is performed. At
this point, the rule \gr{action\_Command}
(see Figure~\vref{fig:cpm:rule_action_command}) is the only rule that has a
match and is highlighted in green. After applying the rule, the graph looks as
depicted in Figure~\vref{fig:cpm:exploration_dphil_nodeadlock_after_command}.

Of course, going through rules using the Simulator is a rather tedious task
that may be a useful tool for developing, testing, and debugging such systems,
but it is not for verification purposes. Fortunately, as we have seen in
Chapter~\ref{chapter:gts}, \groove provides utilities to verify for \ltl
and \ctl formulae. This is where the error rules come into play. To verify
whether the program deadlocks, one can simply try to find a counterexample for
the formula \[
\gr{! F (error\_deadlock\_2Proc\_DiffStates |
error\_Deadlock\_2Proc\_SameState)}\mbox{,}
\]
which states that, starting from the
start graph state, there is no future state where either one of the mentioned
rules matches.

\begin{sidewaysfig}
	\centering
	\begin{sidewaystikzbox}
		\includegraphics{figures/cpm_start_graph_dphil_nodeadlock_application_make.tikz}
	\end{sidewaystikzbox}
	\caption[Dining philosophers start graph, APPLICATION.make part]{Subgraph
		of the dining philosophers start graph in \cpm, consisting of the
		initial processor starting at the root method \gr{APPLICATION.make}.}
	\label{fig:cpm:start_graph_dphil_nodeadlock_application_make}
\end{sidewaysfig}

\begin{sidewaysfig}
	\centering
	\begin{sidewaystikzbox}
		\includegraphics{figures/cpm_start_graph_dphil_nodeadlock_other.tikz}
	\end{sidewaystikzbox}
	\caption[Features of the \eif{PHILOSOPHER} start graph]{Various features of the \eif{PHILOSOPHER} start graph.}
	\label{fig:cpm:start_graph_dphil_nodeadlock_other}
\end{sidewaysfig}

\begin{sidewaysfig}
	\centering
	\begin{sidewaystikzbox}
		\includegraphics{figures/cpm_exploration_dphil_nodeadlock_before_command.tikz}
	\end{sidewaystikzbox}
	\caption[Dining philosophers configuration before command
	action]{Configuration where the processor is about to execute a command
	action. The match is highlighted with dashed lines.}
	\label{fig:cpm:exploration_dphil_nodeadlock_before_command}
\end{sidewaysfig}

\begin{sidewaysfig}
	\centering
	\begin{sidewaystikzbox}
		\includegraphics{figures/cpm_exploration_dphil_nodeadlock_after_command.tikz}
	\end{sidewaystikzbox}
	\caption[Dining philosophers configuration after rule
	application]{Configuration right after an application of the command action
	rule. A new request has been created and is about to be inserted into the
	request queue of the target processor.}
	\label{fig:cpm:exploration_dphil_nodeadlock_after_command}
\end{sidewaysfig}

\myclearpage

\chapter{CPM+OO: An Extension for Objects}
\label{chapter:cpmo}
\glsreset{cpmo}
\cpmo builds on top of \cpm, and aims to bring back object-oriented concepts
that have been intentionally left out from \corescoop and \cpm. While \cpm
focuses on the concurrency aspects of \scoop, it can be difficult to
map real-world \scoop programs, with processors potentially handling multiple
objects, non-separate calls within routine bodies, and other \scoop features
that are not directly modelled in \cpm.
These enhancements allow a more direct
mapping of \scoop programs to the graph model and clear the path for an
automatic translation tool from \scoop programs to \cpmo (cf.
Chapter~\ref{chapter:translation}).

In this Chapter, we discuss various extensions made to the \cpm model. We
explain how the changes affect the behaviour of the system and make informal
arguments for the preservation of soundness and completeness.

\section{Type Graph Overview}
\label{section:cpmo:type_graph}

We start by giving an overview over the changes of various parts of the type
graph, which may seem overwhelming at first, as the type graph has changed
significantly between \cpm and \cpmo. The goal of this section is not to
provide a complete description, but instead relate the updated type graph to
the various added concepts, which will then be explained in detail in
subsequent sections.

\subsection{Processors, Frames, and Objects}

To model local calls and non-separate objects, we introduce the notion of stack
frames and object instances to the model. A subgraph of the \cpmo type graph
with processor and object related nodes can be seen in
Figure~\ref{fig:cpmo:type_graph_processors_and_objects}.

\begin{sidewaysfig}
	\centering
	\begin{tikzbox}
		\includegraphics{figures/cpmo_type_graph_processors_and_objects.tikz}
	\end{tikzbox}
	\caption[Type graph of processors and objects]{Type graph of processors and
		objects.}
	\label{fig:cpmo:type_graph_processors_and_objects}
\end{sidewaysfig}

At the core of the type graph is the \grbf{Processor}. Tied to it is the
\grbf{Queue\_Item} with its parameters and method name that is intended to be
executed.

The basic \grbf{Data} node has been replaced with the \grbf{Object} node by
simple renaming (note that the type graph simply puts syntactic restrictions
on the graphs, different semantics introduced in \cpmo come only with the rule
changes). A processor can now be the handler of multiple objects (whereas \cpm
restricted a processor to be the handler of exactly one data node).

Introducing nested routine calls brings the requirement for some call stack
representation. This is achieved by introducing the \grbf{Frame} node type and
its attached nodes. A processor that is executing a program always works with
a current stack frame, which handles local variables, passed parameters, a
reference to the \eif{Current} object, and the state it needs to return to in
case the current call is a nested call. When creating requests, a queue item
has a stack frame attached, which contains passed parameters.

\subsection{Variables, Parameters, and Results}

\begin{figure}
	\centering
	\begin{tikzbox}
		\includegraphics{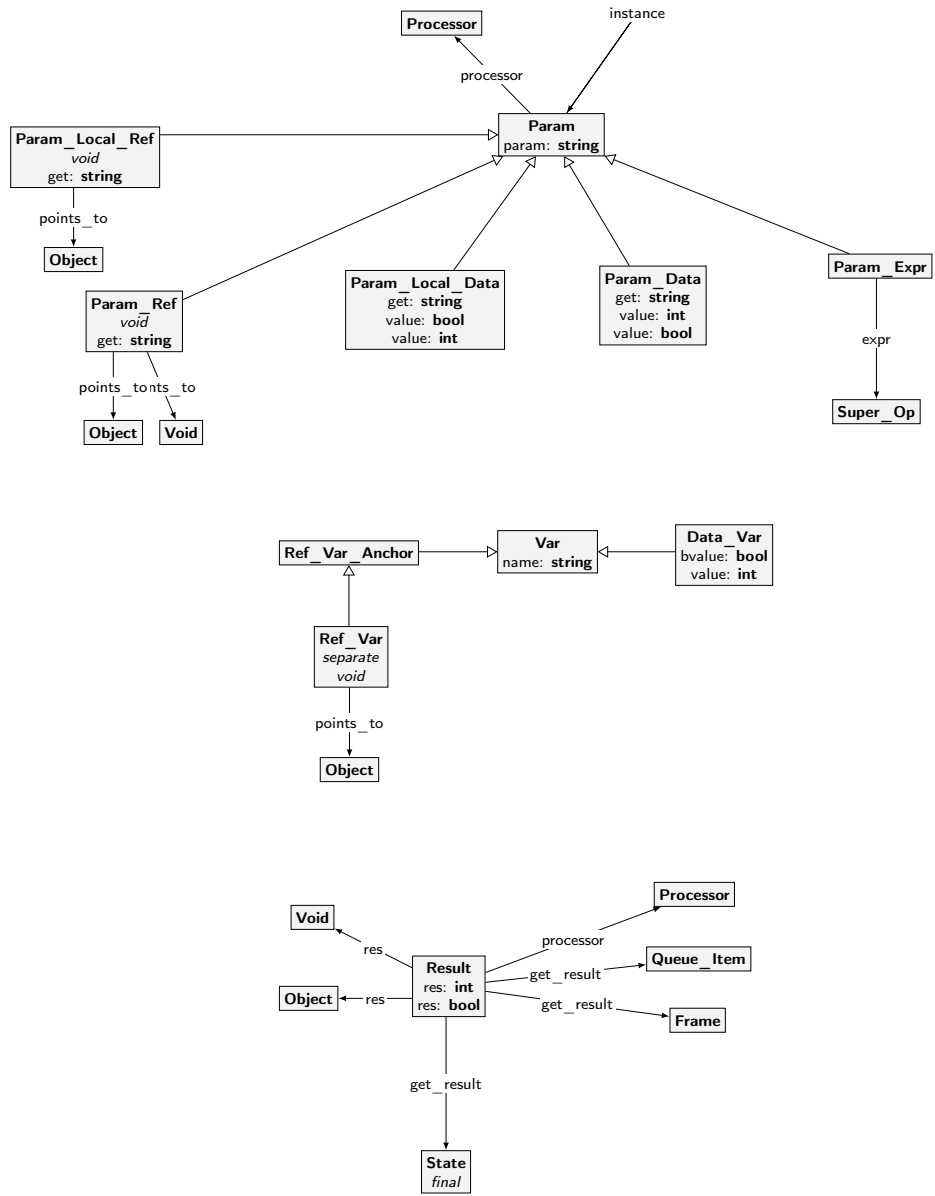}
	\end{tikzbox}
	\caption[Type graph of variable, parameter, and result nodes]{Type graph of
		variable, parameter, and result nodes.}
	\label{fig:cpmo:type_graph_variables_parameters_results}
\end{figure}

The next group of related types is concerned with the representation and
handling of variable declarations and bindings of values to parameters and
query results. Figure~\ref{fig:cpmo:type_graph_variables_parameters_results}
shows the relevant type graph.

Variable declarations are, as in \cpm, divided into three different subtypes:
reference, integer and Boolean variables. While integer and Boolean data
variables did not change, reference variables now point to objects instead of
processors. In addition, reference variables now have a flag denoting whether
it is declared as separate or not.

Parameter nodes represent values passed to commands and queries. Parameters
can be local variables (corresponding to variables in the \eif{local} block)
or attributes of the current object, as well as arbitrary expressions
(\grbf{Param\_Expr} with an \gr{expr} edge to an operation node). Adding
arbitrary expressions (as opposed to local variables and attributes only) as
parameters allows representing complex expressions with a single action node
in \cpmo. This is not possible in \cpm, where helper variables are required to
simulate complex expressions.

\subsection{Actions}

The type graph for actions, shown in Figure~\ref{fig:cpmo:type_graph_actions},
looks similar to the one in \cpm. One important difference is that we use
subtypes for the different kinds of actions as opposed to flags. This has the
advantage that we can not have a single action node with multiple types, a
property which can not be enforced in \groove using flags.

To support arbitrary expressions (e.g.\ \eif{sum := a1.count + a2.count}) as
parameters and operands, we replace a number of string attributes with edges
to nodes of type \grbf{Super\_Op} (and its subtypes). For example, the
unlock action node \grbf{Action\_Assign\_Ref} is pointing to a \grbf{RefOp}
node, which means that we can use arbitrary expressions on the right-hand side
of the assignment.

The type \grbf{Action\_Token} is a supertype of actions that are local to a
processor, or, in the case of queries, potentially local to a processor. This
is later used in a mechanism to mitigate the \statespace explosion problem and
is discussed in detail in Section~\ref{section:cpmo:statespace_optimisations}.

\begin{sidewaysfig}
	\centering
	\begin{sidewaystikzbox}
		\includegraphics{figures/cpmo_type_graph_actions.tikz}
	\end{sidewaystikzbox}
	\caption[Type graph of actions]{Type graph of actions.}
	\label{fig:cpmo:type_graph_actions}
\end{sidewaysfig}

\subsection{Operations}

The group of operation types (cf.
Figure~\vref{fig:cpmo:type_graph_operations}) has undergone a number of
changes during development of the \cpmo model.  Integer, Boolean, and
reference operations (\grbf{Op}, \grbf{BoolOp}, and \grbf{RefOp} types) do now
inherit from the same \grbf{Super\_Op} supertype, which can be matched in
certain rules to cover all kinds of operations.  The different operations
(e.g. ``greater than'' and ``equals'' operations) are represented as unique
types, which is more restrictive than representing the operations with flags
for one single node type. Other additions include types for the handling of
local declarations as well as types for integer and reference type queries.

\subsection{Errors}

\begin{figure}
	\centering
	\begin{tikzbox}
		\includegraphics{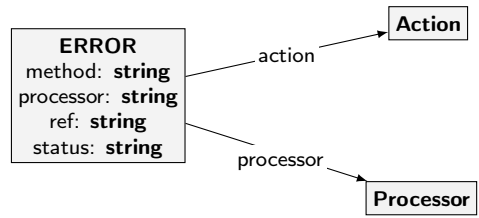}
	\end{tikzbox}
	\caption[Type graph of errors]{Type graph of errors.}
	\label{fig:cpmo:type_graph_errors}
\end{figure}

As in \cpm, the \grbf{ERROR} type (Figure~\ref{fig:cpmo:type_graph_errors})
is used for recording information about detected issues with the program. This
includes both undesirable properties in the behaviour of the program (e.g.\ a
deadlock situation) as well as invalid configurations (e.g.\ multiple handlers
for a single object). The latter kind of error is used to aid the development
and evolution of the model and is designed to catch bugs in the model itself,
not errors in the behaviour of the modelled runtime.

Recording information in error nodes is useful for postprocessing. Since all
rules have an \grbf{ERROR} embargo node, rules can only be applied as long as
there is no error node. Once an error node is created, the system is in a final
state. Our postprocessing tools can then simply go through all the final
states and check whether there is an error node, and if so, generate output
based on the context of the error.

\subsection{Others}

\begin{figure}
	\centering
	\begin{tikzbox}
		\includegraphics{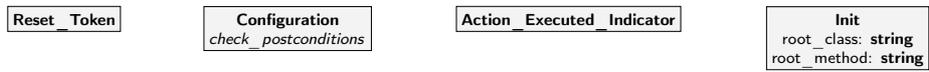}
	\end{tikzbox}
	\caption[Miscellaneous types]{Miscellaneous types.}
	\label{fig:cpmo:type_graph_misc}
\end{figure}

Figure~\ref{fig:cpmo:type_graph_misc} shows the remaining types in \cpmo.
They are used in the model as follows:
\begin{itemize}
	\item \grbf{Reset\_Token} and \grbf{Action\_Executed\_Indicator}: These two
		types are used for an optimisation technique that forces processors that
		are performing non-separate actions to advance as far as possible before
		yielding control to the next processor. We describe the \statespace
		optimisation involving these types in detail in
		Section~\ref{section:cpmo:statespace_optimisations}.
	\item \grbf{Configuration}: A node of this type can be included in a start
		graph to enable special behaviour of the model. In particular, one can add
		the \grit{check\_postconditions} flag to enable postcondition checking.
	\item \grbf{Init}: The initialization type allows specifying the root class
		and procedure.
\end{itemize}

\begin{sidewaysfig}
	\centering
	\begin{tikzbox}
		\includegraphics{figures/cpmo_type_graph_operations.tikz}
	\end{tikzbox}
	\caption[Type graph of operations]{Type graph of
		operations.}
	\label{fig:cpmo:type_graph_operations}
\end{sidewaysfig}

\section{Modelled SCOOP Features}

\subsection{Local and non-separate Calls}

In \cpm, calls are always performed by adding a new queue item to the request
queue of a remote processor. Local calls are not supported, instead one has to
perform inlining of the method bodies where local calls would occur. \cpmo
instead provides mechanisms for local routine calls (i.e.\ where the target is
the current object) and non-separate calls (where the target may be a
different object, but is handled by the current processor).

To achieve this, we use the call stack representation introduced in the type
graph, and introduce rules that handle non-separate calls. The corresponding
rules for separate calls are derived from the \cpm rules that create feature
requests and are enhanced with the notion of a call stack and adapted to the
object representation.

In the following, we first discuss separate calls. We show how we the \cpm
rules to work with objects, frames, and other changes in the \cpmo type graph.
We then present features that are missing in\cpm, such as non-separate calls.

\subsubsection{Separate Calls}

The rules handling creation of feature requests, i.e.\ separate feature calls,
are similar to what we have seen in \cpm (rules \gr{action\_Command} and
\gr{bexp\_Query}, see Figures~\vref{fig:cpm:rule_action_command}
and~\vref{fig:cpm:rule_action_lock_2}).
Figure~\ref{fig:cpmo:rule_action_Command_separate} shows the rule
\gr{action\_Command\_separate} from \cpmo.  While the rule is much larger than
its \cpm counterpart, the semantic behaviour of the two do not differ much.
There are several reasons why the \cpmo rule requires more nodes.  First,
\cpmo supports additional parameter nodes for a command (in particular,
expressions, local references, and local data), resulting in more pairs of
parameters and instances, but they follow the same structure as data and
reference parameters in \cpm. Second, to the lower left, there is a construct
nesting an $\exists$ quantifier inside a $\forall$ quantifier. Since the
parameter node is at the $\forall$ quantifier and its instance at the
$\exists$ quantifier, this expresses that the rule matches, if for all
\gr{Param\_Expr} nodes of the command, there exists an instance of type
\gr{Param}. This construct is used to enforce that the rule is only applied
once all parameter expressions are evaluated (i.e.\ once instance nodes have
been created). Once the requirements are met, a queue item is created, similar
to what the \cpm rule does. But instead of attaching parameters directly to
the queue item, we create a \grbf{Frame} node, representing the prepared stack
frame which can be put on top of the frame stack once the request is handled.
Note that the action does not specify a reference denoting the call target as
a simple string any more, instead it points to a parameter expression via the
\gr{target} edge which allows using complex expressions as targets (e.g.
\eif{foo.get_counter ().count}).

\begin{figure}
	\centering
	\begin{tikzbox}
		\includegraphics{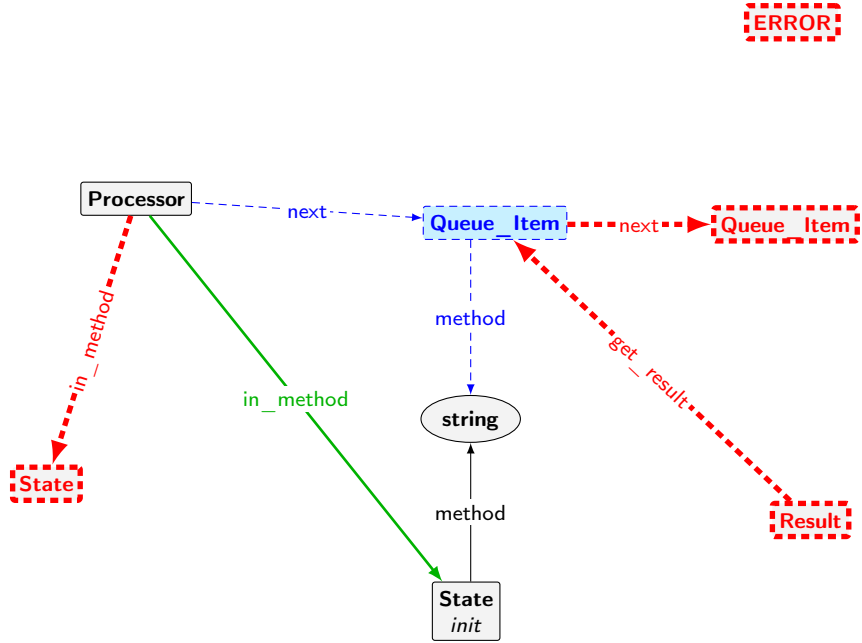}
	\end{tikzbox}
	\caption[Rule \gr{queue\_Remove\_Command\_SingleQueued} in \cpm]{Rule
		\gr{queue\_Remove\_Command\_SingleQueued} in \cpm.}
	\label{fig:cpm:rule_queue_Remove_Command_SingleQueued}
\end{figure}

\begin{figure}
	\centering
	\begin{tikzbox}
		\includegraphics{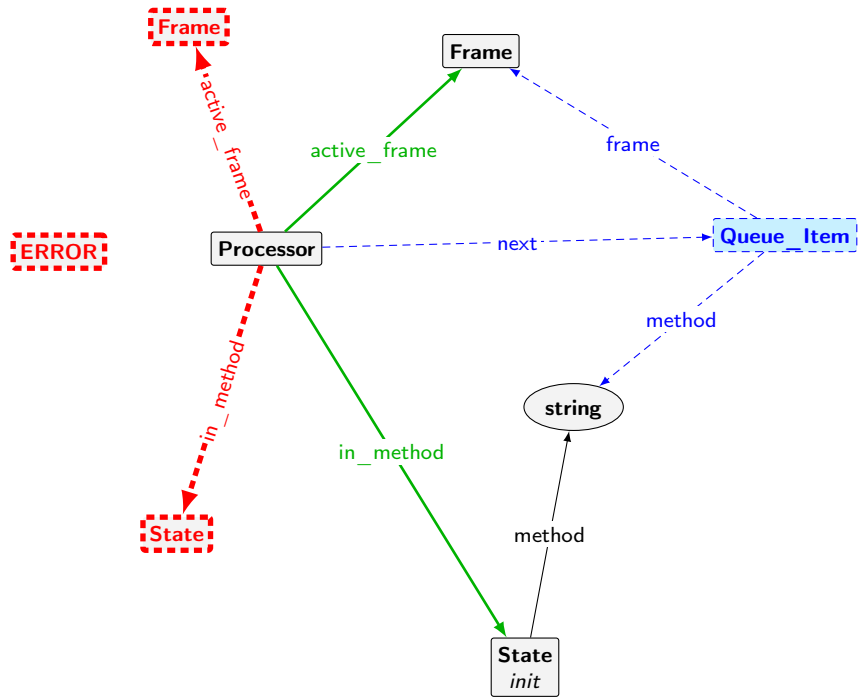}
	\end{tikzbox}
	\caption[Rule \gr{queue\_Remove\_SingleQueued} in \cpmo]{Rule
	\gr{queue\_Remove\_SingleQueued} in \cpmo.}
	\label{fig:cpmo:rule_queue_Remove_SingleQueued}
\end{figure}

Once a processor has queue items attached, the scheduling rules for queues are
applied. These work analogously to the \cpm counterparts. For example,
Figure~\ref{fig:cpm:rule_queue_Remove_Command_SingleQueued} shows the rule
\gr{queue\_Remove\_SingleQueued} in \cpm, while
Figure~\ref{fig:cpmo:rule_queue_Remove_SingleQueued} shows the corresponding
rule from \cpmo. In \cpmo, we not only point the processor to the routine that
should be executed, but also set the active frame edge to the frame attached
to the request, thus providing information about the routine arguments and the
target object (this is needed since the processor may be handling multiple
objects and needs to know to which one \eif{Current} refers to). Note that,
since we attach certain information related to the request type to the frame
instead of directly attaching it to the request, \cpmo requires only two
general \gr{queue\_Remove\_\ldots} rules (as opposed to different rules for
queries and commands in \cpm).

\begin{figure}
	\centering
	\begin{tikzbox}
		\includegraphics{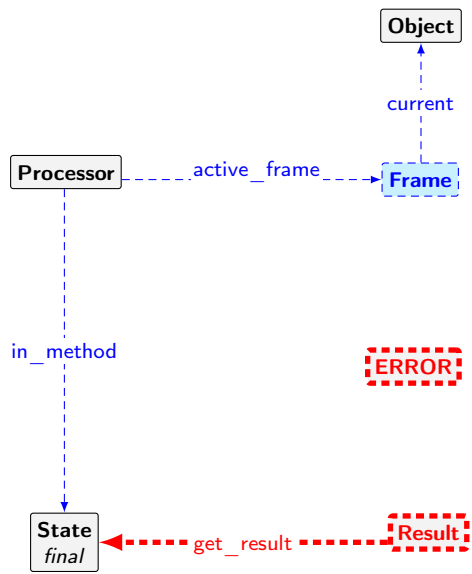}
	\end{tikzbox}
	\caption[Rule \gr{cleanup\_FinalState\_Command\_Empty\_Call\_Stack}]{Rule
		\gr{cleanup\_FinalState\_Command\_Empty\_Call\_Stack}.}
	\label{fig:cpmo:rule_cleanup_FinalState_Command_Empty_Call_Stack}
\end{figure}

After the execution of a request, several rules take care of cleaning up the
frame and handling of possible return values.  For example, the rule
\gr{clea\-nup\_Fi\-nalState\_Commands\_Empty\_Call\_Stack}, shown in
Figure~\ref{fig:cpmo:rule_cleanup_FinalState_Command_Empty_Call_Stack}, shows
how the stack frame is deleted after a command when the processor becomes
idle.

\begin{figure}
	\centering
	\begin{tikzbox}
		\includegraphics{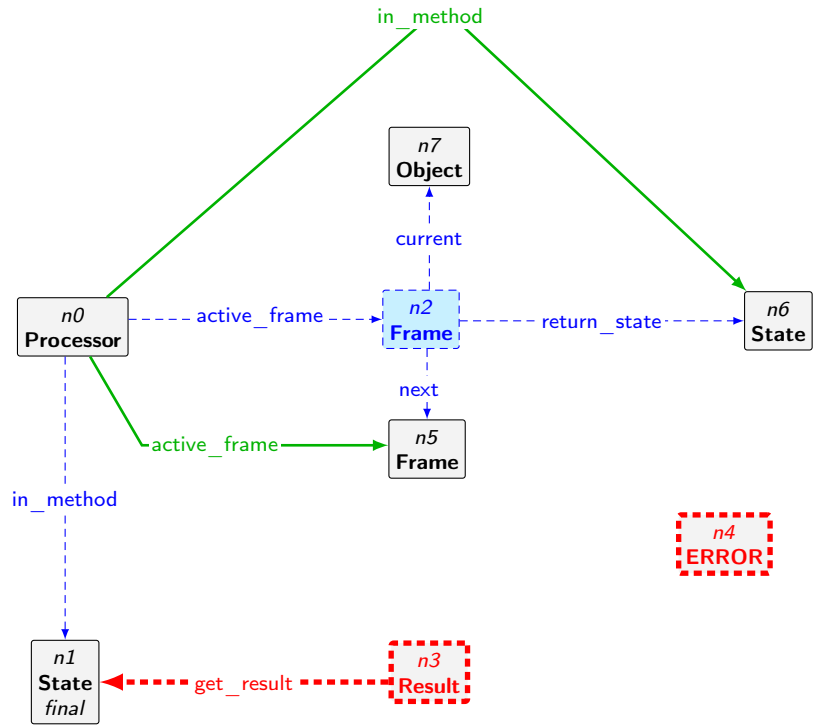}
	\end{tikzbox}
	\caption[Rule \gr{cleanup\_FinalState\_Command}]{Rule
		\gr{cleanup\_FinalState\_Command}.}
	\label{fig:cpmo:rule_cleanup_FinalState_Command}
\end{figure}

\subsubsection{Non-separate Calls}

Thanks to stack frames, we can now also simulate non-separate calls and local
calls (calls with target \eif{Current}). These commands and queries, as per
the formal semantics, do not create a feature request that is enqueued in the
processor's request queue. Instead, they are directly (and sequentially)
executed by the calling processor. We stay with the command example, but the
case for queries is again similar. The rule
\gr{action\_Command\_non-separate}, shown in
Figure~\ref{fig:cpmo:rule_action_Command_non-separate}, handles command calls.
Again, a stack frame is created. But instead of creating a new request, the
processor updates its current frame pointer to the newly created frame, and
starts executing the desired method body. The created frame points to the
current frame via \grgr{next} edge. In addition, the created frame also
includes an edge to the state after the command node, labelled
\grgr{return\_state}. This allows returning to the correct position once the
command finishes and the stack frame is removed.

Once a procedure has been processed and the processor points to a final state
node, several high-priority scheduling rules may be applied, depending on the
state of the call stack and the type of feature (command or query).  These
rules (which in fact are the same ones that handle the analogous case for
separate features) start with the prefix \gr{cleanup\_FinalState\_\ldots} and
pop the current frame from the frame stack.
Figure~\vref{fig:cpmo:rule_cleanup_FinalState_Command} shows the
\gr{cleanup\_FinalState\_Command} rule that deletes a frame and instructs the
processor to continue at the position after the call in the calling procedure,
and resets the active frame edge to point to the original stack frame.

\begin{sidewaysfig}
	\centering
	\begin{tikzbox}
		\includegraphics{figures/cpmo_rule_action_Command_separate.tikz}
	\end{tikzbox}
	\caption[Rule \gr{action\_Command\_separate}]{Rule
		\gr{action\_Command\_separate}.}
	\label{fig:cpmo:rule_action_Command_separate}
\end{sidewaysfig}

\begin{sidewaysfig}
	\centering
	\begin{tikzbox}
		\includegraphics{figures/cpmo_rule_action_Command_non-separate.tikz}
	\end{tikzbox}
	\caption[Rule \gr{action\_Command\_non-separate}]{Rule
		\gr{action\_Command\_non-separate}.}
	\label{fig:cpmo:rule_action_Command_non-separate}
\end{sidewaysfig}

\subsection{Dynamic Object Creation and Variable Names}

When creating processors and associated data nodes in \cpm, the rules
\gr{action\_New\_Void} and \gr{action\_New\_Attached} are applied to create a
fixed number of reference and data variable nodes. To map a certain program to
\cpm, one has to first determine how many variables are required to represent
the program and then adjust these rules to make sure that enough variables are
available for the mapping. If a program involves several classes, \cpm does
not distinguish between them when creating data and handlers. Instead, all
processors get the maximum number of data and reference variables.

\begin{figure}
	\centering
	\begin{tikzbox}
		\includegraphics{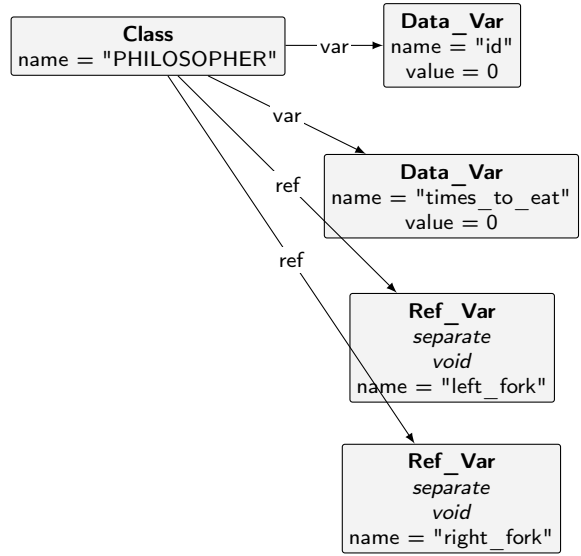}
	\end{tikzbox}
	\caption{Object template for the \eif{PHILOSOPHER} class.}
	\label{fig:cpmo:object_template_philosopher}
\end{figure}

To make direct translations of \scoop programs easier and interpreting
generated start graphs more intuitive, we introduce variable names in \cpmo
that can be directly mapped to the names in the source code. To achieve this,
we encode the reference, Boolean, and integer attributes of classes in the
start graph itself. An example of this can be seen in
Figure~\vref{fig:cpmo:object_template_philosopher}, where we include all
variables relevant to the \eif{PHILOSOPHER} class. We call these constructs
\emph{object templates}.

To make use of object templates, we modify the semantics for object creation
slightly. The rules \gr{action\_New\_From\_Template} and
\gr{action\_New\_Lo\-cal\_From\-\_Template} handle object creation for attributes
and local variables respectively. We attach the class name to
\grbf{Action\_New} nodes, denoting the type of object that we want to create.
Then, the rule matches the template with the corresponding name, copies the
template variables, and attaches them to the newly created object. The created
object now contains attributes corresponding to its declared type.  We can use
arbitrary variable names as opposed to being restricted to generic ones as in
\cpm. In addition, objects now only have variables relevant to their types
attached.

\subsection{Generic Operators}

In \cpm, actions that require arguments, such as assignments or commands, are
limited in that the arguments of these actions, including targets for queries
and commands, can only be local data or reference variables, or simple
operations like addition or negation. As a consequence, expressions that do
not conform to this restriction have to be split up. For example, to
represent the statement
\eif{account.withdraw (account.balance)},
we require two actions, the first one
being a query for \eif{account.balance} that is assigned to a (temporary)
variable, and the second one being the command with this variable as an
argument.

To enable more generic expressions, we change the representation of parameter
nodes (see Figures~\vref{fig:cpm:type_graph_system_state}
and~\vref{fig:cpmo:type_graph_variables_parameters_results} for the relevant
type graphs of \cpm and \cpmo respectively). In \cpmo, we can use the
supertype \grbf{Param} instead of using specific types. We still have the
\grbf{Param\_Ref} and \grbf{Param\_Data} types, as well as analogous
\grbf{Param\_Local\_Ref} and \grbf{Param\-\_Local\-\_Data} types, which can be
used to represent and fetch attributes and local reference and data values. In
addition, and this is where the added flexibility comes from, we also provide
a \grbf{Param\_Expr} type, which has an \gr{expr} edge to the \grbf{Super\_Op}
type. With this addition, we can now use arbitrary operations to fetch
parameter data as opposed to only local and attribute values.  Consequently,
command actions and query operations do not specify the target via string
any more. Instead, as seen in the type graphs in
Figures~\vref{fig:cpmo:type_graph_processors_and_objects}
and~\vref{fig:cpmo:type_graph_operations}, these nodes now have a
\gr{target} edge that points to a \grbf{Param\_Expr} node which specifies the
target. Once targets and parameters are evaluated, the behaviour is the same
as if one would have used the \grbf{Param\_Ref} and \grbf{Param\_Data} types
from \cpm.

\begin{figure}
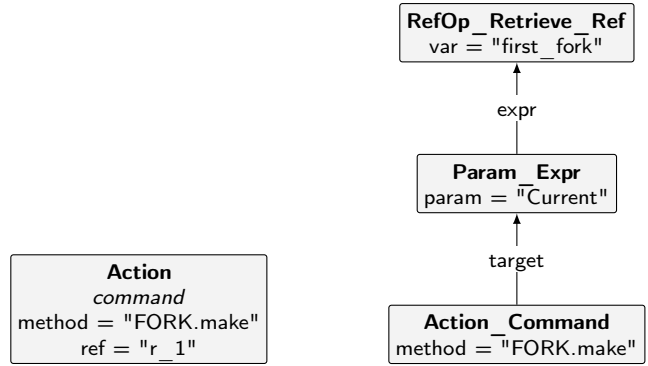

	\centering
	\begin{subfigure}[t]{0.4\textwidth}
		\centering
		\begin{tikzbox}
			\includegraphics{figures/cpmo_cpm_command_example.tikz}
		\end{tikzbox}
		\caption{\cpm command with target specified as an edge
			(\gr{let:ref="r\_1"}).}
	\end{subfigure}
	\begin{subfigure}[t]{0.4\textwidth}
		\centering
		\begin{tikzbox}
			\includegraphics{figures/cpmo_command_example.tikz}
		\end{tikzbox}
		\caption{\cpmo command with generic target via \grbf{RefOp} node.}
	\end{subfigure}
	\caption[Command comparison between \cpm and \cpmo]{Comparison of a command
	node in \cpm and \cpmo.}
	\label{fig:cpmo:command_comparison}
\end{figure}

Figure~\vref{fig:cpmo:command_comparison} shows how a \grbf{RefOp} node can be
used to specify the target, as opposed to a simple string with an attribute
variable name (which is how targets are handled in \cpm).

\subsection{Lock Passing}
\label{section:cpmo:lock_passing}

Lock passing is necessary to avoid deadlock in certain situations. For
example, when an object \eif{a} holds the locks of the handlers of \eif{b} and
\eif{c}, and creates a query request on \eif{b} that in turn requires locking
of \eif{c}, then \eif{b} can not proceed until \eif{a} releases the lock on
\eif{c}, which in turn can not happen until \eif{b} completes the request. No
processor can make progress, and a deadlock has occurred.

To solve this problem, Morandi \cite{Morandi2014} defines a lock passing
mechanism for feature calls. In his thesis, he defines feature calls as
follows \cite[p.19]{Morandi2014} (edited to reflect only relevant parts and
for formatting):

A client $p$ performs the following steps to \textit{call a feature} $f$ with
target expression $e_0$ and argument expressions $e_1, \cdots, e_n$.
\begin{enumerate}
	\item \textit{Target evaluation}. Evaluate the target expression $e_0$ with
		supplier $q$.
	\item \textit{Argument passing}. Evaluate the argument expressions and bind
		them to the formal arguments.
	\item \textit{Lock passing}. Pass all locks to $q$ if a controlled argument
		expression gets attached to an attached formal argument of reference type.
	\item \textit{Feature request}. Generate a feature request to apply $f$ to
		the target.
		\begin{itemize}
			\item If the feature call is non-separate, i.e. $p = q$, then ask $q$ to
				process the feature request immediately using its call stack and wait
				for termination.
			\item Otherwise, add the request to the end of $q$'s request queue.
		\end{itemize}
	\item \textit{Wait by necessity}. If $f$ is a query, then wait for the
		result.
	\item \textit{Lock revocation}. If lock passing happened, then wait for the
		locks to come back.
\end{enumerate}

\begin{figure}
	\centering
	\begin{tikzbox}
		\includegraphics{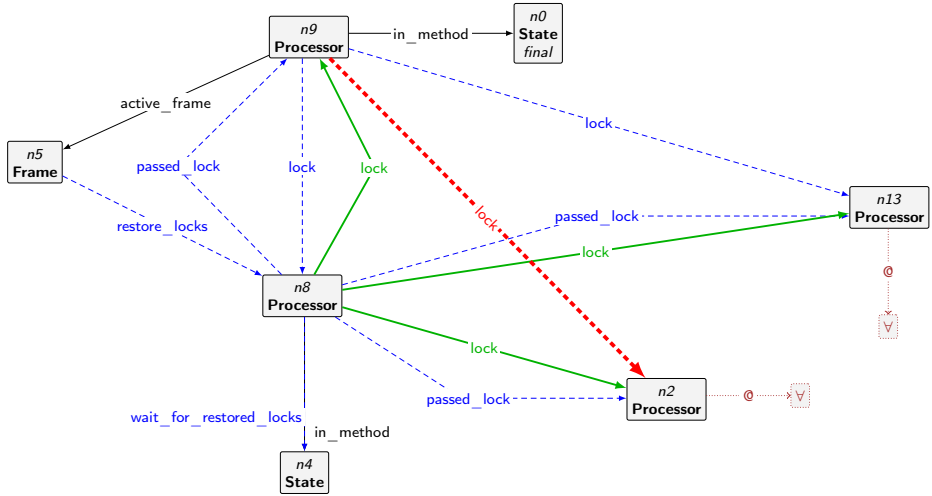}
	\end{tikzbox}
	\caption[Rule \gr{restore\_locks\_b}]{Rule \gr{restore\_locks\_b}.}
	\label{fig:cpmo:rule_restore_locks_b}
\end{figure}

\begin{figure}
	\centering
	\begin{tikzbox}
		\includegraphics{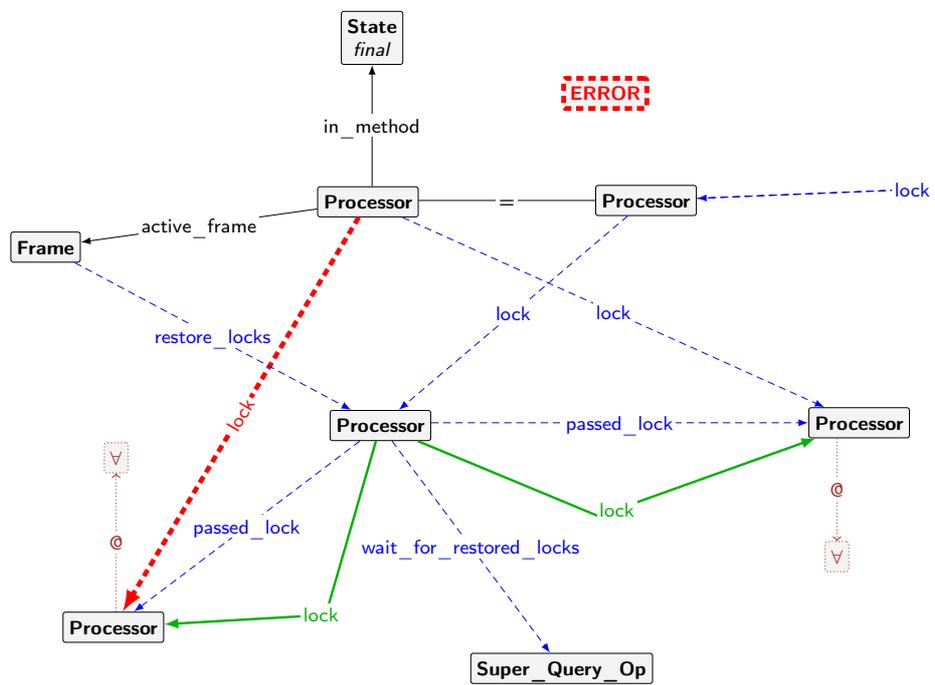}
	\end{tikzbox}
	\caption[Rule \gr{cleanup\_Restore\_Locks\_Query}]{Rule \gr{cleanup\_Restore\_Locks\_Query}.}
	\label{fig:cpmo:rule_cleanup_restore_locks_query}
\end{figure}

To achieve this behaviour, we introduce a number of rules that interact with
each other.
\begin{description}
	\item[\gr{pass\_locks}] This rule, shown in
		Figure~\ref{fig:cpmo:rule_pass_locks}, matches if at least one
		of the attached arguments is controlled, i.e., if the current processor
		(node $n0$) has a lock on some processor (node $n1$) that handles an
		object that is passed to the command. The target processor (node $n2$)
		receives all locks (\grgr{lock} edges) from the client processor. In
		addition, several edges are created for bookkeeping purposes. Edges
		labelled \grgr{passed\_lock} point to processors that the client held a
		lock on and are required to know which locks to restore. The receiving
		processor has an edge \grgr{restore\_locks} that points to the client,
		which allows giving back the locks to the correct processor later.
		Finally, an edge \grgr{wait\_for\_restored\_locks} is added in order to
		make sure that the client does not continue execution before getting the
		locks back (which is ensured by a \grre{wait\_for\_restored\_locks}
		embargo edge in each action rule).
	\item[\gr{pass\_locks\_query}] Analogously to the previous rule, this one
		passes the locks for queries. While the previous one only passes locks if
		the arguments fulfill the requirements, this rule always passes the locks
		regardless of the arguments. This decision follows Clarification~5.4.2 in
		\cite[p. 115]{Morandi2014}.
	\item[\gr{restore\_locks\_\ldots}] These two rules handle restoring rules
		for commands when lock passing has occurred.
		Figure~\ref{fig:cpmo:rule_restore_locks_b} shows one of the rules, where
		the client (node $n8$) gets all the locks that have been passed earlier.
		To do this, edges \grbl{restore\_locks} (which are created when creating
		the queue item with commands and queries that require lock passing) from
		the frame from the original request need to be present, which ensures
		that the locks get restored at the right point in the call stack even if
		the request triggers further feature calls.
	\item[\gr{cleanup\_Restore\_Locks\_Query}] Similar as before, this rule
		(Figure~\ref{fig:cpmo:rule_cleanup_restore_locks_query}) handles restoring
		locks of query requests.
\end{description}

Note that we do not support separate callbacks as described in the original
semantics.

The above rules all have higher priority than action and query
rules, which ensures that whenever the system is in a state where lock passing
or revocation is required, it will apply the corresponding rule. It is
therefore impossible for the system to miss passing the locks or continue
without restoring the locks first. The latter is additionally enforced by
adding embargo edges (\grre{wait\_for\_restored\_locks} in action and query
nodes), which catches the case where locks are not properly restored due to no
restore lock rule being applicable (this situation would indicate a problem
with the \cpmo model, not with the semantics or the inspected program).

\begin{sidewaysfig}
	\centering
	\begin{tikzbox}
		\includegraphics{figures/cpmo_rule_pass_locks.tikz}
	\end{tikzbox}
	\caption[Rule \gr{pass\_locks}]{Rule \gr{pass\_locks}, that matches and is
	applied when at least one of the parameters is controlled.}
	\label{fig:cpmo:rule_pass_locks}
\end{sidewaysfig}

\subsection{Distinguishing Preconditions and Wait Conditions}

In \scoop, a statement in a \eif{require} block involving separate arguments
can either be a precondition or a wait condition, depending on the context from
which it is called. If the processor executing the statement already held
request queue locks to all involved processors of separate arguments before the
call, then no other processor can enqueue requests to the queues of those
handlers. As a result, the outcome of the statement can not change over time,
and therefore it is a precondition. But in the case where the calling processor
does not hold all locks, the result may change over time, and therefore the
statement is a wait condition.

In \cpmo, we distinguish between preconditions and wait conditions. The
\grbf{Action\_Test} nodes that denote the path that is to be followed if a
precondition or wait condition evaluates to \eif{False} are always marked with
a \grit{precondition\_fail} flag. This does not necessarily mean, as described
above, that the statement is in fact a precondition. Instead, the rule
\gr{action\_Test} determines, based on the participating parameters, whether
the statement is a precondition or a wait condition. In the first case, the
rule creates an \grbf{ERROR} node, stating that a precondition has failed, and
as a result, the system is in a final state. But if the statement turns out to
be a wait condition, the rule handles it as such by following the action to the
next state. Eventually, the processor returns the locks (which gives other
processors the possibility to modify the state of the involved objects) before
acquiring them again and evaluating the wait conditions again.

\section{State-Space Optimisations}
\label{section:cpmo:statespace_optimisations}

The \statespace explosion problem is omnipresent in concurrent systems and
therefore it is also present in \cpm and \cpmo. Obviously, this is an issue
that one can not get rid of completely. Still, it can be of practical value to
try and mitigate the problem as much as possible. We implement a number of
optimisations that aim to reduce the \statespace problem by avoiding
unnecessary interleavings. Similar optimisations are already present in \cpm,
such as fine grained rule priorities for certain rules. Of course, we have to
pay attention to the interleavings that are left out with these optimisations.
In the following, we present additional measures taken to decrease the size of
the \statespace and argue why we are confident that they are not problematic
with regards to the properties that can be verified using the \cpmo model.

\subsubsection{Quantifier Usage}

\begin{figure}
	\centering
	\begin{tikzbox}
		\includegraphics{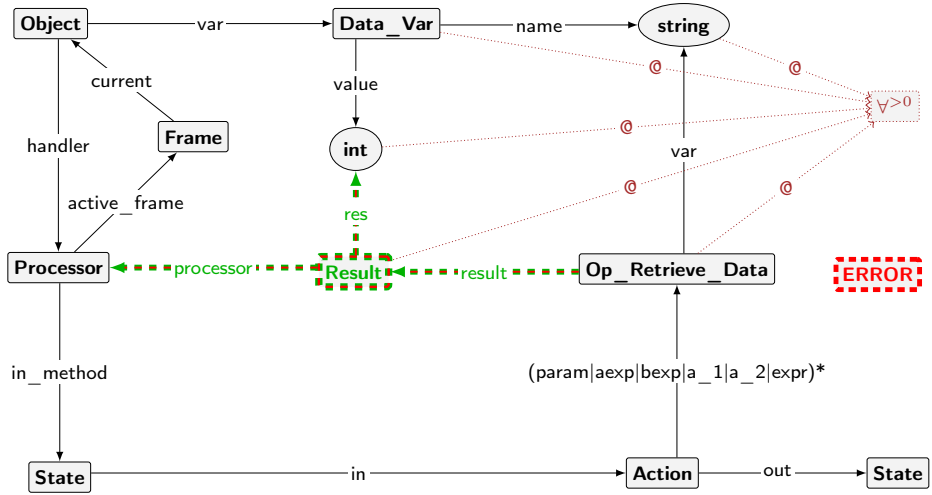}
	\end{tikzbox}
	\caption[Rule \gr{IntOp\_RetrieveData}]{Rule \gr{IntOp\_RetrieveData.}}
	\label{fig:cpmo:rule_IntOp_RetrieveData}
\end{figure}

In \groove, quantifiers can be used to create flexible rules, such as
ones that match a type of node several times, or ones that express a logical
``or'', matching one part of the rule or another. While there are several
situations where quantifiers are used in \cpm, we extended this usage further
in \cpmo to help reduce the size of the \statespace. The rule
\gr{IntOp\_RetrieveData} (which is known as \gr{aexp\_RetrieveData} in \cpm)
is an example, shown in
Figure~\ref{fig:cpmo:rule_IntOp_RetrieveData}, that uses a simple
\gr{$\forall^{>}$} quantifier with several nodes attached, which means that
the rule now creates \grbf{Result} nodes for each \grbf{Op\_Retrieve\_Data}
attached to the action that is matched. In a situation with multiple
\grbf{Op\_Retrieve\_Data}, \cpmo creates all result nodes in one single rule
application, whereas in \cpm, the \statespace diverges at that point,
exploring the different orders of creating these nodes, and converges once all
result nodes have been created (since the resulting state is always the same,
regardless of the order).

\subsubsection{Scheduler Optimisation}

Suppose we have a dining philosophers instance with only two philosophers (and
thus with two forks). We are interested in the different interleavings that
can occur with regards to locking, or more general, we are interested in the
interleavings possible where philosophers and forks and their processors
interact with each other by locking request queues. When simulating the
instance with \cpm or any version of \cpmo, all the relevant interleavings are
captured. Unfortunately, \cpm and \cpmo without scheduler optimisations also
explore a large number of interleavings that we are not interested in, as they
cover the same behaviour with respect to the outcome of the program.  For
example, one execution could execute the complete code of the first
philosopher before the second one starts. In the next interleaving, the second
philosopher might execute a (local) identifier assignment, then the first one
executes everything, before the second one executes the remaining part. With
respect to the interaction of processors and objects, these two interleavings
are equal.

More generally, we can split actions and queries in two groups: non-separate
and separate. In the first case, everything is handled on the current
processor, no other processors are involved. In the latter case, different
processors may be included and therefore, we want to explore all possible
executions (since the outcome of executing \scoop programs is determined by
the order in which requests are enqueued). The idea behind this optimisation
is that as long as a processor is executing locally (e.g.\ a philosopher
initializing by performing integer assignments for the identifier and number
of times to eat attributes) without having an impact on any processor's
request queue, we can advance it as far as possible. Once a
processor is at a point where a non-separate action or query is about to be
executed, it waits until all other processors have reached a similar position
(or have finished executing and are idle). All processors that are not idle
are potentially about to interact with other processors. At this point, it is
important to explore all interleavings, as different orders in locking and
enqueuing requests may result in different situations (e.g.\ in the dining
philosophers example, when the first philosopher has a fork and is about to
pick up the second, but the second philosopher is about to pick up the same
fork, we want to explore both situations where one or the other philosopher
``wins'').

\paragraph{Implementation}

We implement this idea by introducing an \emph{execution token} and organising
the processors in a linked list. In a \cpmo state, at most one processor has a
\grit{token} flag, which denotes that it is allowed to perform non-separate
steps. To achieve this, we give the non-separate rules a higher priority and
add the token as a requirement to match. These rules (e.g.\ rule
\gr{action\_Command\_non-separate}) can only be executed by the processor that
has the token. Once a processor can not perform more non-separate actions, it
passes the token to the next processor. This is repeated, until no processor
can make non-separate progress any more. Once this is the case, the separate
rules (such as \gr{action\_Command\_separate}) can be applied. These do not
require the token, which means multiple rules may be applicable in a given
state, and since these rules all have the same priority, all interleavings are
explored by the system.

\begin{figure}
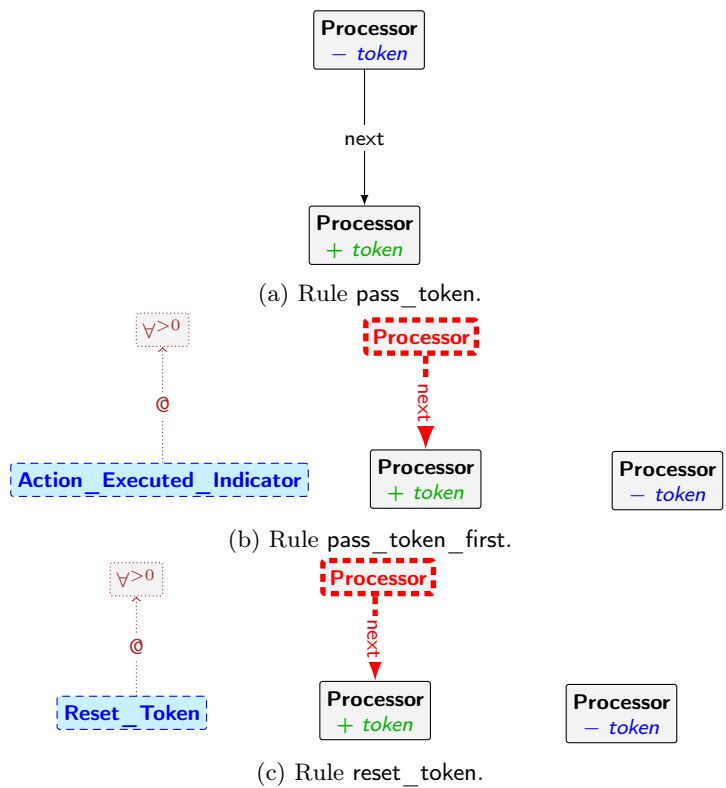

	\centering
	\begin{subfigure}[t]{\textwidth}
		\centering
		\begin{tikzbox}
			\includegraphics{figures/cpmo_rule_pass_token.tikz}
		\end{tikzbox}
		\caption{Rule \gr{pass\_token}.}
	\end{subfigure}\\
	\begin{subfigure}[t]{\textwidth}
		\centering
		\begin{tikzbox}
			\includegraphics{figures/cpmo_rule_pass_token_first.tikz}
		\end{tikzbox}
		\caption{Rule \gr{pass\_token\_first}.}
	\end{subfigure}\\
	\begin{subfigure}[t]{\textwidth}
		\centering
		\begin{tikzbox}
			\includegraphics{figures/cpmo_rule_reset_token.tikz}
		\end{tikzbox}
		\caption{Rule \gr{reset\_token}.}
	\end{subfigure}
	\caption[Token movement rules]{Rules handling token movement.}
	\label{fig:cpmo:rules_pass_token}
\end{figure}

In the following, note that non-separate rules have priority 4 and separate
rules have priority 1.  The rules \gr{pass\_token}, \gr{pass\_token\_first},
and \gr{reset\_token} are relevant for this mechanism. These rules with
priorities 3, 2, and 0 respectively handle the movement of the token flag
along processors and are shown in Figure~\ref{fig:cpmo:rules_pass_token}.
The token is cycled until there is one full cycle where no processor has made
progress. To achieve this, we use a node of type
\grbf{Action\_Executed\_Indicator}. Such a node is created whenever a
\nonseparate action is performed, e.g.\ by the rule \gr{action\_AssignRef}.
When the token is on the last processor, it gets moved to the first one (thus
restarting the cycle) only if there is an indicator node. When the token has
run through the list without an action having been performed, the rule is not
applicable anymore, which enables rules with lower priority, in particular the
separate rules. At this point, all different interleavings between
separate rules are explored as intended. Similar to the
\grbf{Action\_Executed\_Indicator}, separate actions create a
\grbf{Reset\_Token} node that indicates that a separate step has been
performed, which means that some processor may possibly continue with
non-separate steps. If such a node exists, the \gr{reset\_token} rule can be
applied which removes the node and puts the token back on the first processor,
restarting the cycle to perform non-separate steps. In case there is no such
\grbf{Reset\_Token} node, the rule \gr{cleanup\_token} is applied that removes
the token from the processor holding it, ensuring that there are not several
final configurations with the only difference being that the token is on a
different processor.

As we will see later in Chapter~\ref{chapter:evaluation}, this optimisation
results in a huge improvement in the size of the \statespace and allows us to
verify programs where the \statespaces have been too big before.

We argue that this mechanism does not leave out interleavings of interest, as
all the possible sequences in which requests get added to queues are
preserved.  By forcing a certain order for local computations, where the order
of execution has no influence on the outcome, we can avoid exploring a large
number of states.

\section{Rules}

Here, we provide a complete overview of all rules and their
priorities in the \cpmo system. Together with the type graph we have
presented, this makes up the complete \gts of \cpmo. While we discuss certain
rules in detail, we do not show graphs for every rule, but instead refer the
interested reader to the supplementary material repository \cite{onlinerepo}.
In addition, we also present the priorities of the rules as we have done for
\cpm.

We divide the rules in \cpmo into the following categories and discuss them in
the subsequent sections.
\begin{itemize}
	\item Control Flow
	\item System State
	\item Queries and Other Operations
	\item Optimisations
	\item Errors
	\item Configuration
\end{itemize}

\subsection{Control Flow}

Control flow rules handle movement along feature graphs, in particular moving
from a state via an action node to the next state.
Table~\ref{table:cpmo:priorities_control_flow} summarizes these rules and
their priorities. Most rules have direct counterparts in \cpm, although new
variants have been introduced to handle additional features like non-separate
calls. What follows is a short discussion of the rules.

\begin{description}
	\item[\gr{action\_Assign\_Data}] This rule handles an integer or
		Boolean assignment operation where the target is an attribute of the
		current object.
	\item[\gr{action\_Assign\_Local\_Data}] Similarly, this rule handles integer
		and Boolean assignments to variables declared in the \eif{local} block.
		The rule accesses local data stored on the frame instead data stored on
		the object, since local declarations are only valid within the context of
		the current call and therefore of the current call stack frame.
	\item[\gr{action\_AssignRef}] Analogously to the previous two, this rule
		handles reference assignments. While \cpm uses a number of rules for
		reference assignments (\gr{action\_AssignRef\_Ref\_\ldots} rules), the
		\cpmo rule is not injective, and we make additional use of quantifiers to
		express alternatives, which makes it possible to cover all cases with a
		single rule.
	\item[\gr{action\_Assign\_Local\_Ref}] Analogously to
		\gr{action\_Assign\_Local\_Data}, this rule handles the case for reference
		assignments to local variables.
	\item[\gr{action\_AssignResult\_\ldots}] For each data type, \cpmo has a
		rule for assignment to the special \eif{Result} value, representing return
		values in queries. Figure~\ref{fig:cpmo:rule_action_AssignResult_Ref}
		shows the rule for reference values.
	\item[\gr{action\_Command\_non-separate}]
	\item[\gr{action\_Command\_separate}]
	\item[\gr{action\_Command\_separate\_restore\_locks}] The non-separate
		command rule creates a new stack frame, sets it up with parameters, and
		puts it on top of the frame stack of the current processor. In addition,
		the rule points the current processor to the designated feature. This
		represents a local call that is executed immediately. The separate cases
		on the other hand create a feature request and attach the created frame to
		it. The request is attached to the target processor and will then get
		processed by queue management rules. The calling processor can proceed
		since the separate rules only match if the target processor differs from
		the calling processor, which means that the command is asynchronous.
	\item[\gr{action\_CreateRoot}] Since \cpmo allows specifying the root class
		and procedure, we need a rule that creates the initial object, which is
		what \gr{action\_CreateRoot} does. The configuration node with root class
		name and procedure name is deleted in this rule, ensuring that only one
		root object is created. The object is created according to the class
		template, similar to what the rule \gr{action\_New\_From\_Template} does.
	\item[\gr{action\_New\_From\_Template}]
	\item[\gr{action\_New\_Local\_From\_Template}] To instantiate attributes and
		local variables, these rules match \grbf{Action\_New} nodes and their
		context. The created object is then attached to the specified variable,
		where the first rule handles attributes and the second one local
		variables.
	\item[\gr{action\_Lock}] The lock action rule takes, as opposed to the lock
		rules in \cpm, a variable number of references. The rule can only be
		applied if all handlers of the specified objects are not locked. Applying
		this rule implies that all locks are obtained atomically.
		Figure~\ref{fig:cpmo:rule_action_Lock} shows the rule graph.
	\item[\gr{action\_Test}] A form of branching is provided with the test
		action, which works analogously to the \cpm test action.
	\item[\gr{action\_Noop}] This rule simply skips an \grbf{Action\_Noop}
		node and, as the name indicates, performs no real operation.
	\item[\gr{action\_TestPostcondition}] Like the \cpm rule of the same
		name, this one advances a processor in a final state to the first state of
		the postcondition, if the graph is configured to check postconditions.
	\item[\gr{action\_Unlock\_Creator}] 
	\item[\gr{action\_Unlock\_Creator\_non-separate}] Since created objects are
		locked by their creators, they need to have an unlock action as their last
		action in creation procedures, which removes the lock, allowing the creator
		to continue execution (since the creator immediately, by convention, has
		to follow with a pair of lock and unlock actions for the same object that
		was just created). These rules handle the separate and non-separate case.
	\item[\gr{action\_Unlock\_Expr}] If a feature obtains locks at the start,
		there are corresponding unlock actions that release them at the end of the
		feature. This rule handles the unlock actions. Not only does it release
		held locks, but in case the lock is not held (which can happen if several
		passed separate arguments have the same handler) the action becomes an
		empty operation.
\end{description}

Aside from these rules, the rules involved in lock passing belong to this
group. They are not repeated here, instead we refer to
Section~\ref{section:cpmo:lock_passing}, where they are described in detail.

\begin{figure}
	\centering
	\begin{tikzbox}
		\includegraphics{figures/cpmo_rule_action_AssignResult_Ref.tikz}
	\end{tikzbox}
	\caption[Rule \gr{action\_AssignResult\_Ref}]{Rule
		\gr{action\_AssignResult\_Ref}.}
	\label{fig:cpmo:rule_action_AssignResult_Ref}
\end{figure}

\begin{sidewaysfig}
	\centering
	\begin{tikzbox}
		\includegraphics{figures/cpmo_rule_action_Lock.tikz}
	\end{tikzbox}
	\caption[Rule \gr{action\_Lock}]{Rule
		\gr{action\_Lock}.}
	\label{fig:cpmo:rule_action_Lock}
\end{sidewaysfig}

\begin{table}
	\centering
	{\footnotesize
	\begin{tabular}{lr}\toprule
		\textbf{Rule} & \textbf{Priority}\\
		\midrule
		\gr{restore\_locks\_a} & 71 \\
		\gr{restore\_locks\_b} & 70 \\
		\gr{action\_TestPostcondition} & 20 \\
		\gr{prepare\_lock\_wait} & 19 \\
		\gr{pass\_locks\_query\_new} & 18 \\
		\gr{pass\_locks\_} & 17 \\
		\midrule
		\gr{action\_Assign\_Data} & 6 \\
		\gr{action\_Assign\_Local\_Data} & \\
		\gr{action\_Assign\_Local\_Ref} & \\
		\gr{action\_Assign\_Ref}  & \\
		\gr{action\_AssignResult\_Bool} & \\
		\gr{action\_AssignResult\_Int} & \\
		\gr{action\_AssignResult\_Ref} & \\
		\gr{action\_Command\_non-separate} & \\
		\gr{action\_CreateRoot} & \\
		\gr{action\_New\_From\_Template} & \\
		\gr{action\_New\_Local\_From\_Template} & \\
		\gr{action\_Noop} & \\
		\gr{action\_Test} & \\
		\gr{action\_Unlock\_Expr} & \\
		\midrule
		\gr{action\_Command\_separate} & 1 \\
		\gr{action\_Command\_separate\_restore\_locks} & \\
		\gr{action\_Lock} & \\
		\gr{action\_Unlock\_Creator} & \\
		\gr{action\_Unlock\_Creator\_non-separate} & \\
		\bottomrule
	\end{tabular}
	}
	\caption[Control flow rules]{Control flow rules.}
	\label{table:cpmo:priorities_control_flow}
\end{table}

\subsection{System State}

The rules in the system state group, listed in
Table~\ref{table:cpmo:priorities_system_state}, are concerned with queue
management and graph maintenance. The former includes rules that insert queue
items into the request queue and remove them when processing an item. The
latter deal with various graph states with leftover nodes, e.g.\ when a
processor has reached a final state and results need to be discarded.

\begin{description}
	\item[\gr{cleanup\_exp\_DiscardResults\_BoolOp}] 
	\item[\gr{cleanup\_exp\_DiscardResults\_Op}] 
	\item[\gr{cleanup\_exp\_DiscardResults\_RefOp}] 
	\item[\gr{cleanup\_exp\_DiscardResults\_RefOp\_Void}] 
		After evaluating an operation (in the form of \grbf{Super\_Op} nodes),
		it has a \grbf{Result} node attached which is then used by the action to
		process. Once a processor moves past the action node, these rules are are
		applied to remove the \grbf{Result} nodes since they are not used anymore.
	\item[\gr{cleanup\_remove\_Void}] In certain situations, it can happen that
		\grbf{Void} nodes are left without being connected to any part of the
		graph. These are removed by this rule in order to avoid creating several
		states in the \lts that only differ in the amount of unconnected
		\grbf{Void} nodes.
	\item[\gr{cleanup\_Frame\_Remove\_controls}] Frames can have \gr{controls}
		edges to processors which have been controlled prior to the call. This is
		required to determine whether a statement in the \eif{require} block is a
		pre- or wait condition.
	\item[\gr{cleanup\_FinalState\_\ldots}] A number of cleanup rules are
		applied once a processor reaches the final state of the procedure it is
		executing. They perform a range of tasks, including removing the current
		frame or setting the result value such that the calling processor has
		access to it. The following rule exist in this set.
		\begin{itemize}
			\item \gr{cleanup\_FinalState}
			\item \gr{cleanup\_FinalState\_with\_Return\_Value}
			\item \gr{cleanup\_FinalState\_BoolQuery}
			\item \gr{cleanup\_FinalState\_BoolQuery\_with\_next\_frame}
			\item \gr{cleanup\_FinalState\_Command\_Empty\_Call\_Stack} 
			\item \gr{cleanup\_FinalState\_Command}
			\item \gr{cleanup\_FinalState\_IntQuery}
			\item \gr{cleanup\_FinalState\_IntQuery\_with\_next\_frame}
			\item \gr{cleanup\_FinalState\_Local\_Data\_Objects}
			\item \gr{cleanup\_FinalState\_Local\_Ref\_Objects}
			\item \gr{cleanup\_FinalState\_Param\_Data\_Objects}
			\item \gr{cleanup\_FinalState\_Param\_Ref\_Objects}
			\item \gr{cleanup\_FinalState\_RefQuery}
			\item \gr{cleanup\_FinalState\_RefQuery\_with\_next\_frame}
		\end{itemize}
	\item[\gr{queue\_Insert\_EmptyBusy}] 
	\item[\gr{queue\_Insert\_NotEmpty}] When a client creates a request queue
		item, it does not actually insert the item directly into the request
		queue. Instead, rules like \gr{action\_Command\_separate} simply let the
		\grbf{Queue\_Item} point to the target processor via \gr{insert\_into}
		edge. The actual insertion into the queue, depending on whether it is
		currently empty or not, is performed with the
		\gr{queue\_Insert\_EmptyBusy} and \gr{queue\_Insert\_NotEmpty} rules
		respectively.
	\item[\gr{queue\_Remove\_SingleQueued}] 
	\item[\gr{queue\_Remove\_MultipleQueued}]
		Once the request queue has items, these rules are used to remove a
		queue item from the request queue and instruct the processor to start
		execution at the designated procedure. The first one handles the case
		where exactly one item is on the queue, the second one cases with more
		than one item on the queue.
	\item[\gr{prepare\_lock\_wait}] Before a lock action is performed, a
		\gr{wait} edge is inserted from the processor executing the action to the
		processors it intends to lock. These edges are in particular useful for
		detecting deadlock with the rule \gr{error\_deadlock}. These edges are
		deleted once the target processors are locked.
	\item[\gr{remove\_wait\_and\_lock}] When a processor is in a state before a
		lock action, it first creates a \gr{wait} edge that points to the
		processor it intends to lock. The rule \gr{action\_Lock} can only be
		applied if for all target processors, either the lock is already held, or
		a wait edge exists.  In the former case, the graph is not modified any
		further. This rule handles the situation where a processor has both a
		\gr{wait} and a \gr{lock} edge, in which case the \gr{wait} edge simply
		gets deleted.
\end{description}

\begin{table}
	\centering
	{\footnotesize
	\begin{tabular}{lr}\toprule
		\textbf{Rule} & \textbf{Priority}\\
		\midrule
		\gr{cleanup\_Remove\_Void} & 700 \\
		\midrule
		\gr{queue\_Insert\_EmptyBusy} & 600 \\
		\gr{queue\_Insert\_NotEmpty} & 590 \\
		\midrule
		\gr{cleanup\_exp\_DiscardResults\_RefOp\_Void} & 461 \\
		\gr{cleanup\_exp\_DiscardResults\_RefOp} & 460 \\
		\gr{cleanup\_exp\_DiscardResults\_Op} & 450 \\
		\gr{cleanup\_exp\_DiscardResults\_BoolOp} & 440 \\
		\midrule
		\gr{queue\_Remove\_MultipleQueued} & 160 \\
		\gr{queue\_Remove\_SingleQueued} & 159 \\
		\midrule
		\gr{cleanup\_Restore\_Locks\_Query} & 69 \\
		\gr{cleanup\_Frame\_Remove\_controls} & 66 \\
		\gr{cleanup\_FinalState\_IntQuery} & 65 \\
		\gr{cleanup\_FinalState\_IntQuery\_with\_next\_frame} & 64 \\
		\gr{cleanup\_FinalState\_RefQuery\_with\_next\_frame} & 63 \\
		\gr{cleanup\_FinalState\_RefQuery} & 62 \\
		\gr{cleanup\_FinalState\_BoolQuery\_with\_next\_frame} & 61 \\
		\gr{cleanup\_FinalState\_Objects\_with\_Return\_Value} & 60 \\
		\gr{cleanup\_FinalState\_BoolQuery} & 59 \\
		\gr{cleanup\_FinalState\_Local\_Data} & 58 \\
		\gr{cleanup\_FinalState\_Command} & 57 \\
		\gr{cleanup\_FinalState\_Command\_Empty\_Call\_Stack} & 56 \\
		\gr{cleanup\_FinalState} & 55 \\
		\gr{cleanup\_FinalState\_Param\_Data} & 53 \\
		\gr{cleanup\_FinalState\_Param\_Ref} & 52 \\
		\gr{cleanup\_FinalState\_Local\_Ref} & 51 \\
		\midrule
		\gr{remove\_wait\_and\_lock} & 16 \\
		\bottomrule
	\end{tabular}
	}
	\caption[System state rules]{System state rules.}
	\label{table:cpmo:priorities_system_state}
\end{table}

\subsection{Queries and Other Operations}

As in \cpm, we group rules related to queries and operations on integers,
Booleans, and references together.
Table~\ref{table:cpmo:priorities_queries_and_ops} lists all rules in this
group. A description of rules and rule families follows.

\begin{description}
	\item[\gr{BoolOp\_Query\_\ldots}] The Boolean query rules handle separate
		and non-separate queries, where a new frame is created. In the separate
		case, a request queue item is created and attached to the target
		processor, whereas in the non-separate case, the frame is put on the
		current processor's frame stack and the processor is instructed to start
		executing the query.
	\item[\gr{BoolOp\_RetrieveData}] Similar to other \gr{RetrieveData} rules in
		both \cpm and \cpmo, this rule fetches attributes of the current object of
		Boolean types.
	\item[\gr{BoolOp\_\ldots}] Other Boolean operations include constants,
		conjunction, disjunction, equality, and others. These rules, with their
		arguments evaluated, perform the corresponding operation and attach the
		result to the matched \grbf{BoolOp} node.
	\item[\gr{IntOp\_\ldots}] Similar to the rules handling Boolean operations,
		these rules handle various integer operations, such as simple addition.
	\item[\gr{RefOp\_\ldots}] Analogously, a number of rules handle fetching
		references. This includes getting attributes or local references, but also
		creating query requests.
	\item[\gr{getlocal\_\ldots}] The \gr{getlocal\_Data} and \gr{getlocal\_Ref}
		rules prepare instances for local variables, which are attached to  the
		created frame later when a command or query is called.
	\item[\gr{getparam\_\ldots}] Once the values of \grbf{Param} nodes have been
		evaluated, these rules are applied to create instances in the form of
		\grbf{Param\_Data} and \grbf{Param\_Ref} nodes which are then passed to
		the called query or command.
\end{description}

\begin{table}
	\centering
	{\footnotesize
	\begin{tabular}{lr}\toprule
		\textbf{Rule} & \textbf{Priority}\\
		\midrule
		\gr{getparam\_Expr\_Op} & 432 \\
		\gr{getparam\_Expr\_BoolOp} & 431 \\
		\gr{getparam\_Expr\_RefOp} & 429 \\
		\gr{getparam\_Local\_Ref} & 413 \\
		\gr{getparam\_Local\_Data} & 412 \\
		\gr{getparam\_Data} & 411 \\
		\gr{getlocal\_Data} & 410 \\
		\gr{getlocal\_Ref} & 409 \\
		\midrule
		\gr{IntOp\_constant} & 400 \\
		\gr{BoolOp\_constant} & 390 \\
		\gr{RefOp\_RetrieveRef\_Local} & 386 \\
		\gr{RefOp\_RetrieveParam} & 385 \\
		\gr{RefOp\_RetrieveRef\_Void} & 384 \\
		\gr{RefOp\_RetrieveRef} & 383 \\
		\gr{IntOp\_RetrieveParam} & 382 \\
		\gr{IntOp\_RetrieveLocalData} & 381 \\
		\gr{IntOp\_RetrieveData} & 380 \\
		\gr{IntOp\_RetrieveData\_with\_Target} & 379 \\
		\gr{IntOp\_Multiply} & 351 \\
		\gr{IntOp\_Add} & 350 \\
		\gr{IntOp\_Subtract} & 340 \\
		\gr{BoolOp\_RetrieveData} & 334 \\
		\gr{BoolOp\_And} & 333 \\
		\gr{BoolOp\_GreaterEquals} & 332 \\
		\gr{BoolOp\_Equals} & 331 \\
		\gr{BoolOp\_GreaterThan} & 330 \\
		\gr{BoolOp\_Equals\_Ref\_False} & 329 \\
		\gr{BoolOp\_Equals\_Ref\_True} & 329 \\
		\gr{BoolOp\_Equals\_Ref\_Void} & 329 \\
		\gr{BoolOp\_LessEquals} & 321 \\
		\gr{BoolOp\_LessThan} & 320 \\
		\gr{BoolOp\_Not} & 310 \\
		\midrule
		\gr{BoolOp\_Query\_debug\_non-separate} & 0 \\
		\gr{BoolOp\_Query\_separate} & \\
		\gr{BoolOp\_Query\_separate\_restore\_locks} & \\
		\gr{IntOp\_Query} & \\
		\gr{IntOp\_Query\_separate} & \\
		\gr{IntOp\_Query\_separate\_restore\_locks} & \\
		\gr{RefOp\_Query\_non-separate} & \\
		\gr{RefOp\_Query\_separate} & \\
		\gr{RefOp\_Query\_separate\_restore\_locks} & \\
		\bottomrule
	\end{tabular}
	}
	\caption[Query and operation rules]{Query and operation rules.}
	\label{table:cpmo:priorities_queries_and_ops}
\end{table}

\subsection{Optimisations}

Optimisation rules are those involved in handling the execution token
discussed earlier, and are listed along with their priorities in
Table~\ref{table:cpmo:optimisations}. A thorough discussion of the involved
types and rules is given in
Section~\ref{section:cpmo:statespace_optimisations}.

\begin{table}
	\centering
	{\footnotesize
	\begin{tabular}{lr}\toprule
		\textbf{Rule} & \textbf{Priority}\\
		\midrule
		\gr{pass\_token} & 3 \\
		\gr{pass\_token\_first} & 2 \\
		\gr{cleanup\_token} & 0 \\
		\gr{reset\_token} & 0 \\
		\bottomrule
	\end{tabular}
	}
	\caption[\Statespace optimisation rules]{\Statespace optimisation rules.}
	\label{table:cpmo:optimisations}
\end{table}

\subsection{Errors}

In this group of rules, we collect error conditions. This includes properties
such as presence of a deadlock or a void call, which we are interested in when
verifying programs. In addition, we also have rules that aid us during
development and serve as ``sanity checks''. For example, the rule
\gr{debug\_multiple\_handlers} matches, if an object has more than one
handler. Since this situation is not possible according to the \scoop
specification, a
match of this rule means that there is an error in our model. Matching such
``bad states'' was used extensively during development to catch bugs, but the
corresponding rules have been removed from the final \gts.

The
priorities of the current error rules are listed in
Table~\vref{table:cpmo:priorities_errors}, a short description of them
follows.

\begin{figure}
	\centering
	\begin{tikzbox}
		\includegraphics{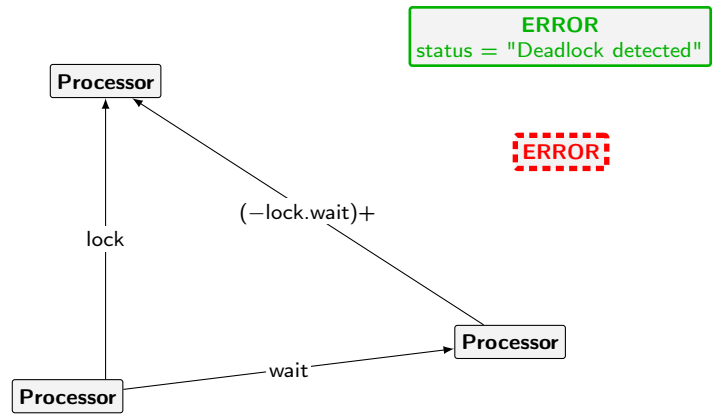}
	\end{tikzbox}
	\caption[Rule \gr{error\_deadlock}]{Rule \gr{error\_deadlock.}}
	\label{fig:cpmo:rule_error_deadlock}
\end{figure}

\begin{description}
	\item[\gr{error\_deadlock}] A large part of the motivation behind this work
		is detecting, amongst other properties, deadlocks in \scoop programs. This
		rule, shown in Figure~\ref{fig:cpmo:rule_error_deadlock}, detects
		deadlocks by matching, if a processor $n1$ has a lock on some processor
		$n4$, but is also waiting on a processor $n2$, which in turn is locked by
		some other processor (not shown, but expressed using the regular
		expression edge \gr{-lock.wait)+}) which again is waiting on $n4$. An
		example configuration where this rule matches is shown in
		Section~\ref{section:case_studies:dining_philosophers}.
	\item[\gr{error\_deadlock\_2} and \gr{error\_deadlock\_3}] Deadlock
		situations can not only occur in the above case where no processor is able
		to acquire locks and make progress. For example, when two processors
		execute the same feature which contains a wait condition that requires the
		other processor to finish this particular feature, then both processors
		can lock the request queue of the other one. They then both wait for the
		other one to handle the query request generated in the wait condition and
		therefore none of them makes progress. The rules \gr{error\_deadlock\_2}
		and \gr{error\_deadlock\_3} handle such situations for two or more
		processors respectively.
	\item[\gr{error\_PostconditionFail}] If a postcondition is evaluated to
		\eif{False}, this rule puts the processor in a special state of type
		\grbf{State\_Postcondition\_Fail} and creates an \grbf{ERROR} node with
		attached information about where the postcondition has failed.
	\item[\gr{error\_Command\_Void\_Target}] 
	\item[\gr{error\_Query\_Void\_Target}] If a target of a command or query has
		been evaluated to a void reference, then the call is invalid, which is
		detected and reported with these two rules.
\end{description}

\begin{table}
	\centering
	{\footnotesize
	\begin{tabular}{lr}\toprule
		\textbf{Rule} & \textbf{Priority}\\
		\midrule
		\gr{debug\_Multiple\_Handlers} & 1000 \\
		\gr{error\_Command\_Void\_Target} & \\
		\gr{error\_Deadlock} & \\
		\gr{error\_Deadlock\_2} & \\
		\gr{error\_Deadlock\_3} & \\
		\gr{error\_Postcondition\_Fail} & \\
		\gr{error\_Query\_Void\_Target} & \\
		\bottomrule
	\end{tabular}
	}
	\caption[Error rules]{Error rules.}
	\label{table:cpmo:priorities_errors}
\end{table}

\subsection{Configuration}

Currently, there is only one rule in this category, the rule
\gr{config\_Check\-Post\-con\-di\-tion}, which is applied if one specifies that
postcondition should be checked. It has a high priority and advances a
processor from a final state to the start of the postcondition, ensuring that
no cleanup rules are applied before the postconditions have been checked. The
processor will evaluate the postconditions and if everything evaluates to
true, end up in another final state where the normal cleanup rules can be
applied. If no such configuration node exists, this rule can not be applied
and the cleanup rules take place, ignoring possible postcondition related
parts of the graph.

\section{Testing}
\label{section:cpmo:testing}

The \cpmo model has been developed in an iterative fashion by adding features
described in this chapter to the \cpm model one by one. Changing the \gts is
error-prone. It is all too easy to alter the behaviour such that it does not
reflect the intended one any more by adding rules that contain bugs, changing
priorities that result in certain rules being applied in a state where we do
not want the rule to be applicable, or altering the type graph and rendering
existing rules useless. To ensure that the model stays true to the intended
behaviour, we use a number of start graphs representing test programs and
specify the expected output. For example, along with evolving the model, we
also evolve the examples of the dining philosophers with both the correct and
the deadlock implementation. Our testing utilities then explore the
\statespaces of these examples and match it against the expected behaviour
which checks properties like \statespace size, the number of final
configurations, and whether \grbf{ERROR} nodes are present in final
configurations.

Once we finalised the type graph for the current \cpmo model, we used this
testing approach in combination with our translation tool, described in
Chapter~\ref{chapter:translation}. This allows us to write \scoop programs and
specify the expected output of our \statespace exploration tool. The testing
utility then first translates the source code to a \cpmo start graph, and then
explores the \statespace and checks whether the actual output matches the
expected output.

\section{Future Work}
\label{section:cpmo:future_work}

With \cpmo at its current state, we are able to simulate a number of \scoop
features directly in the model, as opposed to simulating them using more basic
\cpm constructs. We added rules and types to \cpm that make the model more
expressive and allow start graphs that closely resemble the corresponding
\scoop source code.

To support more \scoop features, one possible way is to extend the \cpmo model
to directly support those features. This has the
advantage that programs that make use of those features can be represented
directly in a compact and readable fashion. Another approach is simulating
these features using existing \cpmo functionality. In our automatic
translation tool, it would require additional work to express features not
directly supported by \cpmo, resembling the work of traditional compilers.

We have several strategies in mind on how to implement certain missing
features. Since they often not only involve considerations regarding the \cpmo
model, but also the translation tool discussed in the next chapter, we
postpone a more thorough discussion of future work until
Section~\ref{section:translation:future_work}.

\myclearpage

\chapter{Translation}
\label{chapter:translation}
With \cpmo, we introduced object-oriented features of \scoop to the \cpm
model. Thanks to that effort, more \scoop programs can now be represented and
simulated using the model. Since both \cpm and the extensions that we
introduce in \cpmo are closely modelled after \scoop, mapping source code to
start graphs becomes a less tedious task. In this chapter, we discuss the
automatic translation tool that translates a subset of \scoop to \cpmo
start graphs.

\section{Overview}

\begin{figure}
	\centering
	\begin{adjustbox}{max height=\figureheight, max width=\figurewidth}
		\includegraphics{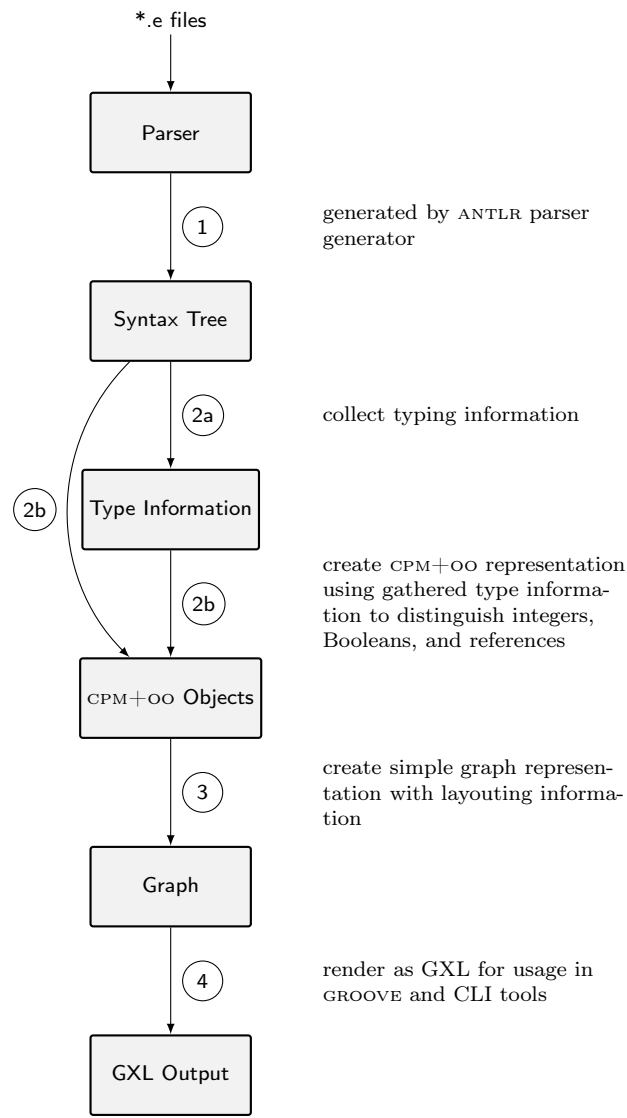}
	\end{adjustbox}
	\caption[Translation overview]{Overview of the steps included in translating
	a set of Eiffel classes to a \cpmo graph rendered as a \gxl file.}
	\label{fig:translation:overview}
\end{figure}

Translating a \scoop program to \cpmo consists of a number of steps, as
depicted in Figure~\ref{fig:translation:overview} where the tool progresses
from top to bottom. In the first step, \scoop source files are parsed and
syntax trees are generated. Using these syntax trees, an internal
representation of the program is created in two steps: First the syntax trees
are walked to gather typing information of features and variables. Then, in a
second pass through the syntax tree, we use typing information to create a
structure that closely relates to the \cpmo type graph.  In the final two
steps, the intermediate representation is transformed to a simple graph
representation, which also contains layout information. Finally, this graph
can be traversed and rendered as an \xml file that can be used in the \cpmo
transformation system.

The tool is implemented in Java and uses a number of libraries, namely the
following.
\begin{description}
	\item[\glsunset{antlr}\gls{antlr} 4.4] \glsreset{antlr}\gls{antlr} is a parser
		generator that, given a grammar in \glsreset{ebnf}\gls{ebnf},
		generates a lexer and parser in Java. The created classes offer a large
		amount of flexibility and implement the visitor pattern, providing a
		natural way to traverse the parse tree.  While this is possible with other
		parser generators as well, the modern and clean nature of the generated
		classes have convinced us to use \gls{antlr} for this project.
	\item[JUnit] The JUnit framework is used to
		automatically test various aspects of the implementation.
	\item[Apache Commons] The commons libraries offer a wide variety of reusable
		software components. This project makes use of the mathematics features,
		in particular for case studies and evaluation purposes.
	\item[\groove] Not only does \groove provide graphical and command-line
		interfaces, but it can also be used as a library in custom software. We
		use the library to perform exploration and verification from within our
		toolchain. This enables us to create more specific output tailored to
		\cpmo as opposed to the generic \gts output provided by the command-line
		interface of \groove.
\end{description}

What follows are more detailed technical descriptions
of the individual steps.

\section{Translating Programs}

With the help of \antlr, the first step of parsing consists of writing a
grammar in the \antlr grammar format. We did not write this from scratch;
instead we adapted the grammar found in EVE \cite{eve} for this usage. During
this process, we modified the grammar to conform to the \antlr file format.
Given this grammar, \antlr is able to generate a lexer and a parser which can
be used in our tool.

It is important to note that at this stage, we consider all \scoop programs.
This means that we are able to parse programs with more advanced features,
such as inheritance and generics. The decision whether a program is
translatable or whether it contains unsupported features is performed when
inspecting the syntax tree (steps
2a and 2b
in Figure~\ref{fig:translation:overview}).

To perform step 2a,
the tool uses a class that implements the \scoop
syntax tree visitor. It keeps track of the class currently inspected and
stores the following type information for each class:
\begin{itemize}
	\item Declared routines and their parameter types and return types (if any).
	\item Declared attributes and their types.
	\item A list of creation procedures.
\end{itemize}
It is necessary to record this typing information in advance, as we need to
know the types of symbols in the next step. In a single-pass approach, it is
possible to encounter symbols from classes that have not yet been analysed,
therefore making it impossible to know whether it is an integer, Boolean, or
reference symbol.

After having gathered the types of declarations, we pass through the parse
tree once again. This time, we create a number of \verb|CPMOGraph| objects,
one for each parsed class. The \verb|CPMOGraph| class and its subclasses are
closely related to the \cpmo type graph. In fact, most types in \cpmo have a
direct representation as a subclass of \verb|CPMOGraph|. For example, we use a
class \verb|BoolConstant| that inherits from \verb|BoolOp|, which in turn
inherits from the class \verb|Op|. Similarly, in \cpmo the
\verb|BoolOp_Constant| type is a subtype of \verb|BoolOp|, which in turn is a
subtype of \verb|Super_Op|.  This direct correlation is useful in a variety of
ways. In particular, translating a \verb|CPMOGraph| to a \cpmo start graph is
straightforward, as we can simply go through the structure and
create a \cpmo graph node (in a generic graph representation) for each
encountered \verb|CPMOGraph| object.
This also means that the ``compiler effort'', i.e.\
relating source code statements (in the form of parts of the syntax tree) to
\cpmo nodes is concentrated in one step, namely the visitor class that
implements step 2b. This second visitor also handles
situations of input programs that use features currently not supported. If our
tool encounters such a feature in step 2b, it either ignores it (in cases
where the feature does not influence the program execution, e.g.\ a \eif{note}
block at the top of a class) or aborts the translation and prints the part of
the source code that resulted in the tool to fail. This way, we have a single
point in the tool where the decision is made whether a program is supported,
and only one point at which translation can fail due to the nature of the
input program.

\section{Supported SCOOP Features}
\label{section:translation:supported_unsupported_features}

For correct translation and simulation in \cpmo, we require a complete \scoop
program to be passed as a set of input files. In particular, all referenced
classes must be part of the input. A number of other restrictions on the input
programs apply for the tools to function correctly. In this section, we give
an overview of the supported features of \scoop and discuss the parts that are
missing.

The translation tool focuses on the basic features of \scoop. The goal is not
to support the complete \scoop language, but enough of the language to allow
writing expressive programs in an object-oriented manner. The following
features are currently supported.
\begin{description}
	\item[Classes and Objects] \cpmo supports objects and classes natively. The
		translation creates object templates for each input class consisting of the
		class name and the names and types of its attributes.
	\item[Feature declarations] Both routines (with the \eif{do} keyword) and
		attributes are translated. While attributes are part of the class
		template, we also create a getter routine (consisting of simply assigning
		the attribute to \eif{Result}) for each attribute. This makes it possible
		to create \grbf{Queue\_Items} that call the getter functions instead of
		accessing the data from other processors directly (which would cause \cpmo
		to misbehave, as the requests would not be served in FIFO order anymore).
	\item[Routine declartions] In routines, we support common constructs such as
		formal arguments, preconditions, wait conditions, and postconditions.
	\item[Local declarations] Local variables of reference, integer, and Boolean
		type are supported. Integer and Boolean types are special native types in
		\cpmo and behave like expanded types. Generic support for expanded types
		is currently not available.
	\item[Instructions] A number of instructions are supported, namely:
		\begin{itemize}
			\item Creation calls with the \eif{create} keyword and an explicit
				creation procedure.
			\item Local calls (with target \eif{Current}).
			\item Qualified calls.
			\item Assignment instruction.
			\item \eif{if-then-else} conditional.
			\item Loop instruction.
			\item Integer and Boolean literals.
		\end{itemize}
	\item[Expressions] We support arbitrarily complex expressions, which are
		translated to a single \grbf{Op} node in the output graph. As a
		consequence of leaving out a number of features such as agents,
		expressions related to those features are not supported (e.g.\ agents
		inside expressions).
\end{description}

Both \cpmo and the translation tool currently lack support for a number of
features. Most prominently, we do not support inheritance. With this, a number
of related \scoop features are not supported either, for example partial
classes, the redefinition of features, arrays, generics, agents, and others.
In addition, we currently leave out a number of other language features, such
as class invariants, old values in postconditions, and others. In
Section~\ref{section:translation:future_work}, we give an overview of the most
important features currently missing and present possible implementation
strategies as future work.

\section{Output}

Once the intermediate representation of the input files is generated (in the
form of \verb|CPMOGraph| objects), the remaining task is outputting the
representation as a \gxl file. To achieve this, we use the visitor pattern
once again: The interface \verb|CPMOGraphVisitor| allows implementing classes
that pass through the \cpmo structure. This is used to create a simple graph
representation using the \verb|output.graph.Graph| class and its subclasses.
These classes implement a straightforward graph representation with nodes and
directed edges. In addition, nodes can also store position values. This allows
us to create start graphs that are ``human readable'' when rendered in
\groove. In particular, we organise routine subgraphs by aligning states and
actions from left to right, while attaching additional nodes, like parameters
and target operations, above them.

Using two separate steps to output a \verb|CPMOGraph| structure to \xml may
seem unnecessary, as we could as well just have generated the \gxl file
directly. But using a separate simple graph representation has the advantage
that we can separate the tasks of creating graphs with layout information and
rendering them in some format (in our case \gxl). This leaves more flexibility
when extending the program, for example when we want to render the graph in
another output format we can traverse a simple graph structure with edges and
positioned nodes. When extending the \cpmo model, we simply have to adjust the
part that generates a graph from \verb|CPMOGraph| objects, but do not have to
adjust anything related to the \gxl output. This leaves us with a well
structured design that cleanly separates concerns and can easily be extended
at various stages.

\section{Testing}

As briefly mentioned in Section~\ref{section:cpmo:testing}, we test our
translation tool in conjunction with the model by providing \scoop programs,
translating them, and exploring their \statespaces. The output is matched
against the expected output using the JUnit framework. With this approach, the
start graph is implicitly tested against the type graph presented in
Section~\ref{section:cpmo:type_graph}. By assuming that the model behaves
correctly at this point, we can test the translation tool by simulating the
generated start graph and analysing the output. In case the output does not
match, we most likely have an error in the translation tool.

We do realise that this is hardly ``unit testing'' in the traditional sense,
instead we test the toolchain as a whole. While this may be suboptimal in
general, we are confident that it is sufficient for the size of this project
and due to the fact that this is a prototype implementation. In addition, we
develop only a single part of the toolchain at a time, i.e.\ we either change
the translation tool or the \cpmo model, which allows us to check the
influence of the changes on the final output.

To make sure that we catch the expected behaviours when translating and
modelling, we use a wide range of test input programs and specify the expected
behaviour. This includes small programs that focus on certain features, e.g.\ 
ones that use a wide range of available query types, as well as larger
example programs that resemble real programs, such as the ones used in the
case studies in Chapter~\ref{chapter:case_studies}.

\section{Future Work}
\label{section:translation:future_work}

In Section~\ref{section:cpmo:future_work}, we briefly discussed features
missing from the \cpmo model, and in
Section~\ref{section:translation:supported_unsupported_features} we named some
\scoop features that are not handled in the translation tool. In this
section, we propose ideas to how certain features could be implemented in the
future. Since this not necessarily only affects the translation to the \cpmo
model, but may require changes in the model itself, we discuss possible
changes to the \cpmo model as well.

In general, supporting additional features can be tackled by either extending
the compiler to translate to the current \cpmo model, which means that the
feature is simulated using more primitive \cpmo constructs, or by
extending \cpmo itself by adding direct support of these features. The
advantage of the latter is that program representations become easier to read
and understand, and a more direct translation can be made from source code to
start graph. While this is a desirable outcome, it also requires careful
reasoning about the model changes, something one can avoid if only the
compiler is extended.

\subsection{Inheritance}

The most important feature towards supporting more complex \scoop programs is
inheritance. The main difficulty in supporting inheritance is the complexity
and feature-richness of the semantics related to inheritance. \scoop offers a
wide range of mechanisms, such as multiple inheritance, redefining,
undefining, and renaming of features, partial classes, and others. As a
result, adding these features to either the translation tool or \cpmo requires
careful analysis of the underlying semantics. Identifying and isolating
individual parts of the inheritance mechanisms and modelling them one by one
(where possible) seems to be the right approach to tackle this task, which
allows us to be confident in the resulting model.

To implement simple inheritance (i.e.\ using the \eif{inherit} keyword), one
strategy would be to ``unfold'' the inheritance structure during translation.
This means that for a class \eif{FOO} that inherits feature \eif{baz} from
class \eif{BAR}, we simply create the feature \eif{baz} for both classes (in
fact, it would suffice to have a feature with two init state nodes, one for
\eif{FOO.baz} and one for \eif{BAR.baz}).  Whether this is a feasible approach
remains to be evaluated.

Extending \cpmo for handling simple inheritance is another possibility.
Implementing the semantics directly would require representing the inheritance
structure in the start graph. When performing queries and commands, rules
would then need to first determine the dynamic type of the target object and
based on this traverse the inheritance structure and select the correct
feature to be applied.

\subsection{Expanded Types}

In \cpmo, we only support integer and Boolean expanded types. A more general
approach would distinguish between expanded types and normal (reference)
types. Supporting expanded types is an important step towards full support of
\scoop, but will require considerable effort and requires extending the \cpmo
model, as expanded types are treated different than normal types in the \scoop
semantics \cite{Morandi2014}, and adding them to \cpmo has implications on
existing parts of the model.

\subsection{Miscellaneous}

A number of other features are currently not supported by our toolchain. This
includes more exotic features of \scoop like non-object calls, assigner calls,
but also basic features like character and floating point number literals or
class invariants. Adding these features to \cpmo have currently lower
priorities as opposed to inheritance and expanded types, but will be
considered once the above is implemented properly.

\myclearpage

\chapter{Case Studies \& Evaluation}

\label{chapter:evaluation}\label{chapter:case_studies}

In the previous two chapters, we discussed the main contribution of this work
that allows us to automatically map a subset of \scoop to \cpmo start graphs
and to verify \statespace properties using \groove.
One part of the motivation behind this work is to provide a ``one-click''
solution that verifies certain properties---e.g.\ deciding whether a deadlock
can occur---for a given input program written in that \scoop subset.

In this chapter, we inspect various \scoop programs as case studies, show how
they are translated to a \cpmo graph using our toolchain, and show the
properties we can verify. We provide metrics for the programs to show how the
model behaves in various situations and present insights about the gained
verification results. We compare our toolchain to \cpm and discuss the
obtained results, before we close this chapter with an outlook on future work.

We use the following abbreviations to denote program configurations in this
chapter.
\begin{description}
	\item[DP($n$, $m$, \{\gr{eat}, \gr{bad\_eat}\})] Dining philosophers with
		$n$ philosophers and $m$ rounds, as presented throughout this thesis. The
		last parameter indicates which implementation is used, where \gr{eat}
		denotes the correct implementation and \gr{bad\_eat} the implementation
		that can result in deadlock. This program is presented as a case
		study in Section~\ref{section:case_studies:dining_philosophers}.
	\item[DS($n$, $m$, $o$, \{\gr{bad}, \gr{good}\})] Dining savages with pot
		size $n$, $m$ savages, and $o$ hunger per savage. The final parameter
		indicates which implementation is used, where \eif{bad} is the one that
		can result in savages being stuck, and \gr{good} the one that always
		terminates. This program is presented as a case study in
		Section~\ref{section:case_studies:dining_savages}.
	\item[CS($n$)] Cigarette smokers problem with $n$ rounds. In this problem, a
		number of cigarette smokers require different ingredients to build
		cigarettes, which are provided by a dealer. This program is discussed as
		a case study in~\ref{section:case_studies:cigarette_smokers}.
	\item[SEPC($n$)] Single-element producer/consumer with $n$ rounds. In this
		program, a producer and a consumer are created. The producer creates $n$
		items that are consumed by the consumer. The producer has a buffer of size
		1, which means that \eif{produce} and \eif{consume} calls have to
		alternate.
	\item[Counter($n$, $m$)] Counter with $n$ counters and $m$ counts per
		counter. This is a simple program that spawns a number of counters ($n$),
		which simply perform the task of incrementing an integer from 0 to $m$.
		While it does not require any synchronisation, it is a small and easy to
		understand example that showcases \scoop features.
	\item[BS($n$, $m$, $o$)] Barbershop with $n$ customers, $m$ chairs, and $o$
		haircuts per customer. This program solves the barbershop
		problem, where a barber serves several customers. The barber can only cut
		one customer's hair at a time. Luckily, a waiting room with a number of
		chairs is available, where customers can wait. Customers can come in the
		barbershop and take a seat in the waiting room if there is an unoccupied
		chair. Otherwise, they leave and come back later.
\end{description}
While the cigarette smokers program is our own implementation, the others are
taken from the EVE source code repository \cite{eve} and adapted to match the
input specification of our toolchain.

The above programs make up the main part of the benchmark programs that we
used during development of \cpmo and the translation tool. In addition, we
have a number of smaller programs that focus on a certain aspects of the model
and translation (e.g.\ one program provides a wide range of statements
involving queries).

\section{Setup}

The values presented in this chapter have been, if not otherwise stated,
obtained using the latest revision of the tools, as described in
Chapters~\ref{chapter:cpm},~\ref{chapter:cpmo} and~\ref{chapter:translation}.
We investigate how the system behaves with different setups (e.g.\ disabling
the token passing mechanism for \statespace reduction) and use the following
two main configurations.
\begin{description}
	\item[Default] Here, all optimisations are turned on, and all rules (in
		particular error rules) are enabled. Pre- and postconditions are checked
		as well.
	\item[No token optimisation] In this configuration, we disable the token
		optimisation, giving all actions and query operations the same priority.
		Error rules and pre- and postcondition checking are still enabled.
\end{description}

Values presented in this Chapter represent the median of five (where
applicable), and are obtained from a workstation with an Intel Core i7-4810MQ
CPU and 16 GB main memory. Runtimes and memory usage are obtained using Java
library classes (for \cpmo measurements) and GNU time 1.7 (for \cpm
measurements in Table~\ref{table:evaluation:cpm_cpmo_comparison} only).

\section{Case Studies}

In this section, we take a closer look at three programs.

We start by we
revisiting the dining philosophers problem one last time and show how the
implementation behaves using our toolchain. We present the implementation and
(parts of) the generated start graph, before discussing evaluation results. In
addition, we compare \cpmo to \cpmo without token optimisations, and we show
how a deadlock can be detected in the bad implementation.

In the second case study, we present the dining savages problem and again
inspect two implementations, where the ``bad'' implementation does not behave
as expected. We point out how our toolchain is not able to detect certain
undesired behaviours.

Finally, we present the cigarette smokers problem, where we show our
implementation and discuss results obtained using both full \statespace
exploration and \ltl formula checking.

\subsection{Dining Philosophers}
\label{section:case_studies:dining_philosophers}

In this section, we conclude our running example by presenting an
implementation in \scoop that is in the subset of programs supported by our
translation tool. We discuss parts of the start graph and take a closer look
at how the program is simulated. Finally, we show how we can detect problems
with the implementation, in particular, we show by example how one can detect
a deadlock situation.

\subsubsection{Source Code}

Listings~\ref{listing:evaluation:dp_application},~\ref{listing:evaluation:dp_philosopher},
and~\ref{listing:evaluation:dp_fork} show the three classes \eif{APPLICATION},
\eif{PHILOSOPHER}, and \eif{FORK} that make up the full working example, with
\eif{APPLICATION.make} as the root procedure. The philosopher not only
contains a correct implementation of the \eif{eat} feature (which gets called
on line 51 in Listing~\ref{listing:evaluation:dp_philosopher}), but also an
implementation called \eif{bad_eat} which, when called instead of \eif{eat},
can result in deadlock. When discussing the analysis we take a look at how
\cpmo handles both variants.

\eiffelfile[
		label={listing:evaluation:dp_application},
		caption={\eif{APPLICATION} class.},
		escapechar=\#,
		inputencoding=latin1
	]{listings/case_studies_dp_application.e}

\eiffelfile[
		label={listing:evaluation:dp_philosopher},
		caption={\eif{PHILOSOPHER} class.},
		escapechar=\#,
		inputencoding=latin1
	]{listings/case_studies_dp_philosopher.e}

\eiffelfile[
		label={listing:evaluation:dp_fork},
		caption={\eif{FORK} class.},
		escapechar=\#,
		inputencoding=latin1
	]{listings/case_studies_dp_fork.e}

In the root procedure (\eif{APPLICATION.make}), we create three philosophers
and forks between each pair of adjacent philosophers. The philosophers get
initialized with an identifier, the two separate forks they need to pick up,
and a round count value, indicating how often they need to eat before
terminating. Note that philosophers are started using the call
\eif{launch_philosopher (a_philosopher)}. This is required since
\eif{a_philosopher} is of separate type and must be controlled.  Passing them
as an argument makes it controlled in the called feature, where we are allowed
to make the call \eif{philosopher.live}. The code of the philosopher's
\eif{make} procedure also uses pre- and postconditions, which we can inspect
later through model checking by our verification tools. Note that the
preconditions are not wait conditions, as no involved variables are of
separate type.

\subsubsection{Start Graph}

We do not include the complete generated start graph here---with 287 nodes and
789 edges it is too large to print. Instead, we refer to \cite{onlinerepo},
where one can find the \cpmo \gts and the start graph
\gr{dining\_philosophers\_3\_phi\-los\-o\-phers\_1\_round\_eat}, which represents
this instance. We focus on highlighting several interesting parts of the graph
and its behaviour under \cpmo here.

Figure~\ref{fig:evaluation:dp_start_graph_philosopher_live} shows the
\eif{live} procedure of the philosopher (nodes have been rearranged manually
for improved readability, but note that the translation tool already performs
basic positioning of the graph nodes). The feature starts at node $n5$. The
first statement in the feature (\eif{until times_to_eat < 1}) is represented
as a pair of Boolean operations with nodes $n9$ and $n11$ and corresponding
\grbf{Action\_Test} nodes that implement the branching. Following the path to
$n19$ means that \eif{times_to_eat < 1} and consequently we end up in node
$n2$, the final state. Otherwise, the other branch is followed where the loop
body is implemented with state nodes $n14$, $n18$, and $n15$. Inside the loop,
two actions represent the statements \eif{eat (left_fork, right_fork)} and
\eif{times_to_eat := times_to_eat - 1}. State $n15$ leads to two test actions
evaluating the \eif{until} part.

\begin{sidewaysfig}
	\centering
	\begin{sidewaystikzbox}
		\includegraphics{figures/evaluation_dp_start_graph_philosopher_live.tikz}
	\end{sidewaystikzbox}
	\caption[\eif{PHILOSOPHER.live} start graph]{Representation of \eif{PHILOSOPHER.live} in the start graph.}
	\label{fig:evaluation:dp_start_graph_philosopher_live}
\end{sidewaysfig}

The creation procedure in the philosopher class, shown in
Figure~\ref{fig:evaluation:dp_start_graph_philosopher_make}, initializes a
number of attributes. In addition, it contains pre- and postconditions which
validate the passed arguments and ensure that the attributes have been
successfully set.  In the graph, the procedure starts at node $n51$ with an
init state. First, the handlers of the separate arguments are locked with an
\grbf{Action\_Lock} node. The following test actions represent preconditions.
Note that nodes $n43$ and $n6$ have a flag \grit{precondition\_fail}. This
denotes that these actions represent the path that is taken when a pre- or
wait condition fails. This does not necessarily mean that the tested statement
is a precondition. Depending on the objects included in the test and whether
their handlers are controlled or not, they are either preconditions or wait
conditions. This can only be detected at runtime, which is handled by the
rule \gr{action\_Test}. When the rule is applied with an action with the
\grit{precondition\_fail} flag and it turns out that it is in fact a
precondition, then the rule creates an \grbf{ERROR} node, which has the effect
that the system is immediately in a final state, which can be analysed by the
postprocessing tools.

Once a processor is in state $n56$, the method body gets executed. When the
processor reaches node $n50$, there are two possibilities: Either
postcondition checking is disabled, in which case the processor returns to the
calling procedure or becomes idle, or postcondition checking is enabled, in
which case the edge to $n20$ is followed.

\begin{sidewaysfig}
	\centering
	\begin{sidewaystikzbox}
		\includegraphics{figures/evaluation_dp_start_graph_philosopher_make.tikz}
	\end{sidewaystikzbox}
	\caption[\eif{PHILOSOPHER.make} start graph]{\eif{PHILOSOPHER.make} start
	graph.}
	\label{fig:evaluation:dp_start_graph_philosopher_make}
\end{sidewaysfig}

\subsubsection{Results}

With the start graph discussed earlier, we can now verify different properties
of the dining philosophers instance. We are not only interested in whether a
deadlock can occur, but we also want to make sure that we never have a
call where the target is a void reference, or that postconditions never fail.

We start with the default configuration with optimisations enabled.
Table~\ref{table:evaluation:dp_results_default} shows results with varying
numbers of philosophers and rounds (number of times each philosopher eats) for
both implementations (\eif{eat} and \eif{bad_eat}). To obtain these results,
we used breadth-first search and explored the full state space. Example output
from our command-line tool for an instance with the \eif{bad_eat}
implementation looks as follows.
\begin{eiffelcode}[language={}]
SCOOP sources: dining_philosophers/dining_philosophers_2_philosophers_1_round_bad_eat
Exploration type: TERMINATION
Start graph size (#nodes / #edges): 326 / 494
Final graph size (#nodes / #edges): 382 / 650 (0.00, 0.00)
Median States: 1282 (0.00)
Median Transitions: 1309 (0.00)
Median wall clock time: 1.38 (0.57)
Median total memory used: 539,321,016 (102,715,198.36)
Median new memory since start: 2,618,731 (204,233.22)
Min/max result states: 2/2
Min/max final states: 2/2

The simulation generated an error node with label: "Deadlock detected".
\end{eiffelcode}
Our tool explores the complete state-space and inspects final graphs. If there
are nodes of type \grbf{ERROR}, the associated information is fetched and
reported. While this kind of output is rather rudimentary, one can use \groove
to save the offending traces and inspect the program execution to find the
root cause of the error.

In cases where no error node is present, our tool first checks whether there
are \gr{in\_method} edges in final states, which means that there are
processors still executing code while no rule can be applied any more. This
indicates that the program is stuck. This situation should not arise, as it
means that the program is stuck without a corresponding error rule, which
can be either a bug in our implementation, or an error situation we did not
define and capture with a rule yet.

Inspecting the numbers in Table~\ref{table:evaluation:dp_results_default}
reveals that we are able to verify deadlock freedom, absence of pre- and
postcondition failures (although only a limited number of such statements are
in the source code) for the correct implementation with up to seven
philosophers, which requires less than \num{150000} states and transitions.
The runtimes are reasonable, with most instances being evaluated in less than
a minute. The numbers for the \eif{bad_eat} implementation are substantially
larger. Due to the fact that locking of the forks is not atomic anymore, more
interleavings are possible. Even with the smallest instance, this results in
an increase of roughly $40\%$ in the amount of states and transitions. With
larger instances, the effect is even bigger, with the one with seven
philosophers having a \statespace of almost \num{3000000} states.  The runtime
of roughly \num{85} minutes is substantially longer than the runtime of the
corresponding correct implementation.

If one is only interested in detecting whether a deadlock can occur or not, an
alternative approach is to use \ltl formula exploration instead of full
\statespace exploration. In this approach, one can instruct \groove to try and
find a counterexample to an \ltl formula. To detect a deadlock, we can use the
formula \gr{! F error\_deadlock}.
Table~\ref{table:evaluation:dp_results_deadlock_detection} shows the
corresponding results. As we can see, the numbers of explored states and
transitions for the correct implementation do not differ from
Table~\ref{table:evaluation:dp_results_default}, as, in order to prove that no
counterexample exists, one has to explore the full \statespace.
For the bad implementation on the other hand, the number of explored states
and transitions are substantially smaller. While this may seem like an
improvement at first,
taking a look at the runtimes reveals that checking the formula comes at a
cost. While the smaller instances of the correct implementation take roughly
the same time in both cases, using \ltl exploration takes longer with the
larger instances. In the bad implementation, finding a counterexample is
faster for small instances, but with larger ones, it takes longer to check the
formula, even though fewer states are explored. In the case with 6
philosophers, finding a counterexample requires (on average) only \num{441416}
states compared to the \num{662009} states of the full \statespace.
Nevertheless, finding the counterexample takes more than twice the time as
compared to exploring the full \statespace.

\begin{table}
	\centering
	{\scriptsize
	\begin{tabular}{rrrrrrr}\toprule
		\textbf{$n$} & \textbf{$i$} & \textbf{Impl.} & \textbf{States} &
		\textbf{Transitions} & \textbf{Time [stddev] (s)} & \textbf{Memory
		[stddev] (GB)}\\
\midrule
2 & 1 & eat & \num{962} & \num{1019} & \num{1.26} [\num{0.49}] & \num{0.59} [\num{0.09}] \\
3 & 1 & & \num{2976} & \num{3134} & \num{3.08} [\num{0.70}] & \num{0.67} [\num{0.18}] \\
3 & 2 & & \num{7974} & \num{8662} & \num{8.23} [\num{0.67}] & \num{0.96} [\num{0.20}] \\
3 & 3 & & \num{16208} & \num{17836} & \num{15.89} [\num{0.70}] & \num{1.96} [\num{0.41}] \\
3 & 5 & & \num{45264} & \num{50410} & \num{45.98} [\num{0.87}] & \num{3.16} [\num{0.72}] \\
4 & 1 & & \num{8326} & \num{8720} & \num{9.38} [\num{0.84}] & \num{1.29} [\num{0.27}] \\
5 & 1 & & \num{21814} & \num{22748} & \num{26.18} [\num{0.87}] & \num{2.98} [\num{0.91}] \\
6 & 1 & & \num{54638} & \num{56788} & \num{75.14} [\num{1.03}] & \num{4.23} [\num{0.27}] \\
7 & 1 & & \num{132518} & \num{137372} & \num{202.84} [\num{3.82}] & \num{5.51} [\num{0.41}] \\
\midrule
2 & 1 & bad\_eat  & \num{1358} & \num{1423} & \num{1.70} [\num{0.44}] & \num{0.49} [\num{0.07}] \\
3 & 1 & & \num{6528} & \num{6888} & \num{6.69} [\num{0.43}] & \num{0.99} [\num{0.19}] \\
3 & 2 & & \num{21130} & \num{22372} & \num{22.12} [\num{1.31}] & \num{1.91} [\num{0.42}] \\
3 & 3 & & \num{47859} & \num{50759} & \num{48.81} [\num{1.69}] & \num{3.82} [\num{0.76}] \\
3 & 5 & & \num{150471} & \num{159855} & \num{155.88} [\num{3.26}] & \num{5.17} [\num{0.24}] \\
4 & 1 & & \num{31105} & \num{32961} & \num{37.89} [\num{1.20}] & \num{3.11} [\num{0.88}] \\
5 & 1 & & \num{144891} & \num{154116} & \num{187.63} [\num{2.51}] & \num{5.25} [\num{0.24}] \\
6 & 1 & & \num{662009} & \num{706430} & \num{963.65} [\num{10.73}] & \num{9.36} [\num{1.04}] \\
7 & 1 & & \num{2972519} & \num{3181087} & \num{5001.01} [\num{21.04}] & \num{12.46} [\num{0.10}] \\
		\bottomrule
	\end{tabular}
	}
	\caption[Dining philosophers results]{Results from full state-space exploration of various instances of
		the dining philosophers program, where $n$ denotes the number of
		philosophers and $i$ the number of times each philosopher eats.}
	\label{table:evaluation:dp_results_default}
\end{table}

\begin{table}
	\centering
	{\scriptsize
	\begin{tabular}{rrrrrrr}\toprule
		\textbf{$n$} & \textbf{$i$} & \textbf{Impl.} & \textbf{States} &
		\textbf{Transitions} & \textbf{Time [stddev] (s)} & \textbf{Memory
		[stddev] (GB)}\\
\midrule
2 & 1 & eat & \num{962} & \num{1019} & \num{1.01} [\num{0.30}] & \num{0.64} [\num{0.14}] \\
3 & 1 & & \num{2976} & \num{3134} & \num{3.21} [\num{0.78}] & \num{0.81} [\num{0.15}] \\
3 & 2 & & \num{7974} & \num{8662} & \num{8.20} [\num{0.74}] & \num{1.21} [\num{0.26}] \\
3 & 3 & & \num{16208} & \num{17836} & \num{17.18} [\num{0.98}] & \num{2.09} [\num{0.42}] \\
3 & 5 & & \num{45264} & \num{50410} & \num{60.23} [\num{1.60}] & \num{3.65} [\num{0.56}] \\
4 & 1 & & \num{8326} & \num{8720} & \num{8.94} [\num{0.75}] & \num{1.36} [\num{0.29}] \\
5 & 1 & & \num{21814} & \num{22748} & \num{28.35} [\num{0.88}] & \num{3.61} [\num{0.85}] \\
6 & 1 & & \num{54638} & \num{56788} & \num{102.30} [\num{1.79}] & \num{4.25} [\num{0.23}] \\
\midrule
2 & 1 & bad\_eat & \num{828} & \num{837} & \num{0.92} [\num{0.55}] & \num{0.46} [\num{0.09}] \\
3 & 1 & & \num{3549} & \num{3686} & \num{3.59} [\num{0.87}] & \num{0.80} [\num{0.16}] \\
3 & 2 & & \num{4718} & \num{4891} & \num{4.52} [\num{1.44}] & \num{1.02} [\num{0.20}] \\
3 & 3 & & \num{3950} & \num{4080} & \num{4.12} [\num{2.83}] & \num{0.94} [\num{0.18}] \\
3 & 5 & & \num{2192} & \num{2218} & \num{2.12} [\num{4.09}] & \num{1.25} [\num{0.24}] \\
4 & 1 & & \num{18549} & \num{19412} & \num{22.03} [\num{0.68}] & \num{2.96} [\num{0.78}] \\
5 & 1 & & \num{85059} & \num{89623} & \num{174.54} [\num{20.11}] & \num{4.23} [\num{0.05}] \\
6 & 1 & & \num{441416} & \num{467285} & \num{2150.79} [\num{112.63}] & \num{5.97} [\num{0.25}] \\
		\bottomrule
	\end{tabular}
	}
	\caption[Deadlock detection results]{Exploration of \ltl formula \gr{! F error\_deadlock}. Exploration
	with more than six philosophers has been aborted after more than
	\num{2500000} states.}
	\label{table:evaluation:dp_results_deadlock_detection}
\end{table}

\paragraph{Disabling Optimisations}

The above numbers are quite promising, as we not only can verify a minimal
dining philosophers program, but also instances with a larger number of
involved processors and rounds. Reducing the runtimes has helped
us immensely during development, as we can test changes made to the system almost
instantly, where we previously had to wait several minutes for a result.  The
story is quite different if we look at earlier revisions of \cpmo. The feature
that has the biggest impact is the optimisation using the execution token
which marks the processor that is allowed to execute sequential actions, and
where the system only processes separate actions once no processor can make
sequential progress any more. If we disable this functionality, we obtain the
numbers presented in Table~\ref{table:evaluation:dp_results_no_token}.
Verifying the instance with three philosophers and a single round already
results in more than \num{250000} states and \num{300000} transitions, a huge
difference compared to the numbers with the optimisation turned on.

The difference can be explained with the fact that without the token
mechanism, large chunks of the program get simulated over and over again in
different interleavings without affecting the outcome. For example, consider
the situation where a
processor is in state $n5$ of
Figure~\ref{fig:evaluation:dp_start_graph_philosopher_live} and has already
evaluated the arguments to the test action nodes. Without the token
optimisation, at this point all other processors could simulate their states
until they are finished. Another execution plan would first advance our
processor to node $n14$ (assuming $n9$ has been evaluated to true), before
simulating the remaining processors all over again. There is no value in
considering both variants, as the outcome of the test action, which is a
purely local step, does not depend on any outside properties of the system (in
particular, it does not depend on the states of other processors). With the
token mechanism, we force the system to take one particular path in these
situations and only allow branching at points where processors can potentially
interact with each other and where different outcomes can originate.

\begin{table}
	\centering
	{\scriptsize
	\begin{tabular}{rrrrrrr}\toprule
		\textbf{$n$} & \textbf{$i$} & \textbf{Impl.} & \textbf{States} &
		\textbf{Transitions} & \textbf{Time [stddev] (s)} & \textbf{Memory
		[stddev] (GB)}\\
		\midrule
2 & 1 & eat & \num{15480} & \num{18265} & \num{15.43} [\num{0.90}] & \num{1.97} [\num{0.32}] \\
3 & 1 & & \num{252112} & \num{304409} & \num{328.91} [\num{19.60}] & \num{6.15} [\num{0.58}] \\
3 & 2 & & \num{711640} & \num{877576} & \num{769.02} [\num{9.55}] & \num{9.18} [\num{1.13}] \\
3 & 3 & & \num{1526582} & \num{1903627} & \num{1690.66} [\num{29.32}] & \num{12.56} [\num{1.85}] \\
\midrule
2 & 1 & bad\_eat & \num{21236} & \num{24417} & \num{20.53} [\num{0.94}] & \num{2.63} [\num{0.49}] \\
3 & 1 & & \num{425983} & \num{499660} & \num{487.62} [\num{14.94}] & \num{7.87} [\num{0.78}] \\
3 & 2 & & \num{1445738} & \num{1710118} & \num{1579.87} [\num{19.29}] & \num{12.52} [\num{1.49}] \\
3 & 3 & & \num{3417959} & \num{4059490} & \num{4065.24} [\num{169.68}] & \num{12.60} [\num{0.19}] \\
		\bottomrule
	\end{tabular}
	}
	\caption[\cpm without token opitimisation results]{Results from verifying the correct implementation without token
		optimisation.}
	\label{table:evaluation:dp_results_no_token}
\end{table}

\paragraph{Detecting Deadlocks}

The ``eat'' implementation behaves as expected. The simulation is unable to
find any issues with it, in particular, we are not able to find a situation
where the program deadlocks. In this section though, we inspect the
alternative implementation of the philosopher's behaviour (i.e.\ the
\eif{bad_eat} method in Listing~\ref{listing:evaluation:dp_philosopher}).
Since we want to find out whether a deadlock can occur or not, we can do so by
using the \ltl formula \gr{! F error\_deadlock}, which states that, starting
from the start graph, there is no future state where the rule
\gr{error\_deadlock} matches. \groove explores the \statespace and reports
whether a counterexample exists for the formula. If so, we have a situation
where a deadlock occurs and we can inspect the trace that leads from the start
graph to that particular state.

When using the \eif{bad_eat} implementation, we can in fact find
counterexamples to the formula. Figure~\ref{fig:evaluation:dp_deadlock} shows
an excerpt of such a state with the involved processors, locks, and states.
Both processors are in the \eif{pickup_right} command and hold their left fork
(which is the other one's right fork). Each processor holds the lock to the
processor the other one is waiting for. As a result, no processor can make
progress, and we have a deadlock situation which is detected by the
\gr{error\_deadlock} rule.

\begin{sidewaysfig}
	\centering
	\begin{tikzbox}
		\includegraphics{figures/evaluation_dp_deadlock.tikz}
	\end{tikzbox}
	\caption[Deadlock situation with 2 philosophers]{Deadlock situation with 2 philosophers.}
	\label{fig:evaluation:dp_deadlock}
\end{sidewaysfig}

\subsection{Dining Savages}
\label{section:case_studies:dining_savages}

Our second case study is the dining savages problem. The premise is that there
are a number of savages that share a single pot that contains their food. A
cook can fill the pot, and each savage can get servings from it. Since the pot
is rather small, the number of servings is limited and only one savage can get
a serving at a time. If a savage is trying to get a serving when the pot is
empty, he notifies the cook to fill it up again, waits until the cook does is
job, and then gets his serving.

\subsubsection{Source Code}
Our implementation of the
program consists of four classes. Apart from the \eif{APPLICATION} class that
initializes and starts the system, there are classes for representing the
cook, a savage, and the pot. In our implementation, we have three
configuration variables, namely the pot size (number of servings the pot can
hold), the number of savages, and the hunger of a savage (which is the number
of servings a savage is going to take before terminating). The program first
creates all objects and then launches the savages.
Listing~\ref{listing:evaluation:ds_savage} shows relevant code of the
savage class. During the lifetime of a savage, it executes the \eif{live}
feature which is a simple loop that executes \eif{step} a number of times. In
a step, a savage calls \eif{fill_pot} which notifies the cook to fill the pot,
if necessary. The program continues, since \eif{fill_pot} can return even if
the pot is empty, as the command that can get called in the routine body is
asynchronous. Afterwards,
\eif{get_serving_from_pot} gets called. Since---in case the pot was empty or
has become empty in the meantime---we can not be sure whether the cook already
filled the pot, we use the wait condition \eif{not my_pot.is_empty}. This
ensures that the savage gets the serving from a non-empty pot. Once the wait
condition is satisfied, the savage has exclusive access to the pot, which
means that it is impossible for the pot to become empty before the savage can
call \eif{my_pot.get_meal}. The final command in a step of the savage is
eating, which simply decreases the hunger value to avoid having an infinite
loop in the \eif{live} procedure.

While the original implementation defines \eif{step} without an argument, we
pass the pot to this feature in our adapted implementation. We do this to
avoid processors being stuck in a wait condition that never gets fulfilled.
Consider the instance where the pot can hold one serving and two savages want
to eat only once. Without passing the pot to the \eif{step} feature call, the
following sequence can occur.
\begin{itemize}
	\item Savage 1 calls \eif{fill_pot}, sees that the pot is full and therefore
		does not ask the cook to fill it.
	\item Savage 2 calls \eif{fill_pot}, sees the same, and does not ask the
		cook either.
	\item Savage 1 calls \eif{get_serving_from_pot}, passes the wait condition,
		and returns. The pot is now empty. Savage 1 has finished its loop and does
		not execute anything any more.
	\item Savage 2 calls \eif{get_serving_from_pot}, but is stuck in the wait
		condition, as the pot is now empty. Since Savage 1 has finished,
		the pot will never get filled again and Savage 2 is stuck forever.
\end{itemize}
By passing the pot as an argument to the \eif{step} routine, we ensure that
for one savage in a single step, all operations involving the pot are executed
without interleaving requests from another savage. This makes the above
interleaving impossible and as a result, savages can not get stuck any more.
Note that the condition in \eif{get_serving_from_pot} is now a precondition,
as the request queue of the handler of the pot is already locked when
\eif{get_serving_from_pot} gets called.
We call this the ``good'' implementation, but also take a look at the
implementation where we do not pass the pot to \eif{step}, which we call
``bad''.

\eiffelfile[
		label={listing:evaluation:ds_savage},
		caption={Savage implementation.}
		escapechar=\#
	]{listings/case_studies_ds_savage.e}

Parts of the source code of the cook is shown in
Listing~\ref{listing:evaluation:ds_cook}. In the \eif{require} block of the
\eif{cook} feature, we use another wait condition to make sure that the pot is
in fact empty when the feature body gets executed.

\eiffelfile[
		label={listing:evaluation:ds_cook},
		caption={Cook implementation.},
		escapechar=\#
	]{listings/case_studies_ds_cook.e}

\subsubsection{Results}

Table~\ref{table:evaluation:ds_results_default} lists results obtained with
\cpmo for a number of instances of the dining savages program. The instances
range from two to four savages and two to six total calls to the
\eif{get_serving_from_pot} routine. Like in the dining philosophers case, the
number of involved processors has the largest impact. Even if we lower the
number of times a savage eats in the last instance (with 4 savages), the
\statespace is by far the biggest in both implementations. This is no
surprise, as with more processors, the number of synchronisation points (i.e.\
situations during the execution where multiple non-separate actions or queries
can be performed) increases, and individual synchronisation points may include
more processors, resulting in more branching in the \lts.

Comparing both implementations paints a similar picture as what we have seen
in the dining philosophers example. In the ``bad'' implementation, we perform
less restrictive locking, thus allow more possible interleavings. While the
impact of the smaller instances is negligible, it becomes obvious with the
larger ones. In the instance with 4 savages, the ``bad'' implementation
takes about three times longer than the ``good'' implementation.
Note that we do full \statespace exploration here, and our
tool does not report an issue with both implementations, i.e.\ no \grbf{ERROR}
nodes get generated. While savages can get stuck in wait conditions in the
``bad'' implementation, these situations are not deadlock situations. A
processor may never proceed past the wait condition, but it can make progress
in the sense that the wait condition is checked over and over again (as
requests are generated for and executed by the handler of the pot).  The
target processor is idle and can execute the requests from the stuck
processor.  In the \lts, this results in a local cycle of states, where there
is no path that ``breaks out'' from this cycle. Currently, we are unable to
detect such situations, and whether it is possible to detect them using \ltl
or \ctl formulae remains to be investigated in future work.

\begin{table}
	\centering
	{\scriptsize
	\begin{tabular}{rrrrrrrr}\toprule
		\textbf{n} & \textbf{m} & \textbf{o} & \textbf{Impl.} & \textbf{States} &
		\textbf{Transitions} & \textbf{Time [stddev] (s)} & \textbf{Memory
		[stddev] (GB)}\\
		\midrule
1 & 2 & 1 & good & \num{3365} & \num{3472} & \num{4.91} [\num{0.54}] & \num{0.70} [\num{0.09}] \\
4 & 2 & 2 & & \num{5710} & \num{5923} & \num{8.58} [\num{0.65}] & \num{0.99} [\num{0.18}] \\
2 & 2 & 2 & & \num{6121} & \num{6340} & \num{9.28} [\num{0.53}] & \num{0.88} [\num{0.16}] \\
2 & 3 & 2 & & \num{66592} & \num{70044} & \num{124.51} [\num{1.63}] & \num{4.51} [\num{0.51}] \\
2 & 4 & 1 & & \num{155578} & \num{165157} & \num{329.14} [\num{5.04}] & \num{5.75} [\num{0.42}] \\
\midrule
1 & 2 & 1 & bad & \num{4193} & \num{4396} & \num{7.07} [\num{0.66}] & \num{0.78} [\num{0.13}] \\
4 & 2 & 2 & & \num{8479} & \num{8999} & \num{13.35} [\num{0.62}] & \num{1.11} [\num{0.21}] \\
2 & 2 & 2 & & \num{9147} & \num{9668} & \num{14.39} [\num{0.82}] & \num{1.26} [\num{0.26}] \\
2 & 3 & 2 & & \num{178493} & \num{191810} & \num{346.94} [\num{2.16}] & \num{5.93} [\num{0.39}] \\
2 & 4 & 1 & & \num{431900} & \num{466498} & \num{962.53} [\num{12.65}] & \num{8.76} [\num{0.94}] \\
		\bottomrule
	\end{tabular}
	}
	\caption[Dining savages results]{Results of various instances of the dining savages program.}
	\label{table:evaluation:ds_results_default}
\end{table}

\subsection{Cigarette Smokers Problem}
\label{section:case_studies:cigarette_smokers}

In our final case study, we implement and evaluate the cigarette smokers
problem. In this problem, there are three cigarette smokers wanting to build
cigarettes and smoke them. Their problem is that each one has only one of the
required three ingredients, namely tobacco, matches, or papers.  Thankfully, a
dealer is available that has an infinite amount of each ingredient. The dealer
randomly makes two of them available at a time, allowing the smoker with the
third ingredient to retrieve them and then build and smoke a cigarette.

The original premise, which we borrow from
\cite{downey:littlebook}, states that both the dealer's supply as well as the
smoker's desire to smoke are infinite. We change this in order to get a
program that terminates. In particular, we now require that all smokers only
retrieve ingredients and smoke $n$ times.  In addition, the dealer puts out
each distinctive pair of ingredients exactly $n$ times. As a result, nobody is
stuck waiting, as once the dealer has put out every pair $n$ times, he can go
home, and all smokers are satisfied as they were able to build and smoke a
cigarette $n$ times.

\subsubsection{Source Code}

Our implementation of the problem consists of three main classes,
\eif{DEALER}, \eif{CLIENT}, and \eif{INGREDIENT_PAIR}. The dealer is a simple
class resembling a semaphore and is used to make sure that no two pairs are
available at the same time (which would imply that all three ingredients are
available, which is not allowed in the problem statement).
Listing~\ref{listing:evaluation:cs_dealer} shows the full source code of the
\eif{DEALER} class. Instead of using a class to represent individual
ingredients, we use one to represent pairs of ingredients. Since we want a
limited amount of each pair, we can use this representation to force each pair
a fixed number of times.  Ingredient pairs
(Listing~\ref{listing:evaluation:cs_pair}) are created with a separate dealer,
and the \eif{put_out} feature is called after creation. A pair then, if the
dealer is not busy, puts itself out, ready to be consumed by a client
(Listing~\ref{listing:evaluation:cs_client}). Once a pair is consumed via the
\eif{consume} feature, it either terminates, or tries to put itself out again.
A client gets initialized with a pair of ingredients, and simply calls
\eif{consume} $n$ times which
is blocked until its ingredient pair is actually out.

\eiffelfile[
		float,
		label={listing:evaluation:cs_dealer},
		caption={\eif{DEALER} class.},
		escapechar=\#
	]{listings/case_studies_cs_dealer.e}

\eiffelfile[
		float,
		label={listing:evaluation:cs_pair},
		caption={\eif{INGREDIENT_PAIR} class.},
		escapechar=\#
	]{listings/case_studies_cs_pair.e}

\eiffelfile[
		float,
		label={listing:evaluation:cs_client},
		caption={\eif{CLIENT} class.},
		escapechar=\#
	]{listings/case_studies_cs_client.e}

With pairs of ingredients as separate objects, we introduce the randomness
specified in the problem statement. The dealer serves as semaphore that can be
occupied by one pair at a time, ensuring that no two pairs are out at the same
time. The source code is, thanks to expressive wait conditions, rather simple
and clear.

\subsubsection{Results}

The generated start graph of the cigarette smokers program consists of more
than 400 nodes and 1100 edges.  We are confident that our implementation works
correctly, and therefore it is no surprise that no \grbf{ERROR} states are
generated when verifying instances of this program.
Table~\ref{table:evaluation:cs_results_default} shows results for various
instances of the program, where we varied the number of times each smoker and
ingredient pair execute their corresponding loops. In the top half of the
table, we explore the full \statespace and report on any \grbf{ERROR} nodes
generated in final states.  While the \statespace grows quickly, we
nevertheless are able to verify in a reasonable amount of time that instances
with up to 5 rounds do not exhibit any of the bad properties we are looking
for.

In the lower half, we used the \ltl formula \gr{!F error\_deadlock} to find
counterexamples for a deadlock error rule. Since we did not find such an error
in the full \statespace exploration above, it is no surprise that the
exploration does not find such a counterexample when using \ltl property
checking either. We can observe once again that formula checking comes at a
cost: all instances take more time in the lower half of the table compared to
their counterparts in the upper half. While we can argue that in these cases
we gain more information and are faster when exploring the full state space,
it is important to note that \ltl checking has value as well. In particular,
when looking at instances where we are no longer able to explore the full
\statespace, i.e.\ when having to rely on bounded model checking, looking for
counterexamples of properties is the only way to gain any valuable
information.

\begin{table}
	\centering
	{\scriptsize
	\begin{tabular}{rrrrrr}\toprule
		\textbf{Rounds} & \textbf{Type} & \textbf{States} &
		\textbf{Transitions} & \textbf{Time [stddev] (s)} & \textbf{Memory
		[stddev] (GB)}\\
		\midrule
1 & default & \num{69130} & \num{75013} & \num{191.99} [\num{4.56}] & \num{4.64} [\num{0.25}] \\
2 & & \num{269497} & \num{291593} & \num{755.23} [\num{16.03}] & \num{6.24} [\num{0.63}] \\
3 & & \num{602402} & \num{649519} & \num{1918.49} [\num{27.66}] & \num{8.57} [\num{0.73}] \\
4 & & \num{1101193} & \num{1184059} & \num{3084.29} [\num{46.23}] & \num{9.74} [\num{1.27}] \\
5 & & \num{1799218} & \num{1930481} & \num{5007.15} [\num{84.28}] & \num{12.45} [\num{1.06}] \\
\midrule
1 & deadlock & \num{69130} & \num{75013} & \num{233.25} [\num{2.62}] & \num{4.65} [\num{0.14}] \\
2 & & \num{269497} & \num{291593} & \num{1376.87} [\num{24.35}] & \num{5.66} [\num{0.25}] \\
3 & & \num{602402} & \num{649519} & \num{4025.06} [\num{54.42}] & \num{5.57} [\num{0.10}] \\
		\bottomrule
	\end{tabular}
	}
	\caption[Cigarette smokers problem results]{Results of various instances of the cigarette smokers program.}
	\label{table:evaluation:cs_results_default}
\end{table}

\section{Comparison with CPM}

In this section, we take a look at how various variants of our models perform.
In \cpmo, we introduced a number of abstractions and except considerable
overhead in the form of a larger \statespace, as compared to \cpm. To reduce
the \statespace size in \cpmo, we used several optimisations, in particular we
used more quantifiers in certain rules and we introduced the token execution
optimisation.

Since there is currently no translation tool that can generate start graphs
for \cpm, we have to create start graphs by hand. This makes translation of
real-world examples tedious and is error-prone. As a result, there are only a
limited number of start graphs for use with \cpm available. Most notably, we
have start graphs for the dining philosophers problem with both
implementations, as well as a start graph for single-element
producer/consumer. It is important to note that start graphs between \cpm and
\cpmo for the same program do differ substantially, making the comparison more
difficult. The graphs for \cpm are simpler in most regards: there are no local
calls, no evaluation of call targets (instead it is directly specified by the
name of an attribute), and other abstractions introduced in \cpmo are missing
as well.
Nevertheless, it is interesting to compare results of those two models.
Additional abstractions in \cpmo have led to less direct action applications,
which resulted in additional rules and interleavings, but using optimisations
has helped reduce the \statespace again.

Table~\ref{table:evaluation:cpm_cpmo_comparison} shows results obtained for
the dining philosophers and the single-element producer/consumer examples with
three models, namely \cpm, \cpmo, and \cpmo without token optimisation. In the
case of \cpm, we use start graphs translated and adapted by hand. In the other
cases, we use our \scoop reference implementations and generate start graphs
with our translation tool.

Comparing the numbers of \cpm to the ones of \cpmo without optimisations shows
that the size of the \statespace of the latter exceeds the size of the
\statespace generated with \cpm.  The DP(3, 2, bad\_eat) instance takes around
27 minutes to verify using \cpmo without the token optimisation, whereas the
same instance verified with \cpm takes less than 4 minutes. Compared to \cpmo
without optimisations, \cpm performs better across all instances. We explain
this with the above arguments, namely that \cpmo increases complexity and adds
more abstractions that require additional computations. In addition, the start
graphs of the \cpm instances have been generated by hand and are optimised for
these examples.

Fortunately, the token optimisation has a huge impact in the performance of
\cpmo. When enabling the optimisations, we can verify each instance in under
30 seconds.  Not only does this outperform \cpmo without optimisations by a
huge margin, but it is also considerably faster and generates smaller
\statespaces than \cpm with optimised start graphs.

We realise that comparing the models using start graphs that differ as much as
they do is not optimal. Still, in our opinion this comparison gives a good impression of
the effect of optimisations and shows that---although \cpmo is inherently more
complex than \cpm---we manage to not only preserve the size of generated
\statespaces, but thanks to optimisations we are even able to produce smaller
\statespaces focusing on the synchronization points of the programs.

\begin{sidewaystab}
	\centering
	{\scriptsize
	\begin{tabular}{rrrrrr}\toprule
		\textbf{Model} & \textbf{Program} & \textbf{States} &
		\textbf{Transitions} & \textbf{Time [stddev] (s)} & \textbf{Memory [stddev] (GB)}\\
		\midrule
		\acr{cpm} & DP(2, 1, bad\_eat) & \num{3923} & \num{4990} & \num{8.89}
		[\num{1.16}] & \num{1.11} [\num{0.08}]\\
		& DP(3, 1, bad\_eat) & \num{41347} & \num{54749} & \num{71.27}
		[\num{1.68}] & \num{3.51} [\num{0.56}]\\
		& DP(3, 2, bad\_eat) & \num{138059} & \num{180173} & \num{231.29}
		[\num{11.12}] & \num{4.93} [\num{0.04}] \\
		& DP(2, 1, eat) & \num{3404} & \num{4281} & \num{8.06} [\num{0.19}] &
		\num{1.10} [\num{0.01}] \\
		& DP(3, 1, eat) & \num{32155} & \num{41793} & \num{56.38} [\num{0.86}] &
		\num{2.70} [\num{0.87}] \\
		& DP(3, 2, eat) & \num{104131} & \num{133304} & \num{185.22} [\num{3.95}]
		& \num{4.80} [\num{0.06}] \\
		& SEPC(5) & \num{17864} & \num{21763} & \num{37.76} [\num{1.51}] &
		\num{2.81} [\num{0.64}] \\
		& SEPC(20) & \num{76949} & \num{93718} & \num{148.51} [\num{5.73}] &
		\num{4.71} [\num{0.09}] \\
		\midrule
\acr{cpmo} (no token) & DP(2, 1, bad\_eat) & \num{21236} & \num{24417} & \num{20.53} [\num{0.94}] & \num{2.63} [\num{0.49}] \\
& DP(3, 1, bad\_eat) & \num{425983} & \num{499660} & \num{487.62} [\num{14.94}] & \num{7.87} [\num{0.78}] \\
& DP(3, 2, bad\_eat) & \num{1445738} & \num{1710118} & \num{1579.87} [\num{19.29}] & \num{12.52} [\num{1.49}] \\
& DP(2, 1, eat) & \num{15480} & \num{18265} & \num{15.43} [\num{0.90}] & \num{1.97} [\num{0.32}] \\
& DP(3, 1, eat) & \num{252112} & \num{304409} & \num{328.91} [\num{19.60}] & \num{6.15} [\num{0.58}] \\
& DP(3, 2, eat) & \num{711640} & \num{877576} & \num{769.02} [\num{9.55}] & \num{9.18} [\num{1.13}] \\
& SEPC(5) & \num{106526} & \num{126392} & \num{152.03} [\num{3.30}] & \num{4.67} [\num{0.22}] \\
& SEPC(20) & \num{462221} & \num{549527} & \num{685.70} [\num{13.13}] & \num{7.07} [\num{0.90}] \\
		\midrule
\acr{cpmo} & DP(2, 1, bad\_eat) & \num{1358} & \num{1423} & \num{1.70} [\num{0.44}] & \num{0.49} [\num{0.07}] \\
& DP(3, 1, bad\_eat) & \num{6528} & \num{6888} & \num{6.69} [\num{0.43}] & \num{0.99} [\num{0.19}] \\
& DP(3, 2, bad\_eat) & \num{21130} & \num{22372} & \num{22.12} [\num{1.31}] & \num{1.91} [\num{0.42}] \\
& DP(2, 1, eat) & \num{962} & \num{1019} & \num{1.26} [\num{0.49}] & \num{0.59} [\num{0.09}] \\
& DP(3, 1, eat) & \num{2976} & \num{3134} & \num{3.08} [\num{0.70}] & \num{0.67} [\num{0.18}] \\
& DP(3, 2, eat) & \num{7974} & \num{8662} & \num{8.23} [\num{0.67}] & \num{0.96} [\num{0.20}] \\
& SEPC(5) & \num{2338} & \num{2412} & \num{3.70} [\num{0.90}] & \num{0.61} [\num{0.10}] \\
& SEPC(20) & \num{9088} & \num{9372} & \num{13.28} [\num{0.79}] & \num{0.98} [\num{0.18}] \\
		\bottomrule
	\end{tabular}
	}
	\caption[Performance comparison results]{Comparison of performance of \acr{cpm}, \acr{cpmo}, and \acr{cpmo} without token
	optimisation.}
	\label{table:evaluation:cpm_cpmo_comparison}
\end{sidewaystab}

\section{Scalability and Future Work}

So far, we have only considered input programs that resulted in \statespaces
that can be fully explored with our toolchain. In our development environment,
we have generated \glspl{lts} with up to \num{4000000} states, at which point
we ran out of memory. The number of states that can be explored depends on a
number of variables though. For example, the chosen exploration strategy can
be a factor, as well as the model and start graph. With larger programs where
full \statespace exploration is not feasible any more, one can consider doing
bounded verification, where one explores only parts of the \statespace. It is
important to note that with bounded verification, it is only possible to
search for instances of errors, but there is no guarantee that an error can be
found, and the absence of errors can not be proved.

\groove offers a range of different exploration strategies. So far, we have
used breadth-first-search and \ltl exploration. For larger \statespaces, this
may not be the optimal choice.  For example, when searching the \statespace
with breadth-first search, one can only reach a particular depth. This may be
undesirable, e.g.\ when the synchronization points that may result in a
deadlock occur only later in a program. By searching for counterexamples with
depth-first search instead, one may be able to actually reach such
synchronization points. Of course, with this approach, not all branches are
explored and one might as well miss the ones that result in a deadlock.
\groove also offers other exploration strategies, such as random linear
exploration, where exactly one path is followed per state, or conditional
exploration with restrictions on the number of edges and nodes in a state.
Custom exploration strategies tailored to \cpm and \cpmo should also be
considered.

A thorough investigation regarding bounded exploration and using various
exploration strategies to gain confidence in the obtained results is out of
scope of this thesis and remains to be done in future work.

\myclearpage

\chapter{Conclusion}
\label{chapter:conclusion}

In this chapter, we conclude the thesis. We start with a review of the
research hypothesis and summarise our efforts and
contributions, before we close the thesis with some final words on future
work.

\section{Contributions}

In Section~\ref{section:introduction:research_hypothesis}, we stated the
following research hypothesis.
\begin{displayquote}
	A subset of valid \scoop programs can be modelled using a graph
	transformation system. These programs can, without modification of the
	source code, be automatically translated to input graphs for the
	transformation system.  Using verification by model checking, it is possible
	to verify a number of properties such as absence of deadlock or absence of
	precondition violations for a given input program.
\end{displayquote}

In our opinion, this thesis satisfies the hypothesis with the following
contributions.

In Chapters~\ref{chapter:cpm} and~\ref{chapter:cpmo}, we described formal
models, implemented in \groove, that can be used to simulate a subset of
\scoop programs. In the former, we discussed \cpm, which focuses on the
concurrency features of \scoop. In the latter, we extend \cpm
by adding object-oriented features from \scoop to obtain the \cpmo model.
By careful (informal) reasoning in individual steps, we were able to preserve
confidence in the correctness and completeness of the model.

In Chapter~\ref{chapter:translation}, we presented a simple compiler that
takes a subset of valid \scoop programs as input and generates input graphs
for our formal model. While we are not able to support the complete \scoop
language, we were able to translate a number of real-world concurrent example
programs, as later discussed in Chapter~\ref{chapter:case_studies}. In
addition, by embedding the \groove binaries, we also created a simple
command-line interface that can be used to verify \scoop programs matching the
input specification with one single command. This supports the research
hypothesis, as the tool works on unmodified \scoop code.

We evaluated our approach in Chapter~\ref{chapter:case_studies} with several
case studies. We investigated a number of implementations of problems suited
for demonstrating concurrent programming, such as the well-known dining
philosophers problem, and we have shown how our translation tool and the
models behave. We discussed various aspects of the current version of the
\cpmo model, but also presented the effects of \statespace optimisations and a
comparison to \cpm. We have seen that our toolchain can verify properties like
absence of deadlock or absence of precondition violations for the inspected
programs, which supports the research hypothesis.

\section{Future Work}

With our tools and model, we are able to translate a number of \scoop programs
and can verify certain properties, such as the absence of deadlock scenarios.
While our input programs already use a number of object-oriented features of
\scoop, we are a long way from supporting the complete language. In
particular, both the translation tool and the model lack support for
inheritance. Our focus in the future is to extend the model and compiler to
support a larger subset of \scoop programs.

We provide a simple tool that works on \scoop source code and prints out
verification results with a single command. The tool outputs simple messages
that state which errors could be found. Ideally, the tool should be integrated
with the EVE~\cite{eve} integrated development environment, which combines a
number of other verification and analysis approaches. Ultimately, the goal
should be to provide a GUI interface that is intuitive and easy to use. The
output should be more verbose, extracting more information about the
situations that occur (e.g.\ stating which features processors are executing
when a deadlock is detected).

In this thesis, we have been using sample input programs that generate small
\statespaces that can be fully explored within minutes or hours. A thorough
investigation and evaluation of our work with respect to larger programs and
bounded verification remains to be done.

While this thesis focuses on the \scoop model, it may be possible to adapt our
approach to other concurrency languages and models, such as \gls{gcd}
\cite{GCD-Reference}. An evaluation remains as future work.

\myclearpage

\pagestyle{plain}

\printglossaries
\myclearpage

\phantomsection
\addcontentsline{toc}{chapter}{\listfigurename}
\listoffigures
\myclearpage

\phantomsection
\renewcommand{\lstlistlistingname}{List of Listings}
\addcontentsline{toc}{chapter}{\lstlistlistingname}
\lstlistoflistings
\myclearpage

\phantomsection
\addcontentsline{toc}{chapter}{\listtablename}
\listoftables
\myclearpage

\begingroup
\sloppy
\printbibliography[heading=bibintoc]
\endgroup

\end{document}